\documentclass[namedreferences]{SolarPhysics}
\usepackage[optionalrh]{spr-sola-addons} % For Solar Physics
\usepackage{epsfig}          % For eps figures, old commands
\usepackage{graphicx}        % For eps figures, newer & more powerfull
\usepackage{amssymb}        % useful mathematical symbols
\usepackage{color}           % For color text: \color command
\usepackage{url}             % For breaking URLs easily trough lines
\usepackage{rotating}
\usepackage{enumerate}
\usepackage{tabularx}
\usepackage[pdfborder={0 0 0 },urlcolor=blue,breaklinks]{hyperref}

\newcommand{\jwlemail}[1]{email: \href{mailto:#1}{\textsf{#1}}}

\begin{document}

\begin{article}

\begin{opening}

\title{Solar Energetic Particles and Associated EIT Disturbances in Solar Cycle 23}

\author{R.~\surname{Miteva}$^{1}$\sep
        K.-L.~\surname{Klein}$^{1}$\sep
        I.~\surname{Kienreich}$^{2}$\sep
        M.~\surname{Temmer}$^{2}$\sep
        A.~\surname{Veronig}$^{2}$\sep
        O.E.~\surname{Malandraki}$^{3}$
       }
\runningauthor{R.~Miteva {\it et al.}}
\runningtitle{Solar Energetic Particles and EIT Waves}

   \institute{$^{1}$ LESIA-Observatoire de Paris, 5 place Jules Janssen, 92195 Meudon, CNRS, UPMC Univ. Paris 06, Univ. Paris-Diderot, France
                     \\\jwlemail{rositsa.miteva@obspm.fr},
                     \\\jwlemail{ludwig.klein@obspm.fr}
              \\$^{2}$ Kanzelh\"ohe Observatory-IGAM, Institute of Physics, University of Graz, Universit\"atsplatz 5, 8010 Graz, Austria
                     \\\jwlemail{ines.kienreich@uni-graz.at},
                     \\\jwlemail{manuela.temmer@uni-graz.at},
                     \\\jwlemail{astrid.veronig@uni-graz.at}
              \\$^{3}$ IAASARS, National Observatory of Athens, GR-15236 Panteli, Greece
                     \\\jwlemail{omaland@astro.noa.gr}
             }

\begin{abstract}

We explore the link between solar energetic particles (SEPs) observed at 1 AU and large-scale disturbances propagating in the solar corona, named after the \textit{Extreme ultraviolet Imaging Telescope} (EIT) as EIT waves, which trace the lateral expansion of a coronal mass ejection (CME). A comprehensive search for SOHO/EIT waves was carried out for 179 SEP events during Solar Cycle 23 (1997\,--\,2006). 87$\,$\% of the SEP events were found to be accompanied by EIT waves. In order to test if the EIT waves play a role in the SEP acceleration, we compared their extrapolated arrival time at the footpoint of the Parker spiral with the particle onset in the 26 eastern SEP events that had no direct magnetic connection to the Earth. We find that the onset of proton events was generally consistent with this scenario. However, in a number of cases the first near-relativistic electrons were detected too early. Furthermore, the electrons had in general only weakly anisotropic pitch-angle distributions. This poses a problem for the idea that the SEPs were accelerated by the EIT wave or in any other spatially confined region in the low corona. The presence of weak electron anisotropies in SEP events from the eastern hemisphere suggests that transport processes in interplanetary space, including cross-field diffusion, play a role in giving the SEPs access to a broad range of helio-longitudes.

\end{abstract}
\end{opening}
%-------------------------------------------------

\section{Introduction} %%%%%%%%%%%%%%%%%%%%%%%%%%%%%%%%%%%%%%%%%%%%% Section 1
     \label{S-Introduction}

Solar energetic particle (SEP) events are transient enhancements of the intensities of electrons above some keV and ions and protons above some hundreds of keV, observed by particle detectors onboard spacecraft or on the ground. The two main candidates for the particle accelerator in the corona are a flare-related reconnection process and a coronal mass ejection (CME) driven shock. Irrespective of the acceleration agent, the particles need to be injected onto magnetic field lines connected to the spacecraft in order to be observed. When the parent activity is in the West, both flares (via rapidly diverging field lines in the low corona: \opencite{2008A&A...486..589K}) and CMEs (via their large-scale shock front: \opencite{2006SSRv..123..127S}) can be connected to the magnetic-field lines that reach Earth. This is more difficult to achieve with increasing distance from the optimal magnetic connection to Earth near 60$^\circ$ western longitude, as predicted by a Parker spiral (PS) model.

Nonetheless, SEP events also occur with activity in the eastern hemisphere. Their origin was discussed in terms of: coronal propagation, diffusive or not, from an accelerator in the associated flare \cite{1991PCS....21..243K}; acceleration at a CME shock in the interplanetary (IP) space \cite{1988JGR....93.9555C,1991PCS....21..243K,2006SSRv..123..217K}; and/or propagation in a transient magnetic-field configuration when the Earth is connected to the Sun by an interplanetary coronal mass ejection (ICME) from a previous eruptive solar event \cite{1991JGR....96.7853R,2012A&A...538A..32M}. In the present work, we investigate whether large-scale traveling disturbances in the low corona, so-called EIT waves, play a role in accelerating SEPs or injecting them into the IP space.

These coronal disturbances were detected for the first time by the \textit{Extreme ultraviolet Imaging Telescope} instrument (EIT: \opencite{1995SoPh..162..291D}) onboard the \textit{Solar and Heliospheric Observatory} (SOHO). On running-difference images, EIT waves \cite{1998GeoRL..25.2465T,1999ApJ...517L.151T}, also called in the past coronal Moreton waves and recently extreme ultra-violet (EUV) waves, appear as bright fronts that emanate from the active region \cite{2012SoPh..281..187P}. In the quiet corona they may propagate up to nearly 360 degrees \cite{2012ApJ...756..143O}, whereas at the boundaries of different magnetic topologies the disturbance undergoes reflections and refractions \cite{2008ApJ...681L.113V,2009ApJ...691L.123G,2013SoPh..286..201K}. Recent observations from the \textit{Solar Terrestrial Relations Observatory} (STEREO) and \textit{Solar Dynamics Observatory} (SDO) spacecraft showed that the EUV structure is closely related to the flanks of the associated CME during its lateral over-expansion \cite{2010ApJ...708.1639M,2011SoPh..273..421T}.

The nature of this coronal disturbance is subject to an ongoing debate (see reviews by \opencite{2008SoPh..253..215V}; \opencite{2009SSRv..149..325W}; \opencite{2010AdSpR..45..527W}; \opencite{2011JASTP..73.1096Z}; \opencite{2012SoPh..281..187P}). There are two opposite interpretations of the observed EUV-intensity fronts. One group interprets them as true fast-mode MHD waves or shocks propagating through the corona and compressing the plasma at its front \cite{1998GeoRL..25.2465T,2001ApJ...560L.105W}. The other group of models considers the bright EUV front as a consequence of the large-scale restructuring of the coronal magnetic field during the CME lift-off \cite{1999SoPh..190..107D,2002ApJ...572L..99C,2007ApJ...656L.101A}. A reconciling interpretation, {\it i.e.} a hybrid of the previous two, was also proposed \cite{2004A&A...427..705Z}. In the present study, we will use the term ``EIT disturbance´´ or ``EIT wave´´ in order to refer to the observational phenomenon without any implication on the actual nature of the disturbance.

The idea of a physical relation between the EIT disturbances and SEP events was already investigated in the past with different authors reaching different conclusions for electrons and protons. In an early study, \inlinecite{1997ESASP.415..207B} compared the propagation of an EIT wave from an eastern active region and the onset of an electron event at SOHO. The EIT wave was just entering the western solar hemisphere when the electron event started, and was therefore not the likely accelerator of the first electrons seen by SOHO. Negative results for electrons were obtained also by \inlinecite{1999ApJ...519..864K}. They performed a detailed velocity-dispersion analysis of near-relativistic electron events and estimated the initial solar release time for each. For 18 of 58 events, EIT disturbances were found in the SOHO/EIT data. In the six events investigated in more detail, the EIT wave speeds were found to be a factor of two too slow to explain the electron release times at the longitude of best magnetic connection from the Sun to the spacecraft. The authors argued that a 3D disturbance could have higher speed at higher altitude (1.5 solar radii) than an EIT wave that is assumed to propagate at low coronal heights \cite{2009SoPh..259...49P,2009ApJ...703L.118K}, and thus might be the particle accelerator. The majority of the events were at western heliospheric longitudes.

The onsets of protons up to 25~MeV were, however, compatible with acceleration as the EIT wave reached the vicinity of the IP field line connected to the spacecraft. This conclusion was reached in the case study of the 24 September 1997 proton event by \inlinecite{1999ApJ...510..460T}, where the initial injection of the $\gtrsim$10 MeV protons was found to be during the period when an eastern EIT wave was traversing the western hemisphere of the Sun. In a recent, very detailed, multi-spacecraft analysis \inlinecite{2012ApJ...752...44R} show that the arrival time of the CME flank and the associated EIT wave at the respective footpoints of the spacecraft-connected Parker spirals could explain the arrival timing of the first energetic electrons and protons at different spacecraft. \inlinecite{2009ApJ...704..469M} argued that both electrons and protons observed in December 2006 by {\it Ulysses} and \textit{Advanced Composition Explorer} (ACE), in association with an active region near the eastern limb, could be attributed to EIT disturbances reaching magnetic field lines connected to the spacecraft.

In summary, the question of whether EIT waves contribute to SEP acceleration has so far no unique answer. The present work addresses the subject by combining a statistical assessment with a timing analysis, where the arrival of EIT waves at the root of the Earth-connected PS in the corona is compared with the onset time of the SEP events near Earth. Starting with a comprehensive list of particle events (Section~\ref{S-All_events_data}) in Solar Cycle 23 (1997\,--\,2006), we performed an extensive search in the SOHO/EIT data for associated large-scale coronal disturbances by visually identifying the wave fronts. Results are summarized in a catalog of EIT waves.

In order to test if an EIT wave accelerates SEPs far from the parent active region, we focus (Section~\ref{S-Eastern}) on events related to eruptive activity in the eastern solar hemisphere. We excluded those events where the SEPs were detected while the spacecraft was in or near an ICME, which could provide a transient direct magnetic-field connection to the eastern solar hemisphere. The onset of the SEP events -- deka-MeV protons and near relativistic electrons observed by \textit{Electron Proton Helium Instrument} (EPHIN) part of the \textit{Comprehensive Suprathermal and Energetic Particle Analyzer} (COSTEP) onboard SOHO -- and the onset and speed of the EIT waves were determined, together with the expected arrival of the EIT wave at the PS footpoint. By comparing this arrival time and the SEP onset one can decide if the EIT waves propagate fast enough to accelerate the particles that arrived first at the spacecraft, Section~\ref{S-Results}. Additional information on the electrons is provided from the observations of the anisotropy characteristics from \textit{Electron, Proton, and Alpha Monitor} (EPAM) onboard ACE. We summarize the results in Section~\ref{S-Summary} and discuss the impact of our findings on scenarios of the SEP origin from eastern activity in Section~\ref{S-Disc}.

\section{Data Analysis} %%%%%%%%%%%%%%%%%%%%%%%%%%%%%%%%%%%%%%%% Section 2
      \label{S-Data}

\begin{table}[ht!]
\caption[]{Solar energetic particle events with origin at western [W] helio-longitudes and associated EIT disturbances, flares and CMEs. We list the SEP event date, the proton onset at 1 AU from \inlinecite{2010JGRA..11508101C}, the EIT wave association and time of first front observation, the SXR flare importance, start time, and longitude and finally the linear speed and angular width [AW] of the associated CME.\\
{\it a}: identified in the SOHO/EIT 304 \AA$\,$ channel; H$\alpha$: flare longitude from the H$\alpha$ reports in the Solar Geophysical Data; I/S/V: SEP events propagating in ICME/Solar wind/Vicinity of an ICME; {\it K}: events from \inlinecite{2011SoPh..269..309K}; {\it n}: next day; {\it u}: uncertain; {\it v}: visual identification; Y/N: Yes/No association.}
\label{T-West_events}
\tiny
%\vspace{0.1cm}
\begin{tabular}{r@{ }lr@{}llll}
\hline
\multicolumn{2}{c}{Event}       & \multicolumn{2}{l}{Proton}   & EIT      & SXR flare    & CME speed,        \\
\multicolumn{2}{c}{date}        & \multicolumn{2}{l}{onset}    & wave     & size         & km$\,$s$^{-1}$    \\
\multicolumn{2}{c}{I/S/V}       & \multicolumn{2}{l}{at 1 AU}  & Y/N      & (onset, UT/  & (first image,     \\
 \multicolumn{2}{c}{}           & \multicolumn{2}{l}{UT}       & (front, UT) & long. W, deg)  & UT/AW, deg)      \\
%(1)    & (2)      & (3)      & (4)          & (5)                   \\
\hline
21 May &1997 S& &21:00 & Y (20:42) & M1.3 (20:08/12) & 296 (21:01/30)    \\
03 Nov &1997 S& &11:00 & Y (10:32) & M4.2 (10:18/22$^{{\rm H}\alpha}$) & 352 (11:11/100)   \\
04 Nov &1997 S& &07:00 & Y (05:57) & X2.1 (05:52/33) & 785 (06:10/110)   \\
06 Nov &1997 V& &12:00 & Y (11:59) & X9.4 (11:49/63) & 1556 (12:11/115)   \\
20 Apr &1998 S& &12:00 & Y (09:35) & M1.4 (09:38/90$^v$) & 1863 (10:07/150)  \\
02 May &1998 I& &14:00 & Y (13:40) & X1.1 (13:31/15) & 938 (14:06/130)   \\
06 May &1998 I& &08:00 & Y (08:10) & X2.7 (07:58/65) & 1099 (08:29/90)    \\
09 Sep &1998 S& &06:00 & no data   & M2.8 (04:52/-)  & no data           \\
30 Sep &1998 S& &14:00 & no data   & M2.8 (13:08/81) & no data           \\
05 Nov &1998 S& $<$&22:00 & N         & M8.4 (19:00/18) & 1118 (20:44/60) \\
22 Nov &1998 S& &08:00 & Y (06:38) & X3.7 (06:30/82) & no data           \\
22 Nov &1998 S& $<$&17:00$^K$ & Y (16:22) & X2.5 (16:10/89) & no data        \\
17 Dec &1998 S& &11:00 & N         & M3.2 (07:40/46) & 302 (08:30/360$^v$)  \\
16 Feb &1999 V& &06:00 & no data   & M1.5 (04:04/-)  & no data           \\
04 Jun &1999 V& &08:00 & Y ($<$07:35) & M3.9 (06:52/69) & 2230 (07:27/80)  \\
27 Jun &1999 V& &11:00 & Y (08:43) & M1.0 (08:34/25) & 903 (09:06/40)     \\
28 Aug &1999 S& &20:00 & N         & X1.1 (17:52/14) & 462 (18:26/110)   \\
28 Dec &1999 I& &02:00 & N         & M4.5 (00:39/56) & 672 (00:54/60)    \\
12 Feb &2000 V& &05:00 & Y ($<$04:14) & M1.7 (03:51/23) & 1107 (04:31/110) \\
02 Mar &2000 V& &09:00 & Y (08:35) & X1.1 (08:20/52$^{{\rm H}\alpha}$) & 776 (08:54/60) \\
03 Mar &2000 S& &03:00 & Y (02:23) & M3.8 (02:08/60) & 841 (02:30/80)     \\
22 Mar &2000 S& &20:00 & Y (18:46) & X1.1 (18:34/57) & 478 (19:31/80)    \\
24 Mar &2000 S& &10:00 & no data   & X1.8 (07:41/82) & no data          \\
04 Apr &2000 S& &16:00 & Y (15:24) & C9.7 (15:12/66) & 1188 (16:33/60)   \\
01 May &2000 S& &11:00 & N         & M1.1 (10:16/61) & 1360 (10:54/20)   \\
04 May &2000 I& &12:00 & Y (11:11) & M6.8 (10:57/90$^{{\rm H}\alpha}$) & 1404 (11:26/70)\\
23 May &2000 V& &22:00 & N         & C9.5 (20:48/43) & 475 (21:30/50)     \\
10 Jun &2000 V& &18:00 & Y (17:12) & M5.2 (16:40/38) & 1108 (17:08/120)  \\
15 Jun &2000 S& &21:00 & N         & M1.8 (19:38/65) & 1081 (20:06/70)   \\
17 Jun &2000 S& &05:00 & no data   & M3.5 (02:25/72) & 857 (03:28/60)    \\
18 Jun &2000 V& &02:00 & no data   & X1.0 (01:52/85) & 629 (02:10/70)    \\
23 Jun &2000 V& &16:00 & Y (14:35) & M3.0 (14:18/72) & 847 (14:54/60)    \\
25 Jun &2000 I& &12:00 & Y (07:48) & M1.9 (07:17/55) & 1617 (07:54/70)   \\
14 Jul &2000 I& &11:00 & Y (10:23) & X5.7 (10:03/7)  & 1674 (10:54/360)   \\
22 Jul &2000 S& &12:00 & Y (11:35) & M3.7 (11:17/56) & 1230 (11:54/80)    \\
12 Aug &2000 I& &11:00 & Y (10:00) & M1.1 (09:45/79$^{{\rm H}\alpha}$) & 662 (10:35/60) \\
09 Sep &2000 I& &10:00 & no data   & M1.6 (08:28/67) & 554 (08:57/70)    \\
12 Sep &2000 S& &13:00 & Y (11:35) & M1.0 (11:31/9)  & 1550 (11:54/100)  \\
19 Sep &2000 I& &11:00 & Y (08:12) & M5.1 (08:06/46) & 766 (08:50/60)    \\
08 Nov &2000 I& &23:00 & Y (23:00) & M7.4 (22:42/77) & 1738 (23:06/120)  \\
24 Nov &2000 S& &06:00 & Y (05:11) & X2.0 (04:55/5$^{{\rm H}\alpha}$)  & 1289 (05:30/360)  \\
24 Nov &2000 S& &16:00 & Y (15:11) & X2.3 (14:51/7)  & 1245 (15:30/360)  \\
\hline
\end{tabular}
\end{table}

\begin{table}[ht!]
\addtocounter{table}{-1}
\caption[]{(continued)}
\label{T-West_events1}
\tiny
%\vspace{0.1cm}
\begin{tabular}{r@{ }lr@{}llll}
\hline
\multicolumn{2}{c}{Event}       & \multicolumn{2}{l}{Proton}   & EIT      & SXR flare    & CME speed,        \\
\multicolumn{2}{c}{date}        & \multicolumn{2}{l}{onset}    & wave     & size         & km$\,$s$^{-1}$    \\
\multicolumn{2}{c}{I/S/V}       & \multicolumn{2}{l}{at 1 AU}  & Y/N      & (onset, UT/  & (first image,     \\
 \multicolumn{2}{c}{}           & \multicolumn{2}{l}{UT}       & (front, UT) & long. W, deg)  & UT/AW, deg)      \\
%(1)    & (2)      & (3)      & (4)          & (5)                \\
\hline
28 Jan &2001 S& &17:00 & Y (15:58) & M1.5 (15:40/59) & 916 (15:54/120)  \\
10 Mar &2001 S& &08:00 & N         & M6.7 (04:00/42) & 819 (04:26/20)   \\
29 Mar &2001 I& &12:00 & Y (10:13) & X1.7 (09:57/19) & 942 (10:26/360)  \\
02 Apr &2001 I& &13:00 & N         & X1.1 (10:58/62$^{{\rm H}\alpha}$) & 992 (11:26/50)\\
02 Apr &2001 I& &23:00 & Y (21:47) & X20 (21:32/70)  & 2505 (22:06/100) \\
09 Apr &2001 V& &16:00 & Y (15:35) & M7.9 (15:20/4)  & 1192 (15:54/360) \\
10 Apr &2001 S& &07:00 & Y (04:59) & X2.3 (05:06/9)  & 2411 (05:30/360) \\
12 Apr &2001 I& &12:00 & Y (10:25) & X2.0 (09:39/43) & 1184 (10:31/120) \\
14 Apr &2001 V& &18:00 & N         & M1.0 (17:15/71) & 830 (17:54/50)   \\
15 Apr &2001 V& &14:00 & Y (13:47) & X14.4 (13:19/85)& 1199 (14:06/110) \\
26 Apr &2001 S& $<$&22:00 & Y (13:11) & M7.8 (11:26/31) & 1006 (12:30/360) \\
20 May &2001 S& &08:00 & Y (06:12) & M6.4 (06:00/90$^{v}$) & 546 (06:26/90) \\
19 Jul &2001 S& &11:00 & Y (10:13) & M1.8 (09:52/62) & 1668 (10:30/40)  \\
12 Sep &2001 V& &23:00 & Y (21:24) & C9.6 (21:05/63) & 668 (22:06/30)   \\
15 Sep &2001 V& &12:00 & N         & M1.5 (11:04/49) & 478 (11:54/80)   \\
19 Oct &2001 S& &01:00 & Y ($<$01:36) & X1.6 (00:47/18) & 558 (01:27/180) \\
19 Oct &2001 S& &16:00 & Y (16:24) & X1.6 (16:13/29) & 901 (16:50/160)  \\
22 Oct &2001 I& &04:00 & N         & M1.0 (00:22/57) & 772 (00:50/20)   \\
25 Oct &2001 V& $>$&15:00 & Y (14:59) & X1.3 (14:42/21) & 1092 (15:26/360) \\
04 Nov &2001 S& &17:00 & no data   & X1.0 (16:03/18) & 1810 (16:35/130) \\
22 Nov &2001 S& &21:00 & Y (20:35) & M3.8 (20:18/67) & 1443 (20:31/120) \\
22 Nov &2001 V& &01:00$^{n}$ & Y (23:11) & M9.9 (22:32/36$^{v}$) & 1437 (23:30/270) \\
26 Dec &2001 S& &06:00 & Y ($<$05:35)& M7.1 (04:32/54) & 1446 (05:30/90)  \\
20 Feb &2002 S& &06:00 & Y (06:33) & M5.1 (05:52/72) & 952 (06:30/50)    \\
15 Mar &2002 S& &24:00 & Y (22:00) & M2.2 (22:09/3) & 957 (23:06/360)     \\
22 Mar &2002 V& &12:00 & Y (10:48) & M1.6 (10:12/90$^{v}$) & 1750 (11:06/130) \\
11 Apr &2002 V& &16:00 & N         & C9.2 (16:16/33) & 540 (16:50/50)     \\
14 Apr &2002 V& &12:00 & N         & C9.6 (07:28/57) & 757 (07:50/50)     \\
15 Apr &2002 S& &03:00 & Y (03:12) & C9.8 (02:46/79) & 674 (03:06/45)     \\
17 Apr &2002 V& &08:00 & Y (08:00) & M2.6 (07:46/34) & 1240 (08:26/70)    \\
21 Apr &2002 I& $>$&00:00 & Y ($<$01:36)& X1.5 (00:43/84) & 2393 (01:27/120)\\
15 Jul &2002 S& &09:00$^{n}$ & Y (20:12) & X3.0 (19:59/1) & 1151 (20:30/100) \\
03 Aug &2002 I& &19:00 & Y ($<$19:26) & X1.0 (18:59/76) & 1150 (19:32/30) \\
14 Aug &2002 S& &03:00 & Y (02:00) & M2.3 (01:47/54) & 1309 (02:30/60)    \\
16 Aug &2002 S& &07:00 & Y (06:00) & M2.4 (05:46/83) & 1378 (06:06/70)    \\
18 Aug &2002 V& &20:00 & Y (21:24) & M2.2 (21:12/19) & 682 (21:54/100)    \\
19 Aug &2002 V& &10:00 & N         & M2.0 (10:28/25) & 549 (11:06/80)     \\
20 Aug &2002 I& &08:00 & N         & M3.4 (08:22/38) & 1099 (08:54/40)    \\
22 Aug &2002 V& &02:00 & Y (02:00) & M5.4 (01:47/62) & 998 (02:06/80)     \\
24 Aug &2002 S& $>$&00:00 & Y ($<$01:36) & X3.1 (00:49/81) & 1913 (01:27/150) \\
09 Nov &2002 S& &15:00 & Y (13:14) & M4.6 (13:08/29) & 1838 (13:32/90)    \\
19 Dec &2002 V& &22:00 & Y (21:50) & M2.7 (21:34/9$^{{\rm H}\alpha}$)  & 1092 (22:06/120) \\
22 Dec &2002 I& $<$&03:00 & Y (02:48) & M1.1 (02:14/42) & 1071 (03:30/80) \\
\hline
\end{tabular}
\end{table}

\begin{table}[ht!]
\addtocounter{table}{-1}
\caption[]{(continued)}
\label{T-West_events1}
\tiny
%\vspace{0.1cm}
\begin{tabular}{r@{ }lr@{}llll}
\hline
\multicolumn{2}{c}{Event}       & \multicolumn{2}{l}{Proton}   & EIT      & SXR flare    & CME speed,        \\
\multicolumn{2}{c}{date}        & \multicolumn{2}{l}{onset}    & wave     & size         & km$\,$s$^{-1}$    \\
\multicolumn{2}{c}{I/S/V}       & \multicolumn{2}{l}{at 1 AU}  & Y/N      & (onset, UT/  & (first image,     \\
 \multicolumn{2}{c}{}           & \multicolumn{2}{l}{UT}       & (front, UT) & long. W, deg)  & UT/AW, deg)      \\
%(1)    & (2)      & (3)      & (4)          & (5)                 \\
\hline
17 Mar &2003 S& &19:00 & Y (19:13) & X1.5 (18:50/39) & 1020 (19:54/50)   \\
18 Mar &2003 S& &18:00 & Y (12:00) & X1.5 (11:51/46) & 1042 (13:54/80)   \\
23 Apr &2003 S& &00:00 & Y ($<$03:36)& M5.1 (00:39/25) & 916 (01:27/70)  \\
24 Apr &2003 S& &12:00 & Y ($<$13:36)& M3.3 (12:45/39) & 609 (13:27/45)  \\
27 May &2003 S& &22:00 & Y$^a$ (23:12) & X1.3 (22:56/17) & 964 (23:50/360)\\
28 May &2003 S& $>$&00:00 & Y$^a$ (00:24) & X3.6 (00:17/22$^{{\rm H}\alpha}$)  & 1366 (00:50/220) \\
31 May &2003 I& &03:00 & Y$^a$ (02:23) & M9.3 (02:13/65) & 1835 (02:30/150) \\
10 Jun &2003 S& &14:00 & Y ($<$14:28)  & M3.6 (13:54/90) & no data       \\
19 Aug &2003 I& &u     & Y (08:00) & M2.0 (07:38/63) & 412 (08:30/40)     \\
26 Oct &2003 V& &18:00 & Y (17:33) & X1.2 (17:21/38)& 1537 (17:54/130)  \\
29 Oct &2003 I& &22:00 & Y (20:48) & X10 (20:37/2)   & 2029 (20:54/360)  \\
02 Nov &2003 V& &18:00 & Y (17:33) & X8.3 (17:03/56) & 2598 (17:30/130)  \\
03 Nov &2003 S& $>$&03:00$^K$ & Y (01:17) & X2.7 (01:09/83) & 827 (01:59/65)\\
03 Nov &2003 S& &12:00$^K$ & Y (09:58) & X3.9 (09:43/77) & 1420 (10:06/100) \\
04 Nov &2003 S& &21:00 & Y (19:48) & X28 (19:29/83)  & 2657 (19:54/130)  \\
20 Nov &2003 V& &07:00 & Y (07:48) & M9.6 (07:35/8)  & 669 (08:06/90)    \\
04 Feb &2004 S& &11:00 & Y (11:24) & C9.9 (11:12/48) & 764 (11:54/20)    \\
11 Apr &2004 S& &05:00 & Y (04:33$^u$)& C9.6 (03:54/47) & 1645 (04:30/90)\\
13 Jul &2004 S& &01:00 & Y (00:24) & M6.7 (00:09/45) & 607 (00:54/60)    \\
25 Jul &2004 V& &16:00 & Y (14:00) & M1.1 (14:19/33) & 1333 (14:54/130)   \\
19 Sep &2004 I& &18:00 & no data   & M1.9 (16:46/58) & no data           \\
30 Oct &2004 S& &07:00 & Y (06:24) & M4.2 (06:08/21) & 422 (06:54/90)     \\
30 Oct &2004 S& &13:00 & Y (11:48) & X1.2 (11:38/18) & 427 (12:30/90)    \\
30 Oct &2004 S& &17:00 & Y (16:36) & M5.9 (16:18/20) & 690 (16:54/90)    \\
07 Nov &2004 V& &17:00 & Y (15:59) & X2.0 (15:42/17) & 1759 (16:54/150)  \\
09 Nov &2004 V& &19:00 & Y (17:11) & M8.9 (16:59/51) & 2000 (17:26/130)  \\
10 Nov &2004 I& &03:00 & Y (02:11) & X2.5 (01:59/49) & 3387 (02:26/120)  \\
02 Dec &2004 S& $<$&03:00$^n$ & Y ($<$00:00$^{n}$) & M1.5 (23:44/2)& 1216 (00:26/360)\\
15 Jan &2005 V& &23:00 & Y (22:36) & X2.6 (22:25/8)  & 2861 (23:06/130)  \\
17 Jan &2005 V& &10:00 & Y (07:12) & X3.8 (06:59/25) & 2094 (09:30/110)  \\
19 Jan &2005 I& $>$&08:00$^K$ & Y (08:11) & X1.3 (08:03/51) & 2020 (08:29/360) \\
20 Jan &2005 V& &06:00 & Y (06:48) & X7.1 (06:36/58) & 882 (06:54/80)     \\
06 May &2005 S& &03:00 & N         & C9.3 (03:05/74) & 1120 (03:30/20)   \\
06 May &2005 S& &12:00 & Y (11:35) & M1.3 (11:11/76) & 1144 (11:54/30)   \\
11 May &2005 S& &20:00 & Y ($<$19:50) & M1.1 (19:22/47) & 550 (20:13/70) \\
16 Jun &2005 I& &20:00 & no data   & M4.0 (20:01/90) & no data           \\
09 Jul &2005 V& &24:00 & Y (22:00) & M2.8 (21:47/28) & 1540 (22:30/65)   \\
12 Jul &2005 V& &19:00 & Y (16:12$^u$) & M1.5 (15:47/67) & 523 (16:54/80)\\
13 Jul &2005 S& &05:00 & Y (02:47) & M1.1 (02:35/82) & 759 (03:06/40)    \\
13 Jul &2005 S& &14:00 & Y (14:23) & M5.0 (14:01/90) & 1423 (14:30/70)   \\
14 Jul &2005 S& &11:00 & Y (10:35) & X1.2 (10:16/90) & 2115 (10:54/80)   \\
22 Aug &2005 S& &01:00 & Y (01:10) & M2.6 (00:44/54) & 1194 (01:32/160)  \\
22 Aug &2005 S& &18:00 & Y (17:08) & M5.6 (16:46/65) & 2378 (17:30/100)  \\
%050915 I& u$^K$ & no data   & X1$^u$ (08:30/14) & no data         \\
06 Jul &2006 S& &09:00 & Y (08:31) & M2.5 (08:13/34) & 911 (08:54/160)   \\
13 Dec &2006 S& &02:00 & Y (02:25) & X3.4 (02:14/23) & 1774 (02:54/180)  \\
14 Dec &2006 V& &22:00 & Y (22:12) & X1.5 (21:07/46) & 1042 (22:30/70)   \\
\hline
\end{tabular}
\end{table}

\begin{table}[ht!]
\caption[]{Solar energetic particle events with origin at eastern [E] helio-longitudes and associated EIT disturbances, flares and CMEs. Table columns and abbreviations as for Table~\ref{T-West_events}.\\
{\it d}: doubtful flare association, {\it M}: Moreton wave reported.}
\label{T-East_events}
\tiny
%\vspace{0.1cm}
\begin{tabular}{r@{ }lr@{}llll}
\hline
\multicolumn{2}{c}{Event}       & \multicolumn{2}{l}{Proton}   & EIT      & SXR flare    & CME speed,        \\
\multicolumn{2}{c}{date}        & \multicolumn{2}{l}{onset}    & wave     & size         & km$\,$s$^{-1}$    \\
\multicolumn{2}{c}{I/S/V}       & \multicolumn{2}{l}{at 1 AU}  & Y/N      & (onset, UT/  & (first image,     \\
 \multicolumn{2}{c}{}           & \multicolumn{2}{l}{UT}       & (front, UT) & long. W, deg)  & UT/AW, deg)      \\
%(1)      & (2)       & (3)      & (4)          & (5)              \\
\hline
01 Apr &1997 S &  &16:00   & Y (14:00) & M1.9 (13:43/16) & 312 (15:19/40)   \\
24 Sep &1997 S &  &04:00   & Y (02:29) & M5.9 (02:43/19) & 532 (03:38/70)   \\
29 Apr &1998 S &$<$&24:00  & Y (16:19) & M6.8 (16:06/20) & 1374 (16:59/90)  \\
18 Aug &1998 V &  &24:00 &no data$^{M}$& X4.9 (22:10/87) & no data          \\
19 Aug &1998 V &  &23:00 &no data$^{M}$& X3.9 (21:35/75) & no data          \\
24 Aug &1998 S &  &23:00 &no data$^{M}$& X1.0 (21:50/9)  & no data          \\
20 Sep &1998 S &$<$&03:00$^{n}$&no data& M1.8 (02:33/62$^{{\rm H}\alpha}$)& no data \\
23 Sep &1998 I & $<$&10:00 & no data   & M7.1 (06:40/9)  & no data          \\
03 May &1999 S & $<$&20:00 & Y (05:47) & M4.4 (05:36/32) & 1584 (06:06/110) \\
29 Jun &1999 V & $<$&12:00 & N         & M1.4 (05:01/7)$^{d}$ & 589 (05:54/360)\\
17 Nov &1999 S & $<$&12:00 & no data   & M7.4 (09:47/21) & no data          \\
18 Jan &2000 S & &18:00    & Y (17:11) & M3.9 (17:07/11) & 739 (17:54/120)  \\
17 Feb &2000 S & &21:00    & Y (20:23) & M1.3 (20:17/7)  & 728 (21:30/360)  \\
10 May &2000 S &$<$&03:00$^{n}$& N$^u$ & C8.7 (19:26/20) & 641 (20:06/130)  \\
06 Jun &2000 I &$<$&18:00  & Y (15:11) & X2.3 (14:58/18) & 1119 (15:54/180) \\
10 Jul &2000 V & &24:00    & Y (22:11) & M5.7 (21:05/49) & 1352 (21:50/130) \\
29 Oct &2000 I & $<$&12:00 & Y (01:51) & M4.4 (01:28/35) & no data          \\
25 Nov &2000 V & $<$&18:00 & Y (01:13) & M8.2 (00:59/50) & 2519 (01:31/120) \\
20 Jan &2001 S &$<$&05:00$^{n}$& Y (18:47)& M1.3 (18:33/40) & 839 (19:32/130) \\
%       &            & Y (21:22) & M7.7 (21:06/46) & 1507 (21:30/130) & 21:12 \\
25 Mar &2001 S &$<$&22:00  & N         & C9.0 (16:25/25)$^{d}$ & 677 (17:06/360) \\
15 Jun &2001 S & &11:00    & Y (10:12) & M6.3 (10:01/41) & 1090 (10:32/50)  \\
17 Sep &2001 S & $<$&12:00 & Y (08:24) & M1.5 (08:18/4)  & 1009 (08:54/50)  \\
24 Sep &2001 I & &11:00    & Y (10:25) & X2.6 (09:32/23) & 2402 (10:30/120) \\
09 Oct &2001 S & $<$&14:00 & Y (10:55) & M1.4 (10:46/8)  & 973 (11:30/120)  \\
22 Oct &2001 I & &15:00    & N$^{u}$   & M6.7 (14:27/18)$^{d}$ & 1336 (15:06/140) \\
17 Nov &2001 S & &07:00    & N$^{u}$   & M2.8 (04:49/42)$^{u}$ & 1379 (05:30/160) \\
28 Nov &2001 S & $<$&20:00 & Y (16:35) & M6.9 (16:26/16) & 500 (17:30/90)   \\
20 May &2002 I & &15:00    & Y (15:24) & X2.1 (15:21/65) & 553 (15:50/30)   \\
16 Aug &2002 S & $<$&15:00 & Y (11:36) & M5.2 (11:32/20) & 1585 (12:30/160) \\
21 Apr &2003 S & &17:00    & no data   & M2.8 (12:54/2)  &  784 (13:36/120) \\
25 Apr &2003 S & $<$&18:00 & Y (05:36) & M1.2 (05:23/79) &  806 (05:50/90)  \\
15 Jun &2003 I & $<$&24:00 & Y (23:58) & X1.3 (23:25/80) & 2053 (23:54/130) \\
17 Jul &2003 S & &11:00    & Y (08:22) & C9.8 (08:17/21) &  531 (08:54/360) \\
26 Oct &2003 V & &07:00    & Y (06:21) & X1.2 (05:57/41) & 1371 (06:54/70)  \\
28 Oct &2003 V & &12:00    & Y$^{M}$ (10:36)& X17.2 (09:51/8) & 2459 (11:30/360) \\
18 Nov &2003 S & &11:00    & Y (07:47) & M3.2 (07:23/18) & 1223 (08:06/80)  \\
07 Jan &2004 S & $<$&14:00 & Y (10:24) & M8.3 (10:14/69) & 1822 (10:30/70)  \\
12 Sep &2004 S & $<$&04:00 & no data   & M4.8 (00:04/42) & 1328 (00:36/140) \\
04 Nov &2004 S &$<$&05:00$^{n}$ & Y (22:22)& M5.4 (22:53/18)& 1055 (23:30/140)\\
14 Jan &2005 S & &11:00    & N$^{u}$   & C8.9 (10:08/15) &  396 (11:30/20)  \\
15 Jan &2005 S & &06:00    & Y (05:59) & M8.6 (05:54/6)  & 2049 (06:30/90)  \\
13 May &2005 S & &17:00    & Y (16:37) & M8.0 (16:13/11) & 1689 (17:12/360) \\
03 Jun &2005 S &$<$&05:00$^{n}$& no data & M1.0 (11:51/90)& 1679 (12:32/80) \\
07 Sep &2005 S & $<$&20:00 & no data   & X17 (17:17/77)  & no data          \\
13 Sep &2005 I & &22:00    & no data   & X1.5 (19:19/10) & 1866 (20:00/130) \\
06 Nov &2006 S & &15:00    & Y (17:47) & C8.8 (17:43/90$^{v}$)& 1994 (17:54/70) \\
05 Dec &2006 S & $<$&15:00 & no data   & X9.0 (10:18/68) & no data          \\
06 Dec &2006 S & $<$&22:00 & no data$^{M}$ & X6.5 (18:29/63) & no data      \\
\hline
\end{tabular}
\end{table}

The present work started with a compilation of an event list of solar energetic particle events observed during Solar Cycle 23 (1997\,--\,2006). The event selection is based on two previously published works, the comprehensive SEP catalog given by \inlinecite{2010JGRA..11508101C} and a few events that were added from \inlinecite{2011SoPh..269..309K}. We selected SEP events associated with flares with peak soft X-ray (SXR) flux $\gtrsim 9\times 10^{-6}$~W$\,$s$^{-2}$ (GOES class $\gtrsim$C9) which were located on the solar disk including limb events ({\it i.e.} with heliolongitude within $\pm$90 degrees). The final list in the present study comprises 179 SEP events.

\subsection{Properties of SEP-Associated Coronal Activity}
 \label{S-All_events_data}

\subsubsection{Flares and CMEs}
 \label{S-Fl_CMEs}

We adopted the proposed solar origin for each SEP event from \inlinecite{2010JGRA..11508101C}, but cross-checked the flare and CME characteristics. We used the Solar Geophysical Data reports at
\href{ftp://ftp.ngdc.noaa.gov/STP/SOLAR\_DATA/SGD\_PDFversion/}{\textsf{ftp.ngdc.noaa.gov/STP/SOLAR\_DATA/SGD\_PDFversion/}},
the online GOES flare list (under IDL--SolarSoft), and/or the information provided at \href{http://www.SolarMonitor.org}{\textsf{www.SolarMonitor.org}}. In case the flare location was not reported in the SXR catalog, the position of the nearest H$\alpha$ flare was used instead (denoted with a superscript H$\alpha$). We found that 131 particle events from our list are associated with coronal activity at western helio-longitudes (given chronologically in Table~\ref{T-West_events}) and 48 have eastern coronal origin (Table~\ref{T-East_events}). The properties of the SEP-associated CMEs, {\it e.g.} CME linear (projected) speed and first appearance in the C2 coronagraph of the \textit{Large Angle and Spectrometric Coronagraph Experiment} (LASCO-C2) field of view are taken from the SOHO/LASCO CME catalog at
\href{http://cdaw.gsfc.nasa.gov/CME_list/}{\textsf{cdaw.gsfc.nasa.gov/CME\_list/}},
whereas the angular width [AW] is adopted from \inlinecite{2010JGRA..11508101C}.

\subsubsection{EIT Disturbances}
      \label{S-EIT}

With the so-compiled list of SEP events we performed a comprehensive search in the SOHO/EIT data for large-scale coronal disturbances associated with the coronal origin of the SEP events. For this purpose we prepared running-ratio images in the SOHO/EIT 195 \AA$\,$ channel (23/179 events were in a SOHO data gap). Three different observers independently completed a visual identification of the disturbances. Yes/No [Y/N] identification for each event is reported when at least two observer identifications agreed, listed in column~3 of Table~\ref{T-West_events} and \ref{T-East_events}. For a positively identified EIT wave we also give the time when the disturbance was first observed.

\subsubsection{IP Conditions}
      \label{S-IP}

In order to characterize the conditions in the IP medium, we used the \inlinecite{2010SoPh..264..189R} catalog of global magnetic disturbances, {\it i.e.} ICMEs. In their catalog, for each ICME the times of the following magnetic boundaries were identified: shock-arrival time, start and end times of the ICME as measured near Earth. We used these times in order to classify the SEP events into the following categories: ``ICME events'' [I] when the particle onset at 1 AU [$t_{\rm 1AU}$] is temporarily between the ICME start and end times; ``sheath'', when $t_{\rm 1AU}$ is in the sheath region, {\it i.e.} between the shock and the ICME start and/or ``vicinity events'', when $t_{\rm 1AU}$ is less than one day apart from the reported shock or the ICME end (both denoted with V); and ``solar-wind events'' [S], when $t_{\rm 1AU}$ is more than one day apart from any boundary of an ICME. The abbreviations I/S/V are given after the event date in all tables.

\subsection{Detailed Analysis of Eastern Events}
      \label{S-Eastern}

\subsubsection{Eastern SEP Events}
      \label{S-eastSEPs}

\begin{table}[t!]
\caption[]{Properties of all 29 eastern particle events associated with EIT waves. We list the SEP date, onset time at 1 AU [$t_{\rm 1AU}$] and rise time [$t_{\rm r}$] for protons and electrons, onset time [$t_{\rm III}$] of the Type-III radio burst from {\it Wind}/WAVES spacecraft, electron anisotropy (beam, E'contaminated, irregular, isotropic, moderate) and connection distance. For the events given in italic font no EIT wave speed could be determined due to single/uncertain wave-front identification.  \\
{\it h}: high-energy channel 0.67\,--\,3 MeV; {\it l}: low-energy channel 0.25\,--\,0.7 MeV; {\it n}: next day; \mbox{{\it p}: previous event}; {\it sp}: spike in the particle profile; {\it V}: from \inlinecite{2013JSWSC...3A..12V}; {\it u}: uncertain}
\label{T-SEP_data}
\tiny
\vspace{0.5cm}
\begin{tabular}{r@{ }ll@{}cl@{}ccl@{}c}
\hline
\multicolumn{2}{c}{Event} &  \multicolumn{2}{c}{SOHO/EPHIN protons} & \multicolumn{2}{c}{SOHO/EPHIN electrons} & Onset         & Anis. & Conn.  \\
\multicolumn{2}{c}{date}  &  $t_{\rm 1AU}$, UT & $t_{\rm r}$        & $t_{\rm 1AU}$, UT  & $t_{\rm r}$         & $t_{\rm III}$ & ACE/  & dist.  \\
\multicolumn{2}{c}{I/S/V} & (unc. range, min)  & min                 & (unc. range, min) & min                 & UT            & EPAM  & deg    \\
%(1)       & (2)                & (3)         & (4)               & (5)                        & (6)           & (7)   &  (8)            \\
\hline
01 Apr &1997 S& 16:58 ($-$132,+138)  &  15    & 13:50$^h$ ($-$87,+60)  & 17  & 13:41 & not av. & 74 \\
24 Sep &1997 S& 03:58 ($-$23,+17)    &   3    & 03:00$^h$ ($-$8,+4)    & 1   & 02:49 & irreg.$^V$  & 84 \\
29 Apr &1998 S& 00:08$^{n}$ ($-$118,+108)& 39 & 20:30$^h$ ($-$65,+59)  & 103 & 16:07 & moder.  & 92 \\
{\it 03 May} &{\it 1999 S}& 13:59 ($-$155,+325) & 54 & 11:04$^l$ ($-$91,+110) & 74 & 05:40 & moder.  & 79\\
18 Jan &2000 S& 18:18 ($-$4,+10)     &   2    & 17:14$^h$ ($-$25,+16)  & 4   & 17:12 & irreg.$^V$  & 84 \\
17 Feb &2000 S& 21:11 ($-$3,+10)     &   2    & 20:42$^l$ ($-$8,+6)    & 2   & 20:26 & beam$^V$    & 67 \\
06 Jun &2000 I& 17:33 ($-$35,+103)   &  18    & 16:47$^h$ ($-$67,+33)  & 11  & 15:07 & E'con.$^V$  & 61 \\
10 Jul &2000 V& 23:20 ($-$9,+39)     &   9    & 20:44$^h$ ($-$15,+22)  & 11  & 21:23 & beam$^V$    & 98 \\
29 Oct &2000 I& {\it u}              &   -    & 06:25$^l$ ($-$43,+76)  & 48  & 01:47 & moder.  & 97 \\
25 Nov &2000 V&  weak                &   -    & {\it p}                & -   & 01:04 & -       & 110 \\
20 Jan &2001 S& 00:04$^{n}$ ($-$90,+98)&  26  & {\it u,sp}             & -   & 18:42 & -       & 115 \\
15 Jun &2001 S& 10:59 ($-$20,+32)    &   6    & 10:48$^l$ ($-$8,+6)    & 7   & 10:06 & moder.  & 107 \\
17 Sep &2001 S& weak                 &   -    & 10:27$^l$ ($-$27,+37)  & 47  & 08:21 & isotr.  & 56 \\
24 Sep &2001 I& 11:29 ($-$5,+11)     &   2    & 10:40$^h$ ($-$4,+4)    & 1   & 09:30 & beam    & 75  \\
09 Oct &2001 S& 15:47 ($-$155,+201)  &  80    & 15:08$^h$ ($-$27,+70)  & 298 & 10:45 & moder.  & 63 \\
28 Nov &2001 S& weak                 &   -    & weak                   & -   & 16:24 & -       & 72  \\
20 May &2002 I& 15:59 ($-$4,+14)     &   3    & 15:37$^l$ ($-$1,+1)    & 1   & 15:23 & beam    & 116 \\
16 Aug &2002 S& {\it p}              &   -    & {\it p}                & -   & 12:06 & -       & 60 \\
25 Apr &2003 S& weak                 &   -    & 11:41$^l$ ($-$52,+98)  & 107 & 05:22 & moder.  & 124 \\
15 Jun &2003 I& {\it u}              &   -    & {\it u}                & -   & 23:42 & -       & 123 \\
17 Jul &2003 S& weak                 &   -    & 09:18$^l$ ($-$34,+26)  & 6   & 08:18 & irreg.  & 73 \\
26 Oct &2003 V& {\it u}              &   -    & {\it u}                & -   & 06:20 & -       & 99  \\
{\it 28 Oct} &{\it 2003 V}& 11:59 ($-$4,+2)&   1    & 11:16$^h$ ($-$1,+1)    & 4   & 10:57 & beam$^V$    & 45  \\
18 Nov &2003 S& 09:23 ($-$139,+71)   &  14    & 08:43$^h$ ($-$17,+19)  & 29  & 07:23 & isotr.  & 52  \\
{\it 07 Jan} &{\it 2004 S}& {\it u}        &   -    & {\it u}                & -   & 10:21 & -       & 104 \\
04 Nov &2004 S& {\it u}              &   -    & 23:47$^l$ ($-$10,+15)  & 24  & 23:01 & isotr.  & 80 \\
%041202 S&    u                &   -    & 00:34$^{l,n}$ ($-$54,+45)& 56  & 23:49 & mod.   & 56  \\
15 Jan &2005 S& 06:53 ($-$6,+10)     &   2    & 06:33$^l$ ($-$1,+2)    & 2   & 06:07 & moder.  & 46 \\
13 May &2005 S& 16:44 ($-$59,+64)    &  11    & 17:11$^l$ ($-$1,+1)    & 2   & 16:40 & isotr.$^V$  & 57  \\
06 Nov &2006 S& 18:46 ($-$69,+51)    &   9    & 18:02$^l$ ($-$2,+4)    & 2   & 17:38 & moder.  & 160\\
\hline
\end{tabular}
\end{table}

\begin{figure}[t!]
\centerline {
\includegraphics[width=0.5\textwidth,clip=]{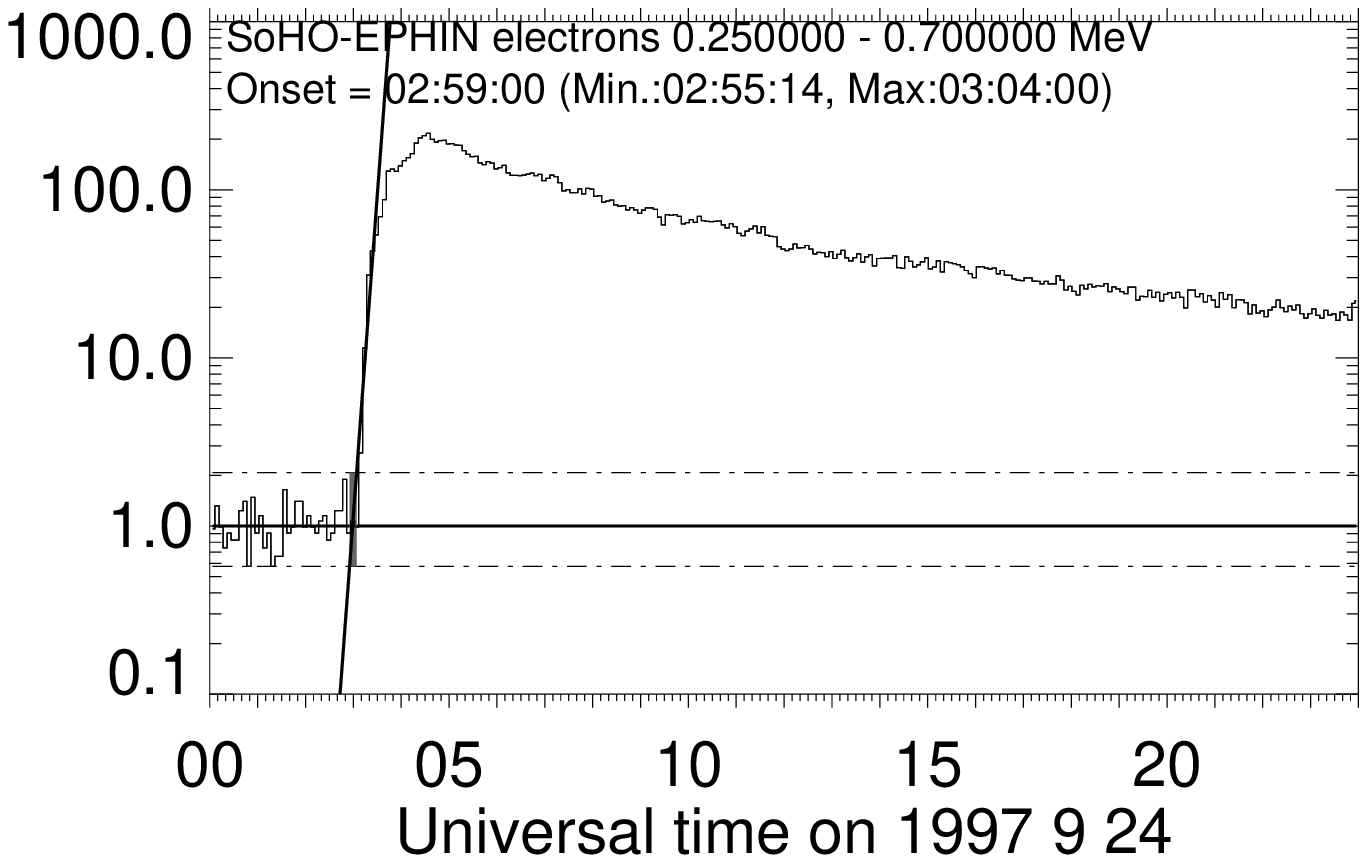}% SoHO-EPHIN_onset_19970924_0  % final Fig1a
\includegraphics[width=0.5\textwidth,clip=]{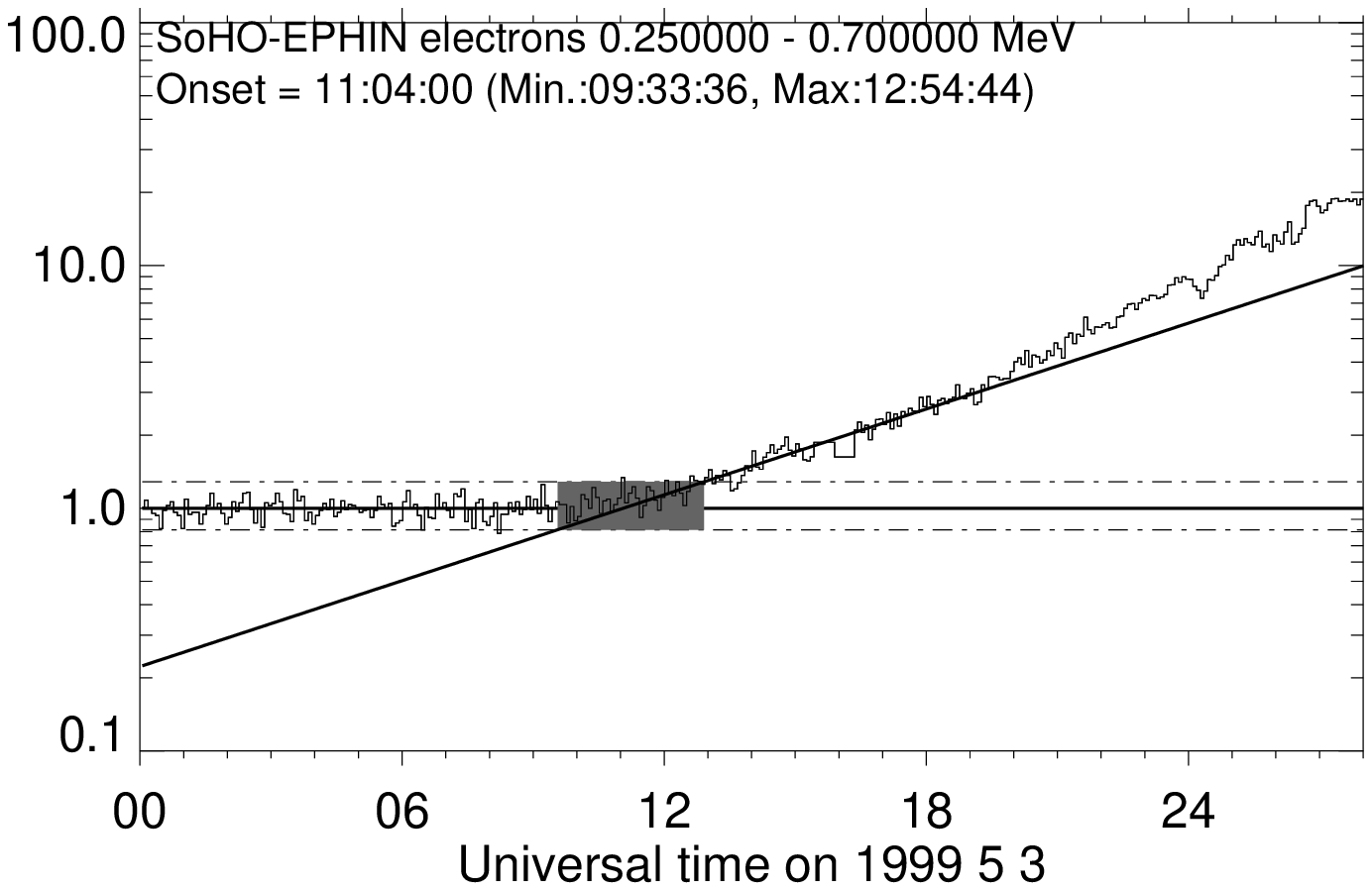}% SoHO-EPHIN_onset_19990503_0  % final Fig1b
}
\centerline {
\includegraphics[width=0.5\textwidth,clip=]{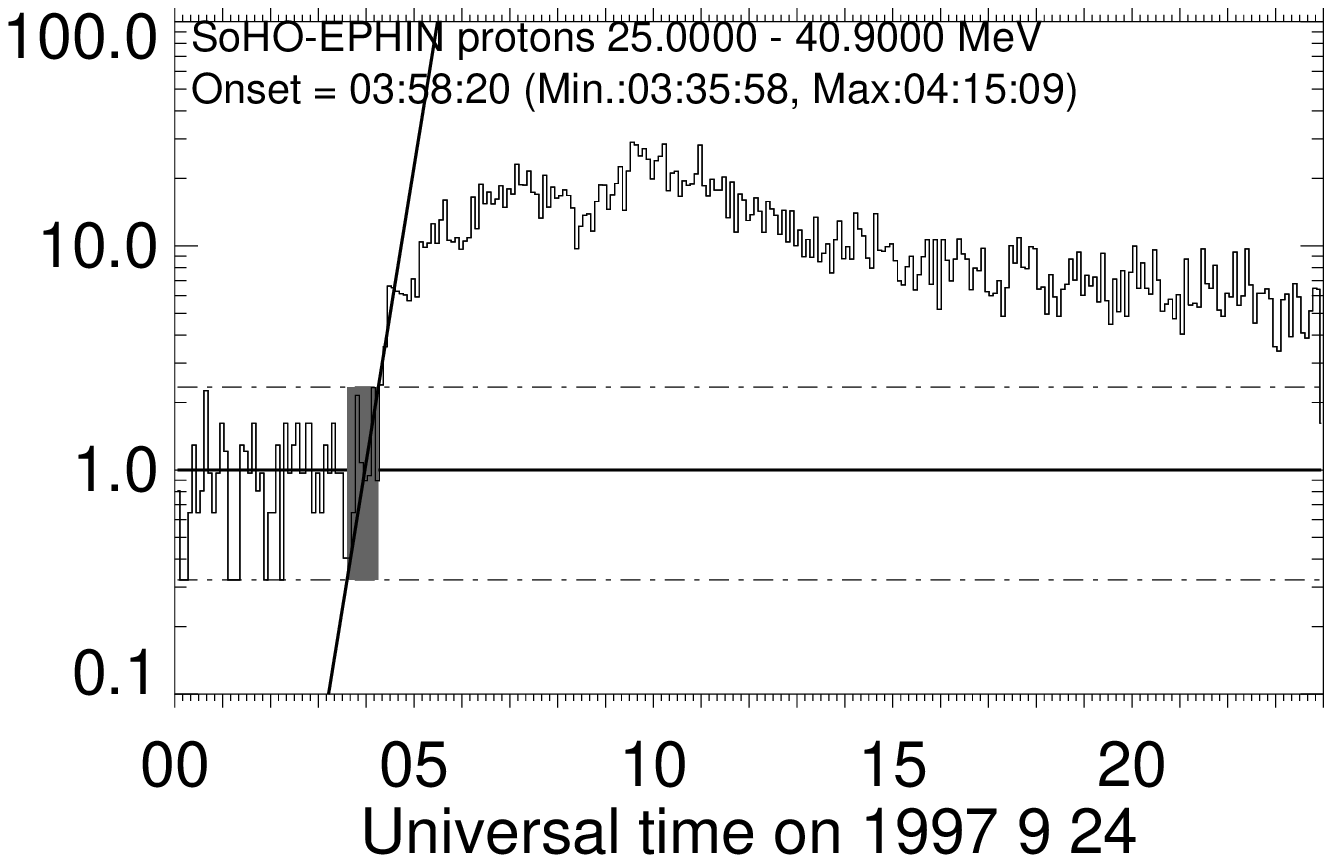}% SoHO-EPHIN_onset_19970924_2  % final Fig1c
\includegraphics[width=0.5\textwidth,clip=]{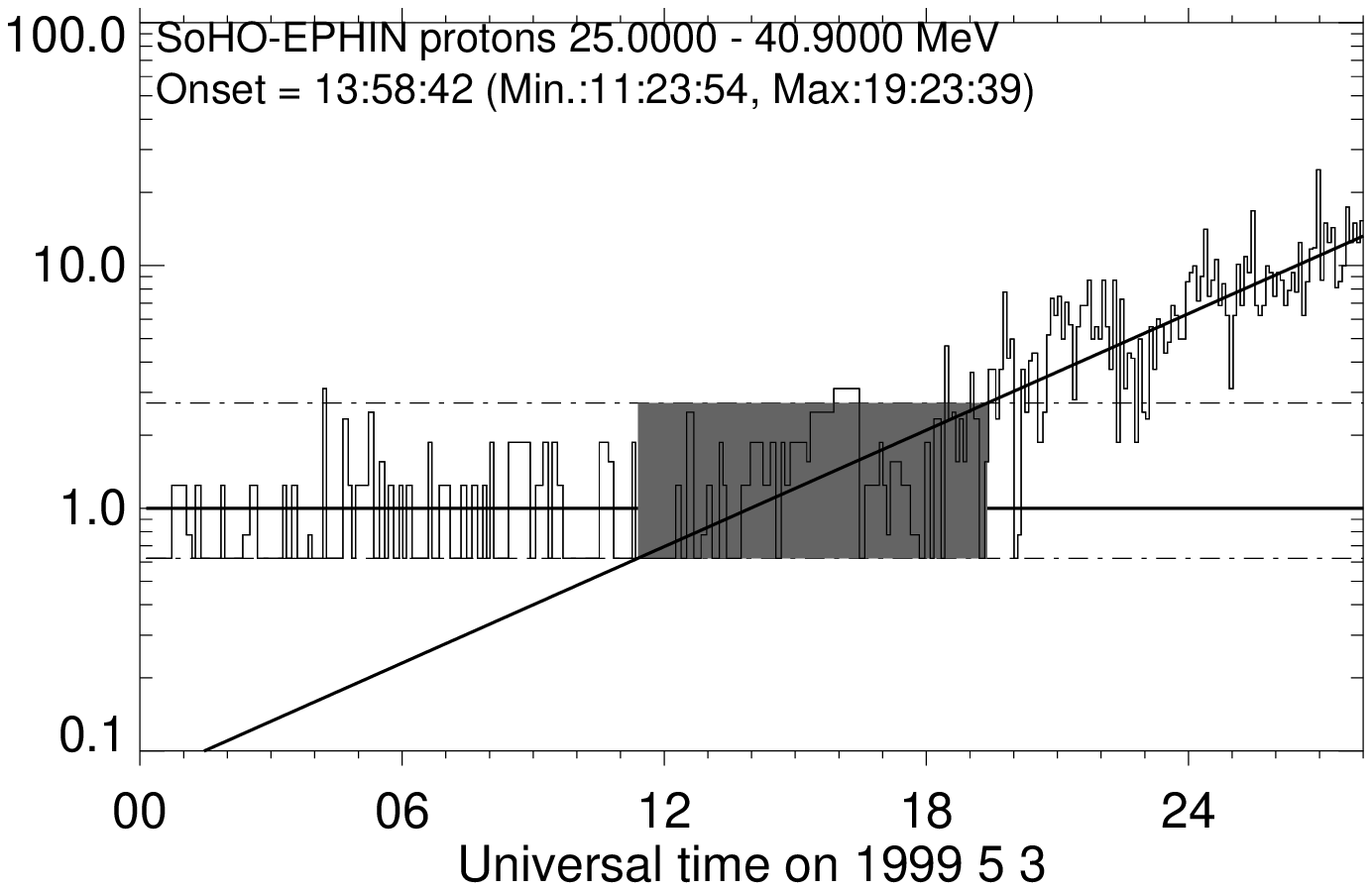}% SoHO-EPHIN_onset_19990503_2  % final Fig1d
}
\caption[]{Time histories of near-relativistic electrons (0.25\,--\,0.70 MeV) top plots, and deka-MeV protons (25\,--\,41 MeV) bottom plots, of two SEP events (SOHO/EPHIN observations). Horizontal lines give the pre-event background (solid), evaluated as the average intensity during a given time interval, and the background $\pm 3 \sigma$ levels (dashed--dotted). The ordinate displays the intensity normalized to the background. The inclined line is the linear fit to the logarithm of the SEP intensity during the early rise of the event. The onset time is the time of intersection of this line with the background, the grey rectangle designates the uncertainty as defined by the times of intersection of the fitted profile with the background $\pm 3 \sigma$ levels.}
\label{Fig_EPHIN}
\end{figure}

A detailed analysis of the particle onset times at 1 AU, estimated from the temporal behavior of the observed flux, was performed for the events in the East that were also associated with EIT waves (Table~\ref{T-SEP_data}). {We used the EPHIN instrument, part of the \textit{Comprehensive Suprathermal and Energetic Particle analyser} (COSTEP) onboard SOHO \cite{1995SoPh..162..483M}. For the proton events we used the 25\,--\,41 MeV channel (column~2) whereas the electron onset time is estimated either from the 0.25\,--\,0.7 MeV or 0.67\,--\,3 MeV channel, as given with superscript ``l'' or ``h'' in column~4, respectively}. We fitted a straight line to the logarithm of the intensity profiles of electrons and protons during the first hours of the rise phase. The intersection of the fitted line with the pre-event background, determined by averaging over a suitable interval before the rise, is considered as the onset time of the event [$t_{\rm 1AU}$]. The earliest and latest possible onset times were determined by the intersection with the background $\pm 3 \sigma$ level. The differences of these values with the onset time (in minutes) are given within brackets in columns~2 and 4. Figure~\ref{Fig_EPHIN} illustrates the procedure for two cases: the 24 September 1997 event (left column) has a rapidly increasing time profile, allowing for an accurate determination of the onset time. The 03 May 1999 event, on the other hand, rises slowly. The onset time has a large uncertainty, indicated by the grey rectangle in the figure, especially for SEP events with a shallow rise. The time interval where the straight line is fitted to the intensity profile is determined by eye. It covers the interval from the event onset to the instant when the time profile visibly deviates from the exponential rise ({\it e.g.} after 21:00 UT for the electrons on 03 May 1999). The slope of the fitted line is to present the early rise of the event. The time constant, referred to as rise time $t_{\rm r}$ in the following, is listed in columns~3 and 5 of Table~\ref{T-SEP_data}.

In addition, we characterized the electron anisotropy for all eastern events associated with EIT waves (in 7/22 cases the anisotropies were adopted from \inlinecite{2013JSWSC...3A..12V}) through a visual inspection of the pitch-angle distributions (PAD) measured by ACE/EPAM \cite{1998SSRv...86..541G} in the rise phase of the electron event (five- or ten-minute average). Usually this is done in the 0.18\,--\,0.31 MeV channel. Lower energies are used when the data in this specific channel were not of sufficient quality. The PADs were identified as i) beam, when the PAD was strongly peaked along the anti-sunward direction of the IMF, without particles from the sunward direction; ii) moderate, when the PAD had its maximum in the antisunward direction, but was not strongly peaked; iii) isotropic, when no anisotropic features were present; iv) bad coverage, when all sectors were very close to each other in pitch-angle and no distribution function could be obtained; v) irregular, when the PAD displayed strong fluctuations; vi) E` contaminated when EPAM's E` channels were contaminated by ions so that PADs could not be formed; vii) no data, when no sectored data were available.

Illustrations of such PADs are given, {\it e.g.} by \inlinecite{1994GeoRL..21.1747A} and \inlinecite{2002JASTP..64..517M} for beamed distributions or \inlinecite{2012SoPh..281..333M} for moderate anisotropy. Examples of EPAM PADs can be visualized in the SEPServer catalog \url{server.sepserver.eu} \cite{2013JSWSC...3A..12V}. The results for the eastern events are given in column~7, Table~\ref{T-SEP_data}.

Finally, the absolute value of the connection distance, {\it i.e.} the offset between the flare longitude and the PS longitude, is identified (column~8 of Table~\ref{T-SEP_data}). We used the PS longitude defined at the solar-wind source surface (at 2.5 solar radii) that is a function of the solar-wind speed only. For the solar-wind speed we used data from the \textit{Proton Monitor} (PM: \href{http://umtof.umd.edu/pm/}{\textsf{umtof.umd.edu/pm/}}) part of the \textit{Charge, Element, and Isotope Analysis System} (CELIAS)
instrument onboard SOHO \cite{1995SoPh..162..441H} averaged over 12 hours before the first particles arrival at the spacecraft.

\subsubsection{Eastern EIT Waves} %%%%%%%%%%%%%%%%%%%%%%%%%%%%%%%%%%%%%%%%
      \label{S-eastEITs}

We focus now on 26 eastern SEP events and perform a detailed analysis on their associated EIT signatures. Among all 48 eastern SEP events, EIT disturbances in the running-ratio SOHO/EIT images were identified for 29 of them, whereas for six events no clear signatures of an EIT disturbance could be agreed upon (No-association), see Table~\ref{T-East_events}. EIT data were not available for 13 events, although a Moreton wave was reported for four of them. In order to determine an average speed of the propagating disturbance $[v_{\rm av,EIT}]$, measurements of at least two wave fronts are needed (namely, the wave front could be followed in two successive EIT images). Hence, the events with a single or uncertain front identification (03 May 1999, 28 October 2003 and 07 January 2004) were dropped from further timing analysis. The velocities of the EIT waves are always derived from the averaged distances of the first two fronts of the disturbance, assumed to propagate along the solar surface, divided by the elapsed time. For the uncertainties in the speed estimation we will adopt the value of $\pm$50 km$\,$s$^{-1}$ reported by \inlinecite{2010AdSpR..45..527W} due to an uncertainty in the wave-front identification of about 20 Mm.

In summary, for 26 events the wave front could be identified in at least two images, see Figure~\ref{F-Fronts}, and an averaged speed of the disturbance is estimated, column~3 in Table~\ref{T-EIT_waves}. In Figure~\ref{F-Fronts} the fronts are overlayed on the ratio-image of the first EIT front that could be identified. All wave fronts or part of them are directed westward, except for the events on 18 January 2000, 17 February 2000 and 13 May 2005 for which the identified wave fronts propagate to the East. However, we assume that part of the disturbance travels to the West as indicated by loop activation at western helio-longitudes.

For the timing analysis, the onset time and origin of the disturbance are important parameters. However, both are to some degree uncertain due to the low temporal cadence of the SOHO/EIT instrument (of the order of 12 minutes, but occasionally longer). By subtracting half of the time to the previous, undisturbed EIT image, we get a proxy value for the EIT wave onset time [$t_{\rm EIT,on}$], column~2 in Table~\ref{T-EIT_waves}. The origin of the EIT wave was approximated with the flare longitude, which may introduce an error up to about 200 Mm \cite{2001ApJ...560L.105W}. We give more details on the properties of the EIT waves and their relationship to other coronal phenomena in the Appendix.

To estimate the arrival time of the EIT wave at the footpoint of the PS, we assume that the disturbance propagates along the solar surface at constant speed, traversing an arc of the length of the connection distance in longitude. We ignore the travel time in latitude. By neglecting a possible deceleration of the EIT wave and any latitudinal travel, we obtain a lower limit for the arrival time [$t_{\rm EIT}$]. Since this value was obtained from remote sensing observations near 1~AU, the photon travel time to Earth must be subtracted. From the range of EIT wave speed and the error in position of 200 Mm, we calculated the uncertainty in the wave travel time and consequently for $t_{\rm EIT}$, column~5 in Table~\ref{T-EIT_waves}. The uncertainties for $t_{\rm EIT}$ range from $\approx$20 minutes in the case of slow waves down to five minutes for the fastest wave. On average, the uncertainty on the EIT wave timings is below 20 minutes and often less than the uncertainties of the particle onset times [$t_{\rm 1AU}$].

\begin{figure}[ht!]
\centerline{\hspace*{0.03\textwidth}
            \includegraphics[width=0.53\textwidth,clip=]{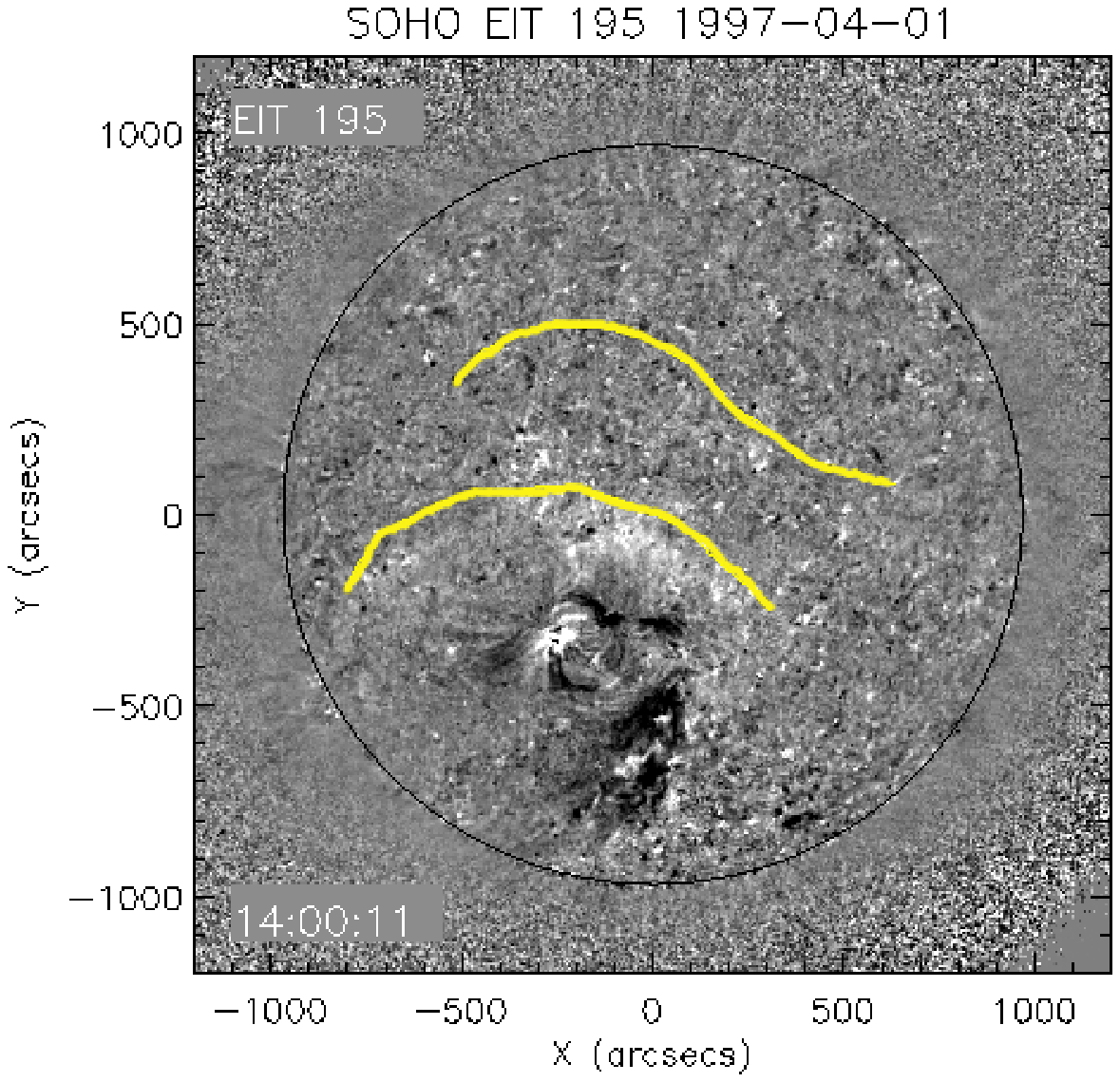}   % final Fig2
            \hspace*{-0.08\textwidth}
            \includegraphics[width=0.53\textwidth,clip=]{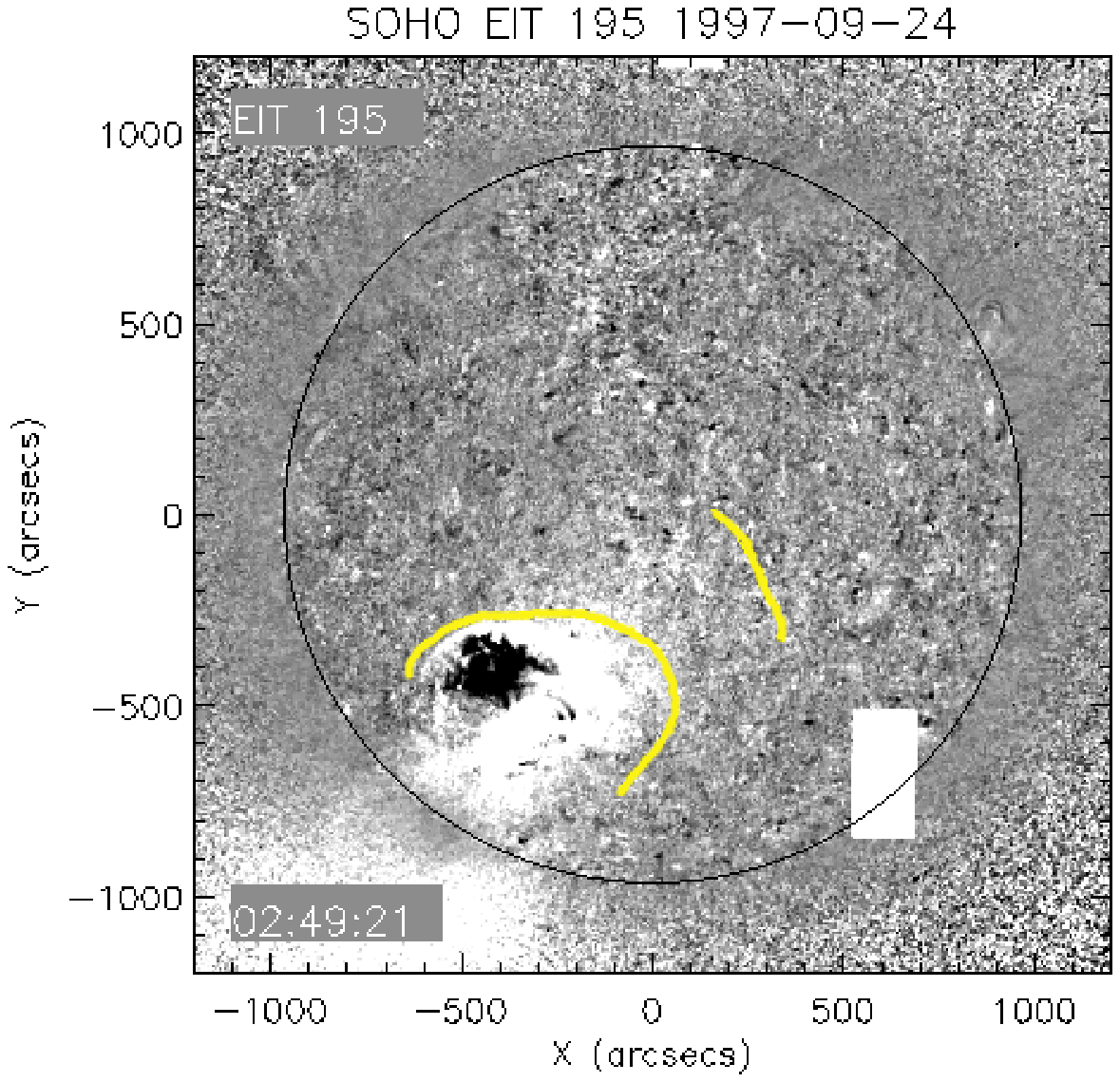}}
            \vspace*{-0.06\textwidth}
\centerline{\hspace*{0.03\textwidth}
            \includegraphics[width=0.53\textwidth,clip=]{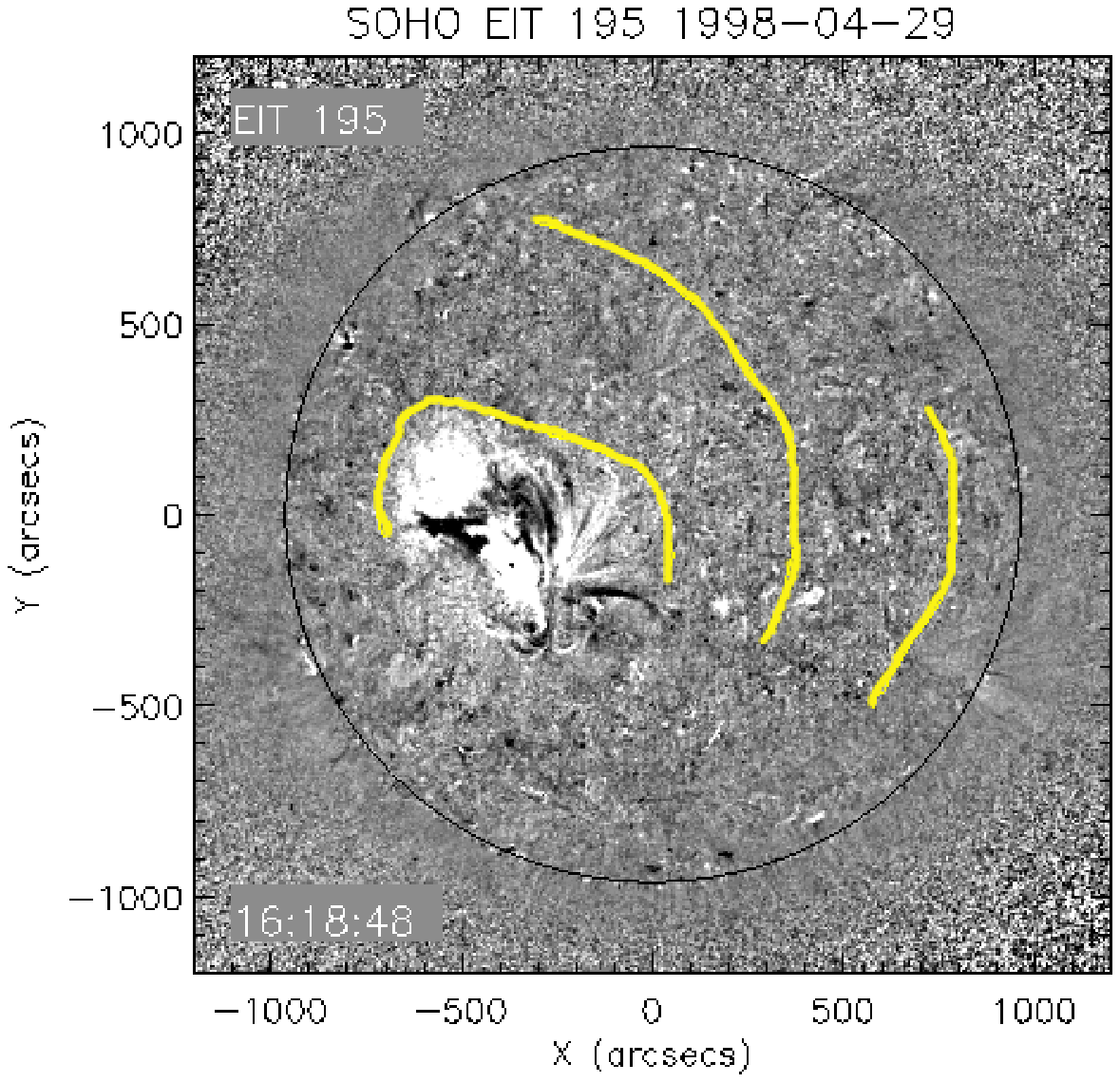}
            \hspace*{-0.08\textwidth}
            \includegraphics[width=0.53\textwidth,clip=]{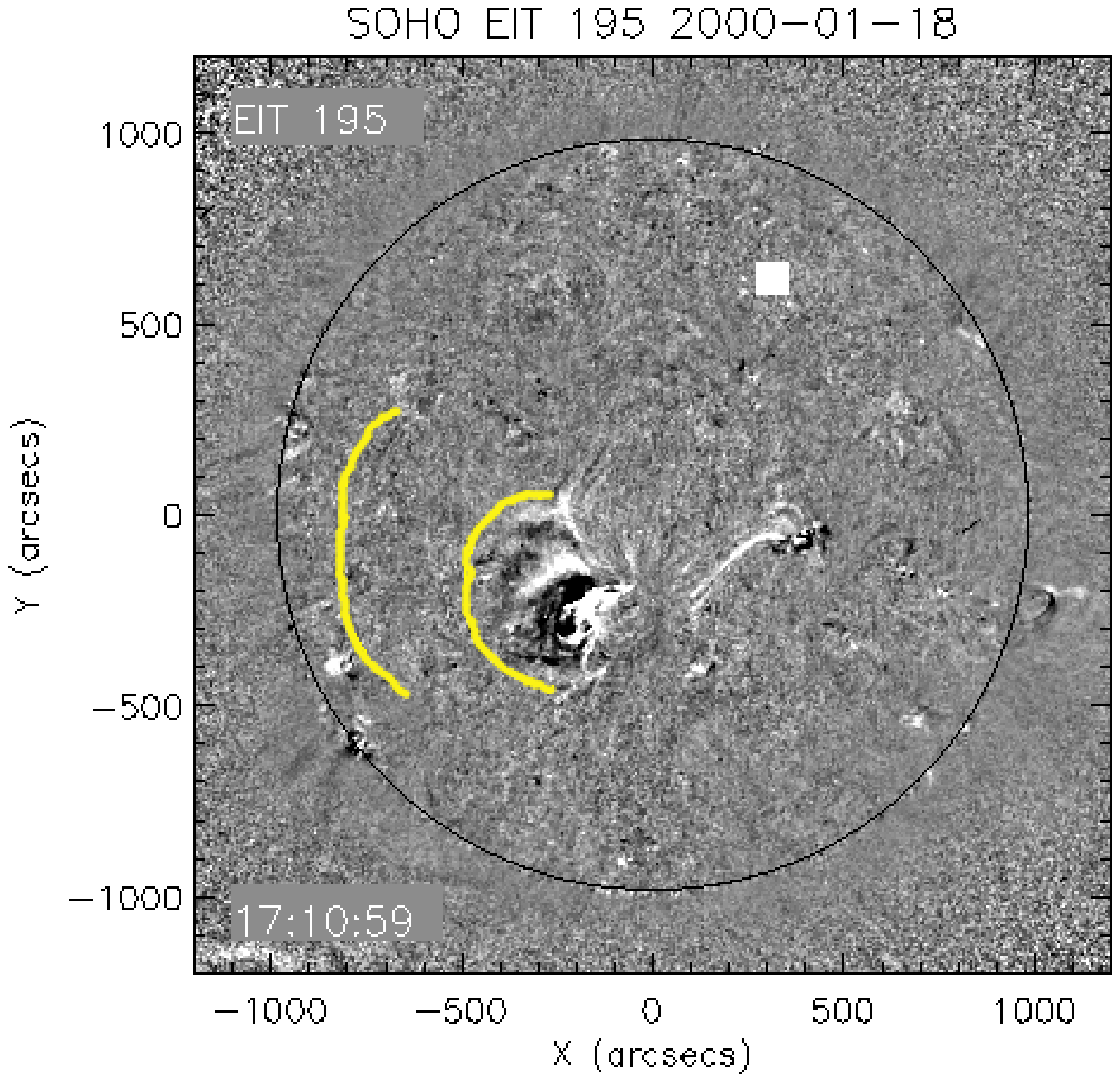}}
            \vspace*{-0.06\textwidth}
\centerline{\hspace*{0.03\textwidth}
            \includegraphics[width=0.53\textwidth,clip=]{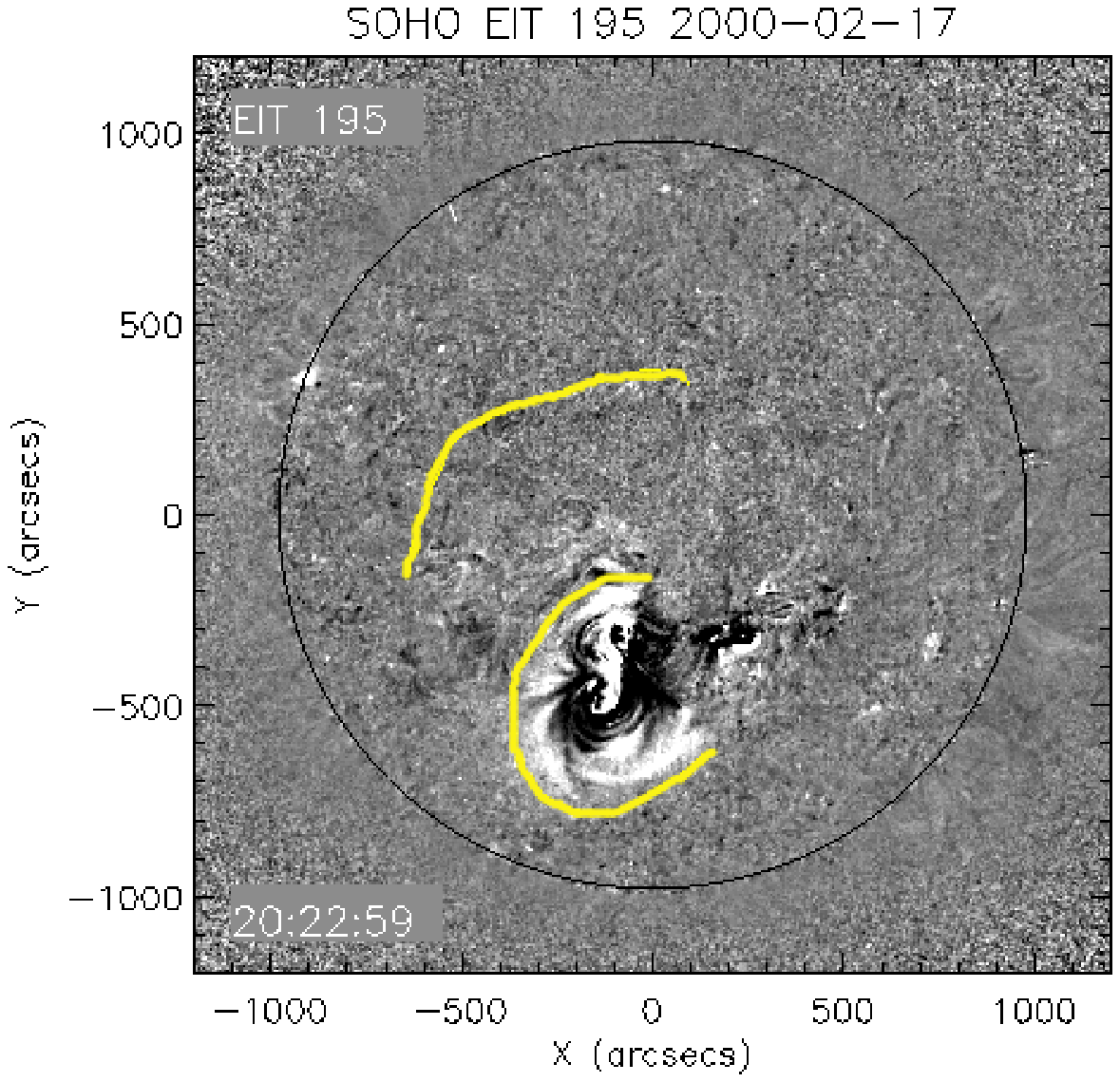}
            \hspace*{-0.08\textwidth}
            \includegraphics[width=0.53\textwidth,clip=]{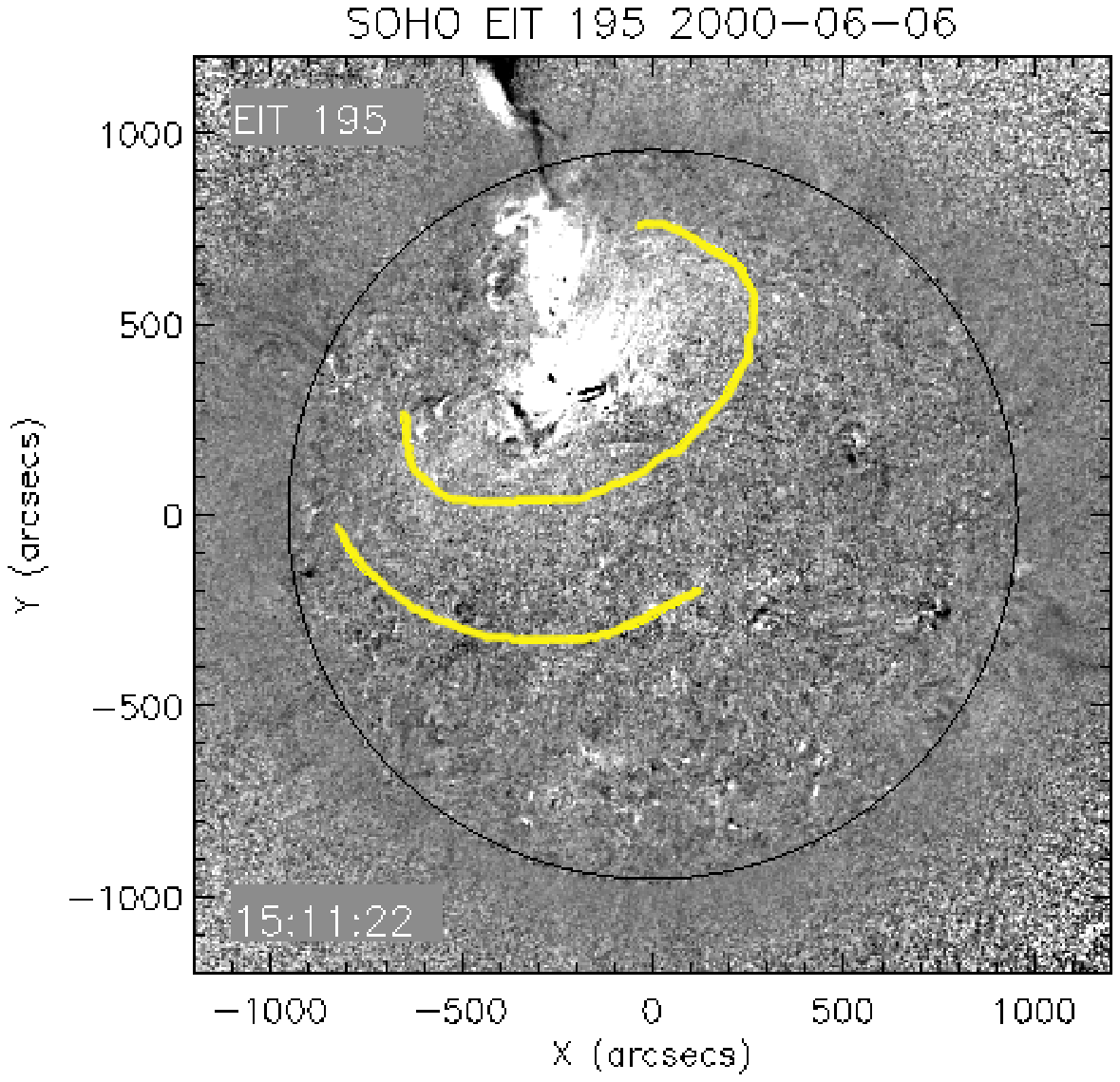}}
            \vspace*{-0.03\textwidth}
\caption{EIT wave fronts identified from running-ratio images in 195$\,$\AA $\,$from the SOHO/EIT instrument. The wave fronts are overplotted on the running-ratio image of the first identified front.}
   \label{F-Fronts}
   \end{figure}

 \begin{figure}[ht!]
\centerline{\hspace*{0.03\textwidth}
            \includegraphics[width=0.53\textwidth,clip=]{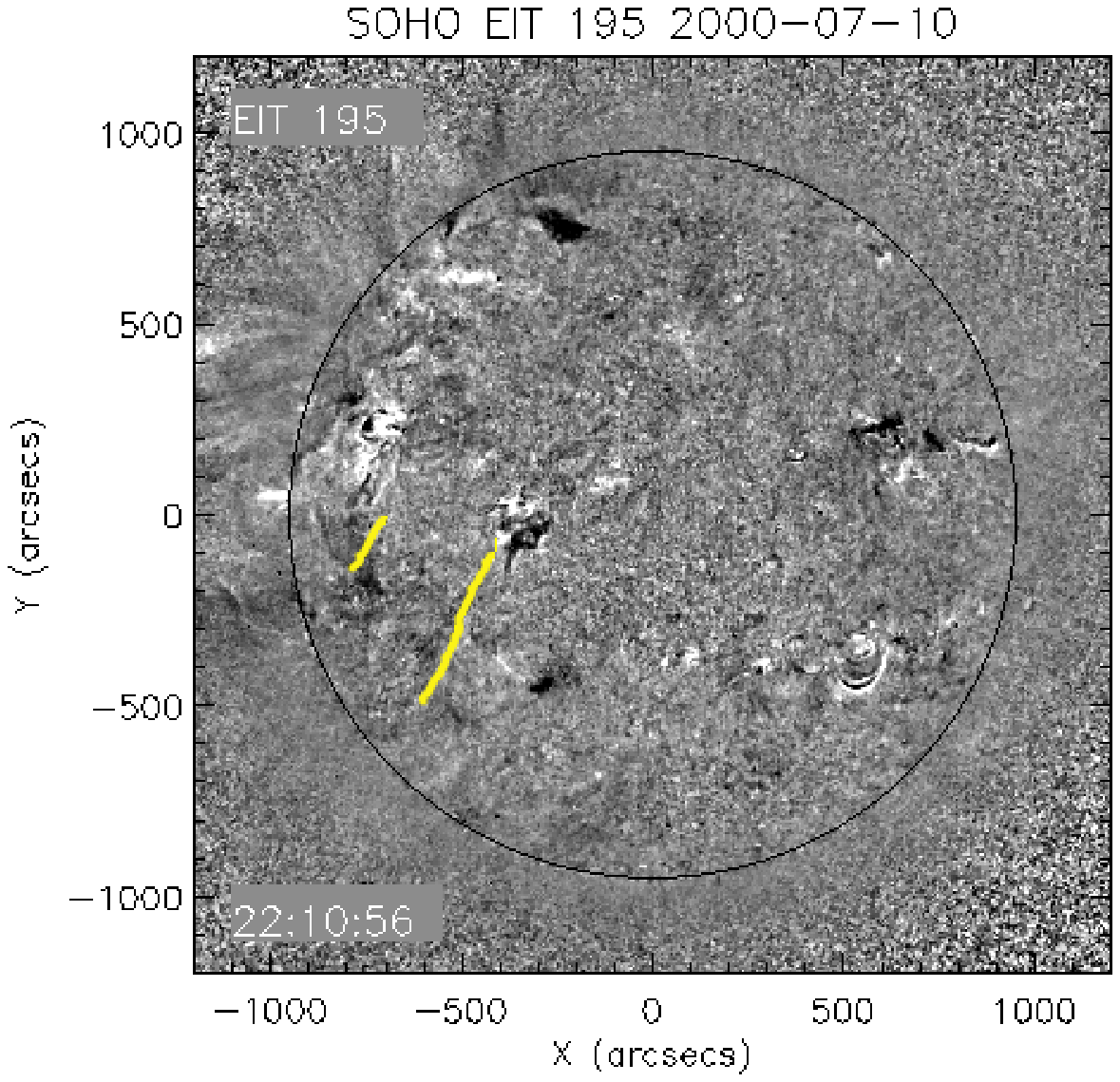}
            \hspace*{-0.08\textwidth}
            \includegraphics[width=0.53\textwidth,clip=]{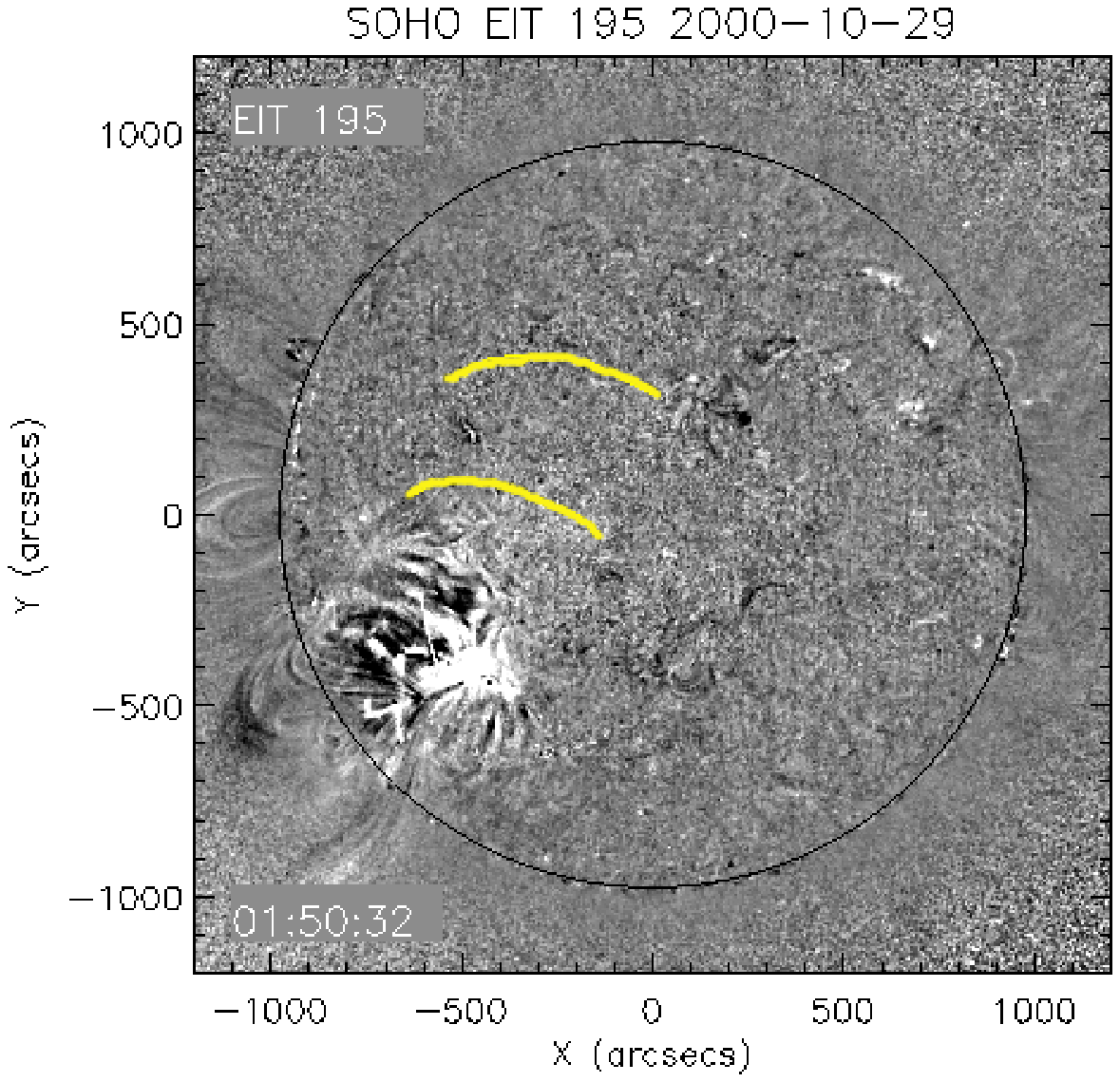}}
            \vspace*{-0.06\textwidth}
\centerline{\hspace*{0.03\textwidth}
            \includegraphics[width=0.53\textwidth,clip=]{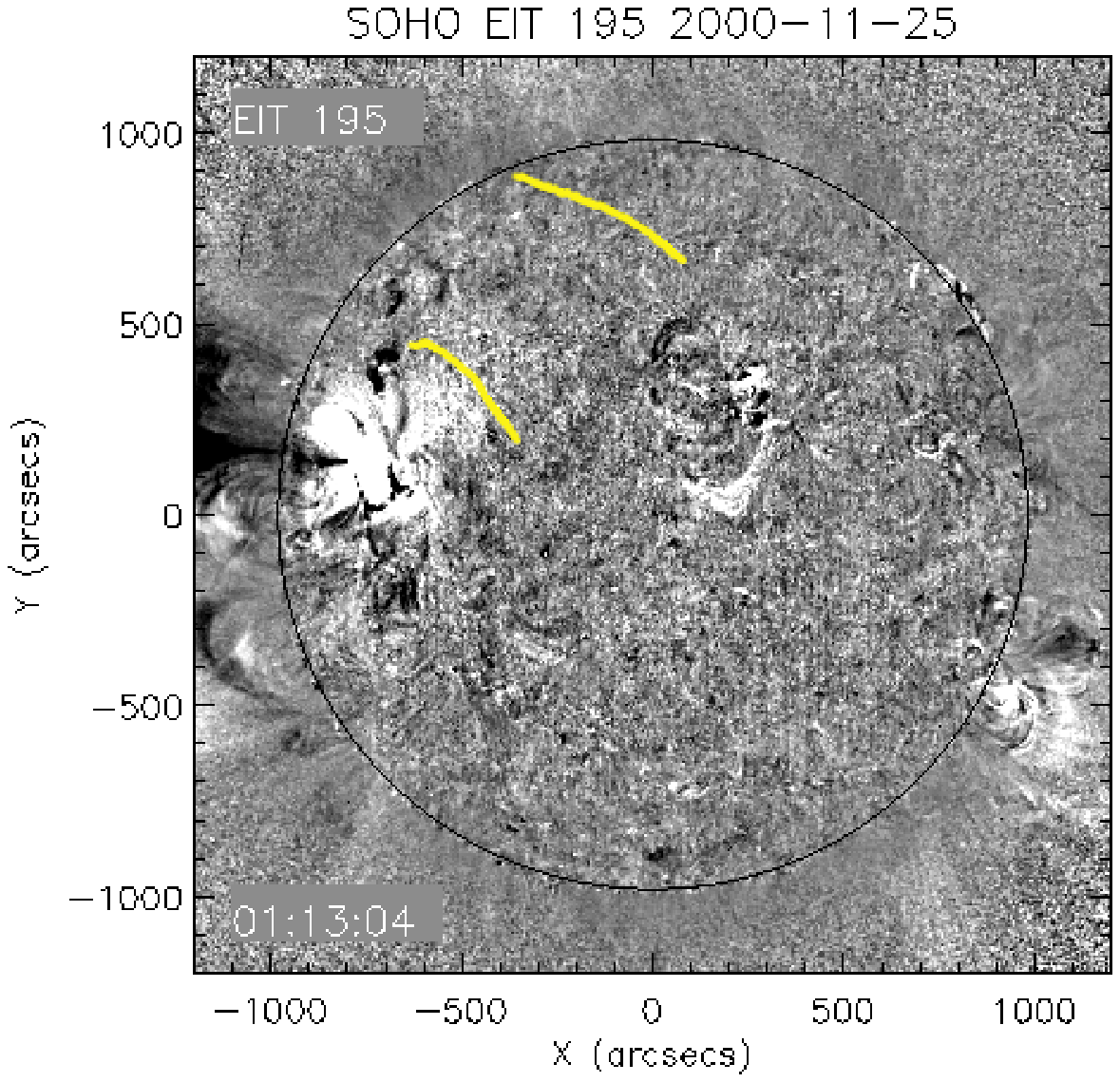}
            \hspace*{-0.08\textwidth}
            \includegraphics[width=0.53\textwidth,clip=]{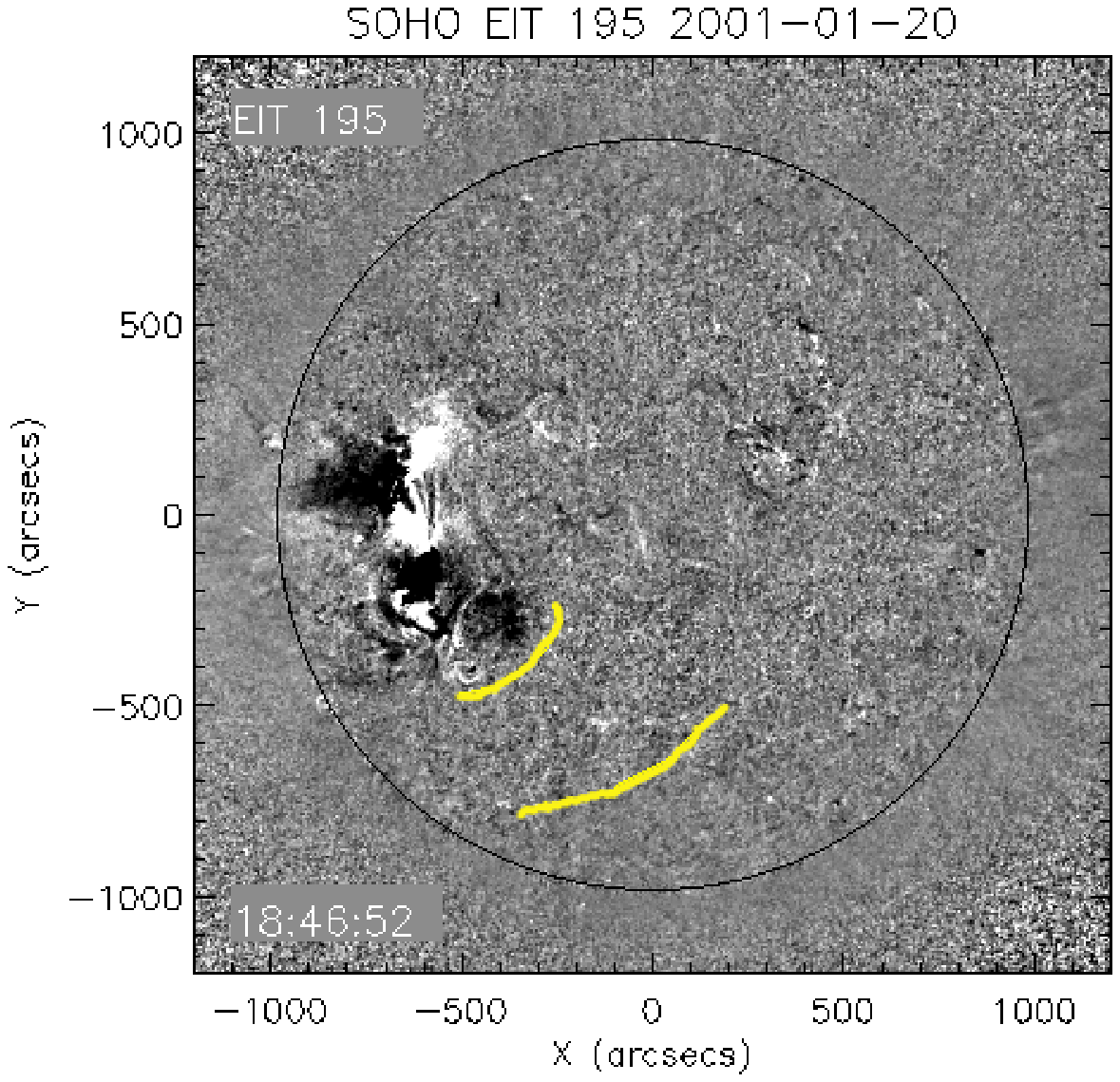}}
            \vspace*{-0.06\textwidth}
\centerline{\hspace*{0.03\textwidth}
            \includegraphics[width=0.53\textwidth,clip=]{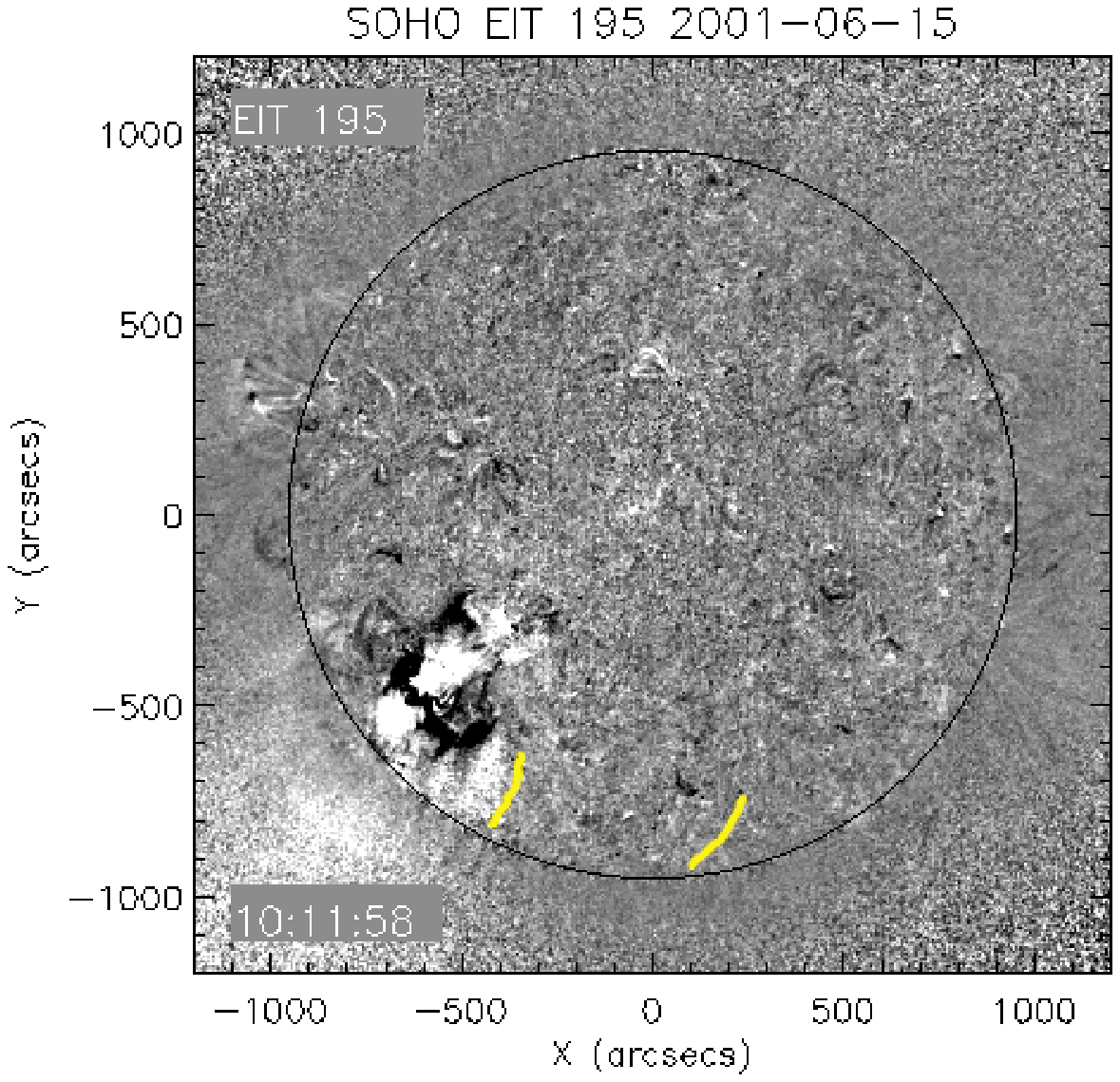}
            \hspace*{-0.08\textwidth}
            \includegraphics[width=0.53\textwidth,clip=]{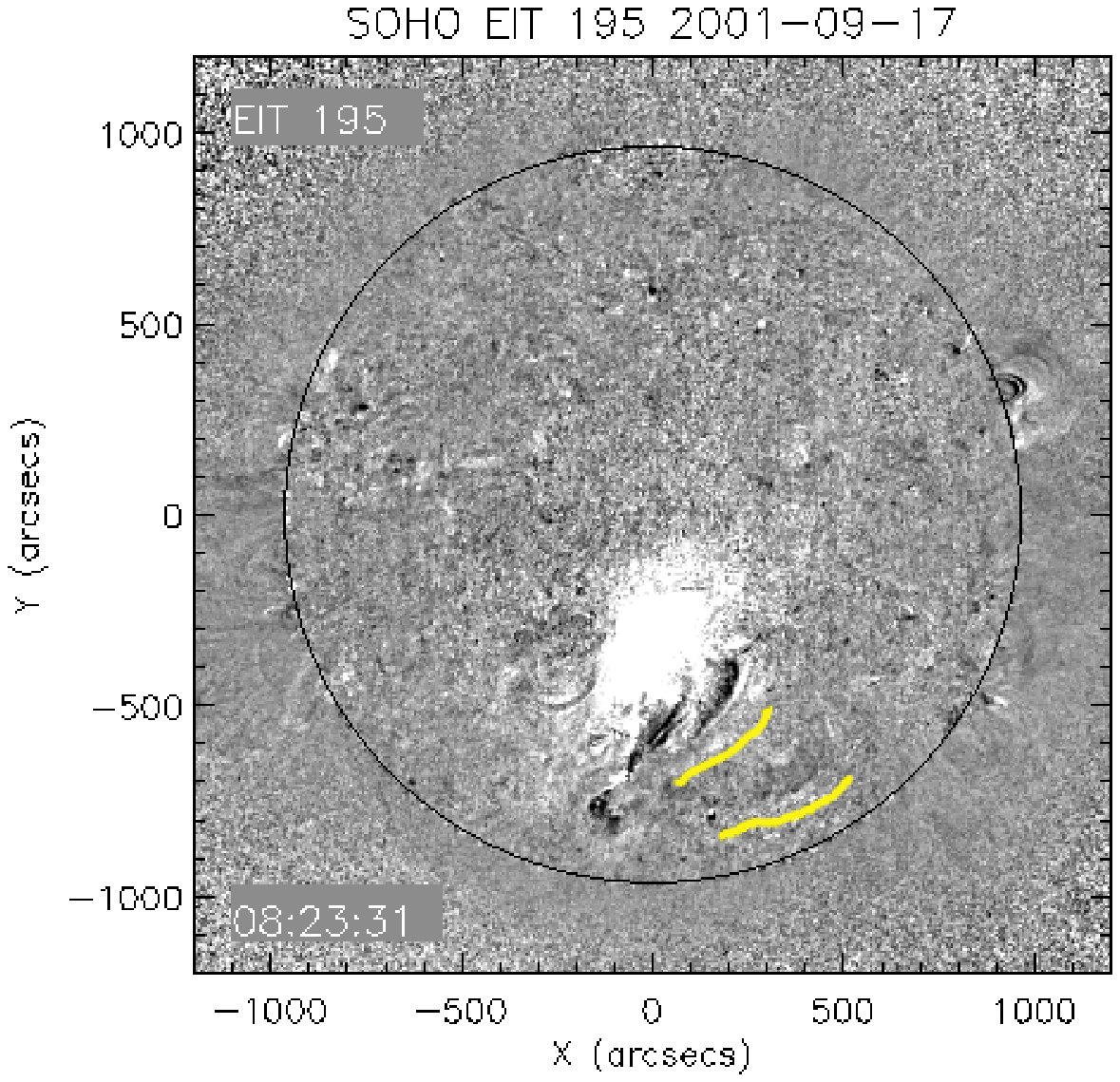}}
            \vspace*{-0.03\textwidth}
\caption{(continued)}
            \vspace*{0.05\textwidth}
\addtocounter{figure}{-1}
   \end{figure}

 \begin{figure}[ht!]
\centerline{\hspace*{0.03\textwidth}
            \includegraphics[width=0.53\textwidth,clip=]{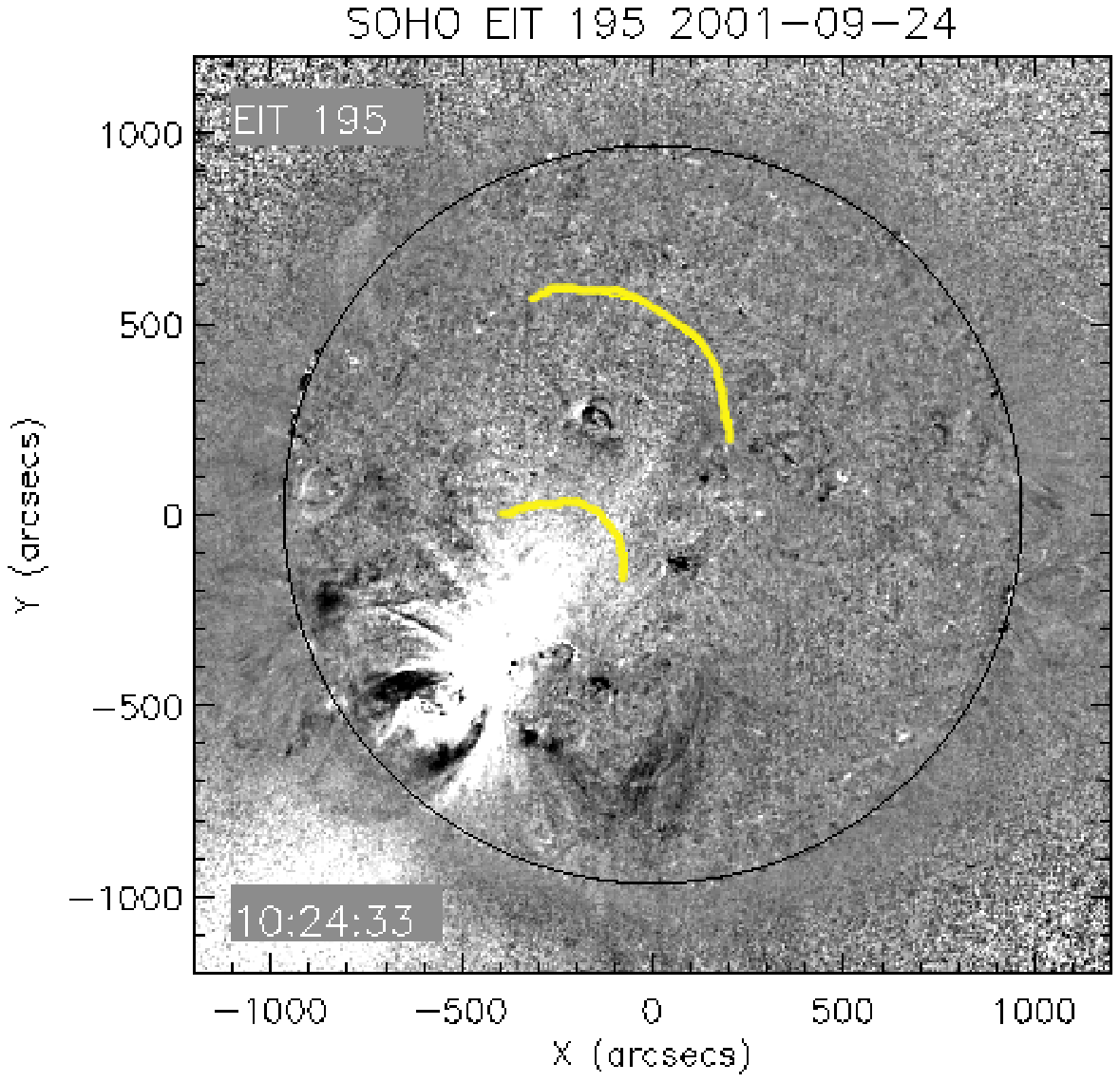}
            \hspace*{-0.08\textwidth}
            \includegraphics[width=0.53\textwidth,clip=]{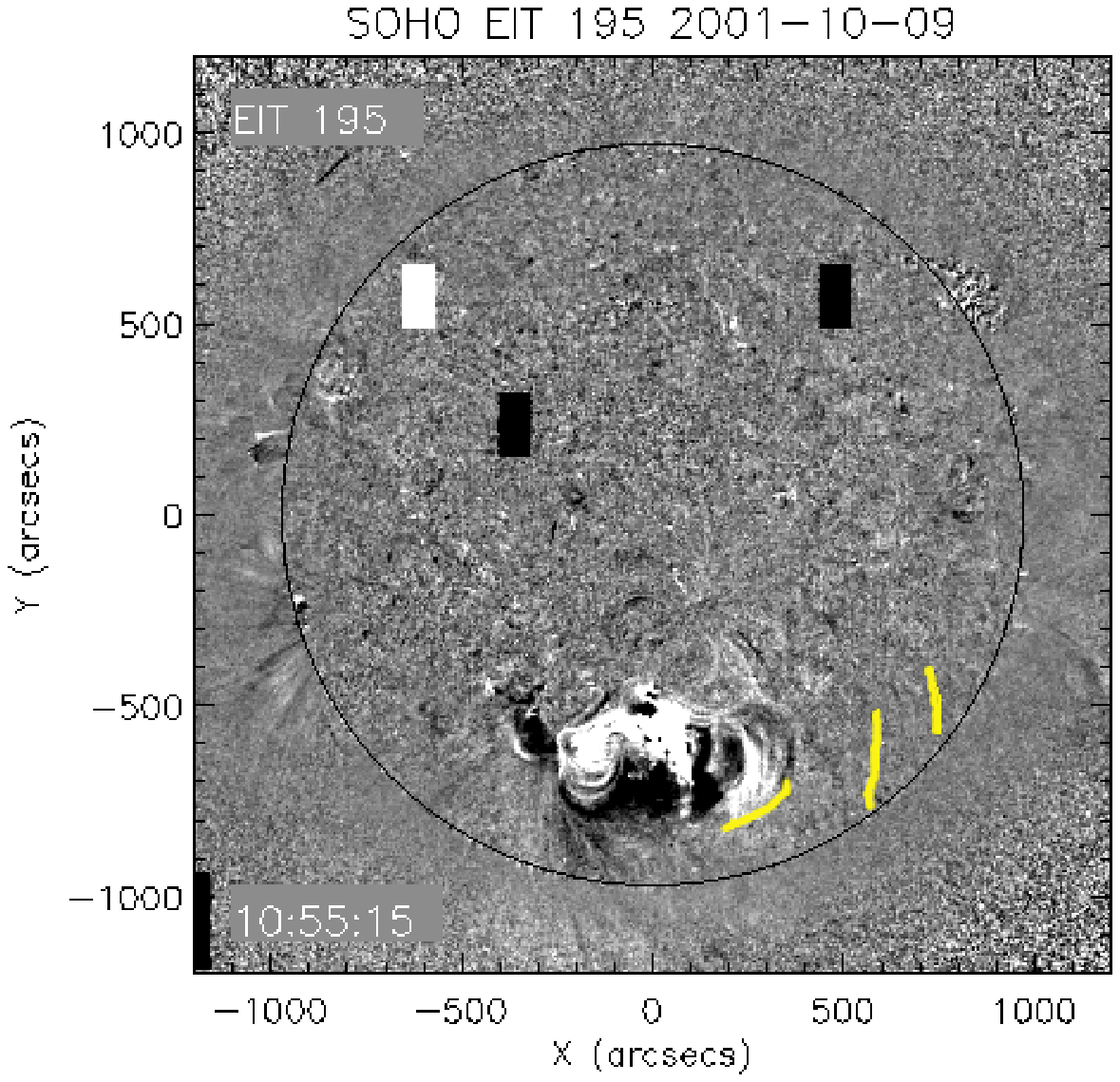}}
            \vspace*{-0.06\textwidth}
\centerline{\hspace*{0.03\textwidth}
            \includegraphics[width=0.53\textwidth,clip=]{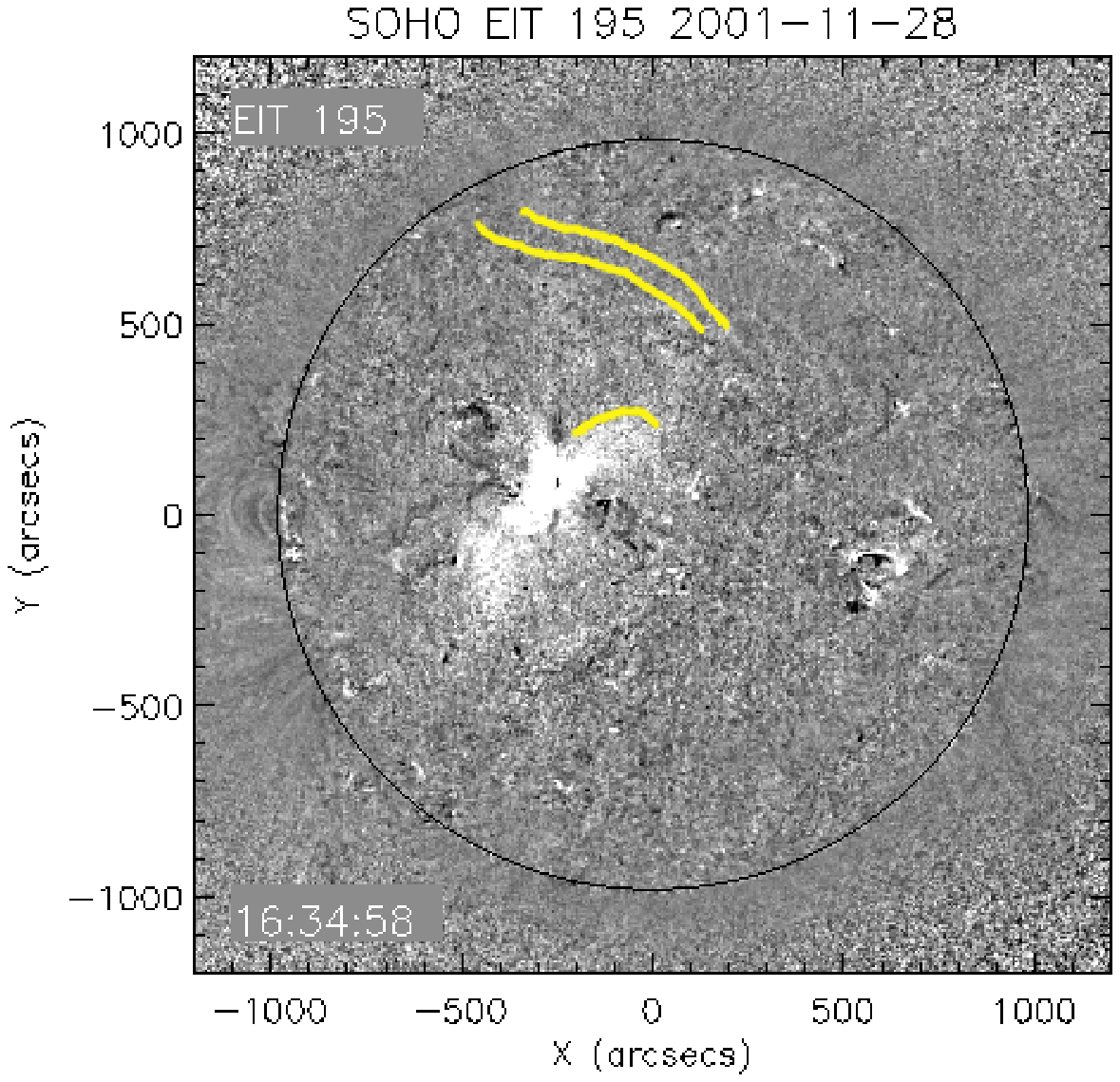}
            \hspace*{-0.08\textwidth}
            \includegraphics[width=0.53\textwidth,clip=]{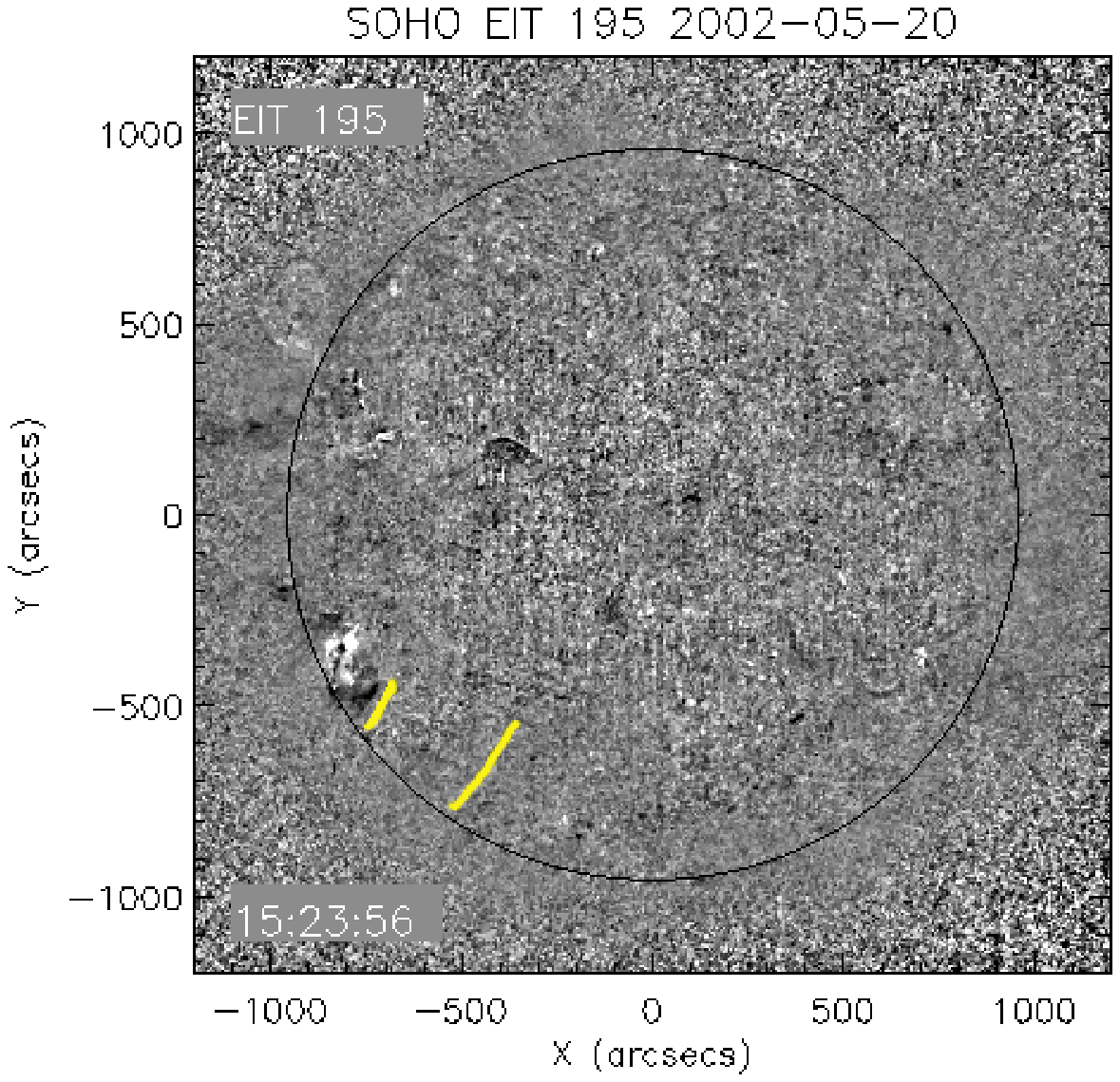}}
            \vspace*{-0.06\textwidth}
\centerline{\hspace*{0.03\textwidth}
            \includegraphics[width=0.53\textwidth,clip=]{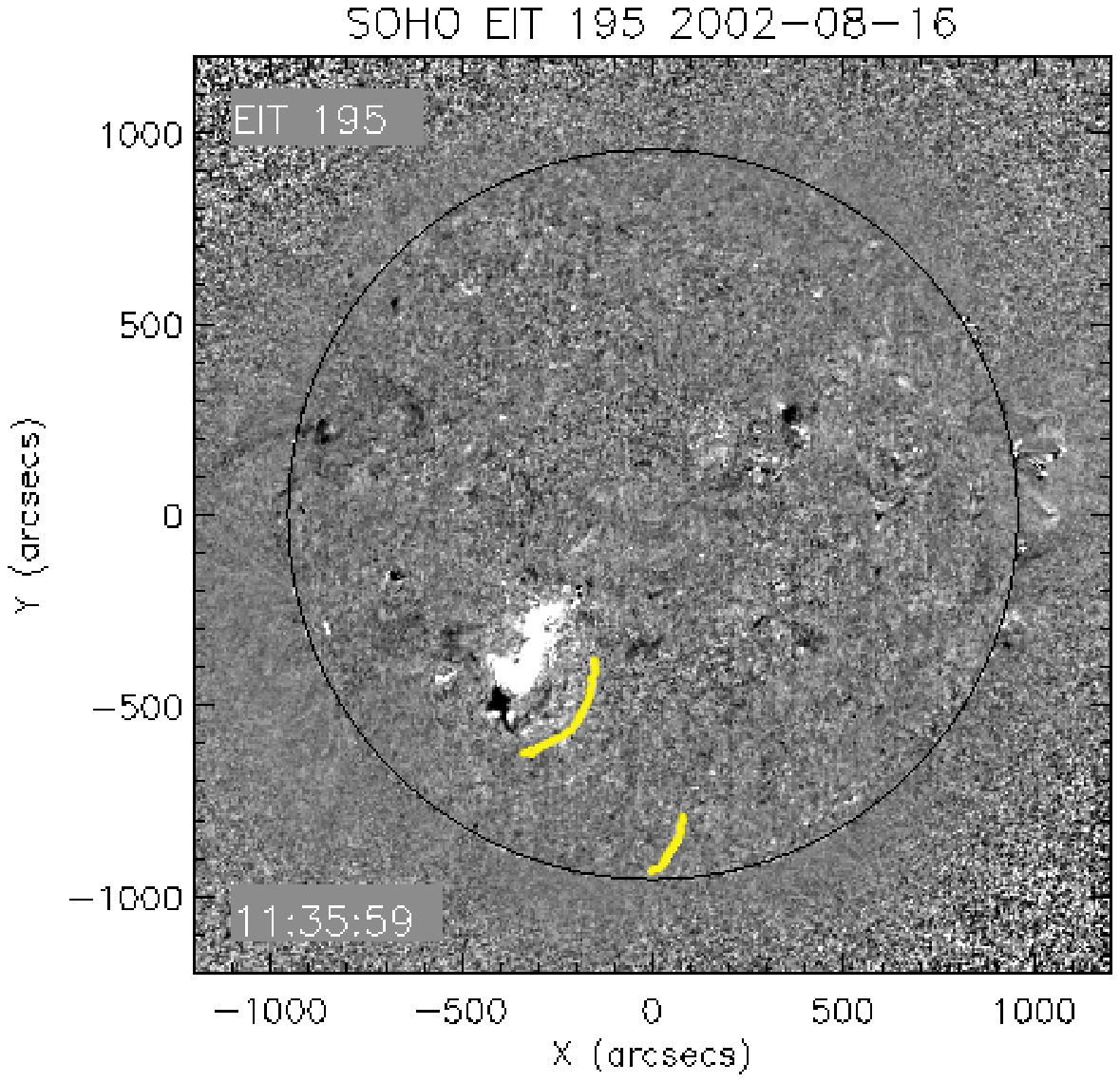}
            \hspace*{-0.08\textwidth}
            \includegraphics[width=0.53\textwidth,clip=]{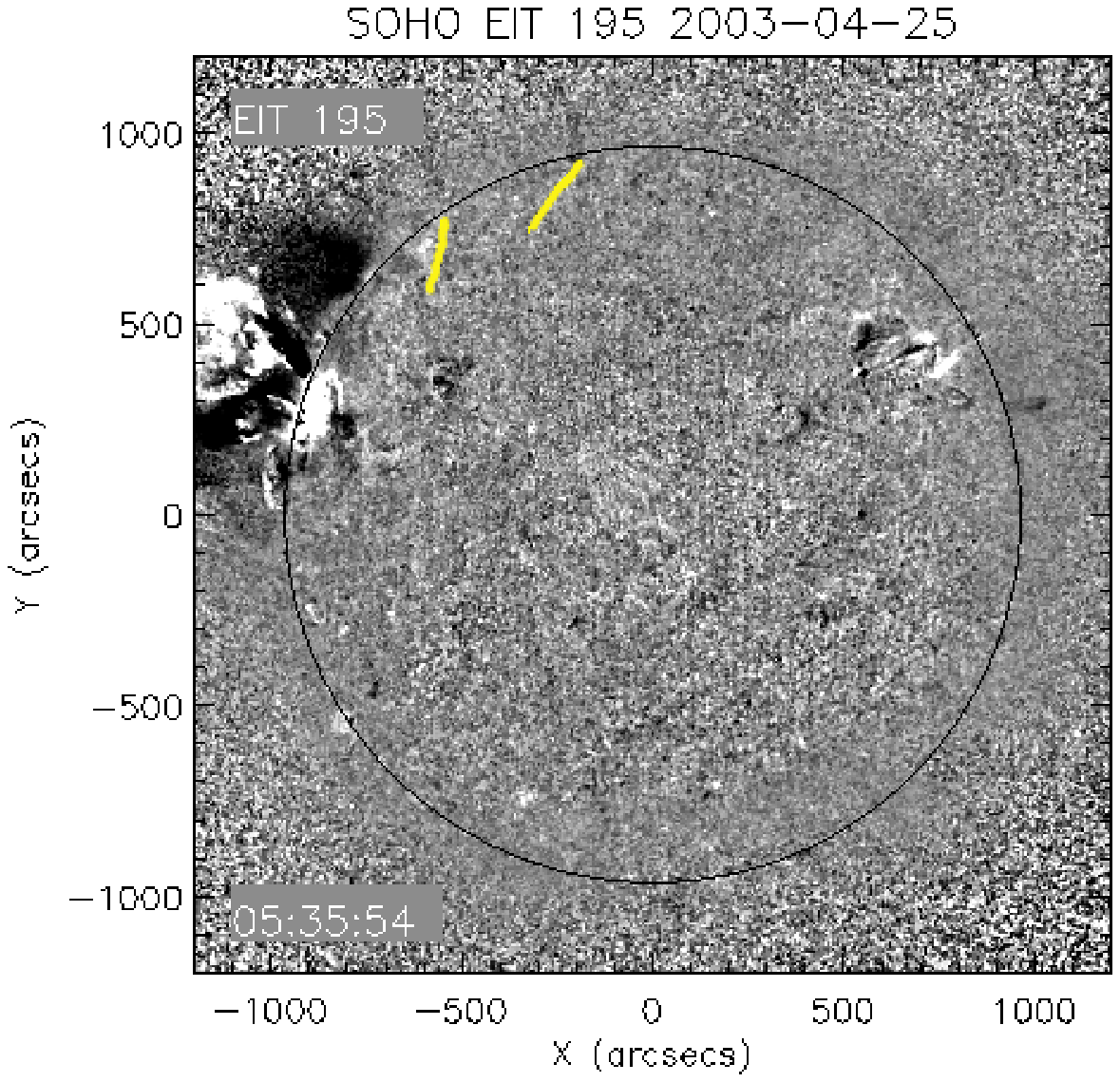}}
            \vspace*{-0.03\textwidth}
\caption{(continued)}
            \vspace*{0.05\textwidth}
\addtocounter{figure}{-1}
   \end{figure}

\begin{figure}[ht!]
\centerline{\hspace*{0.03\textwidth}
            \includegraphics[width=0.53\textwidth,clip=]{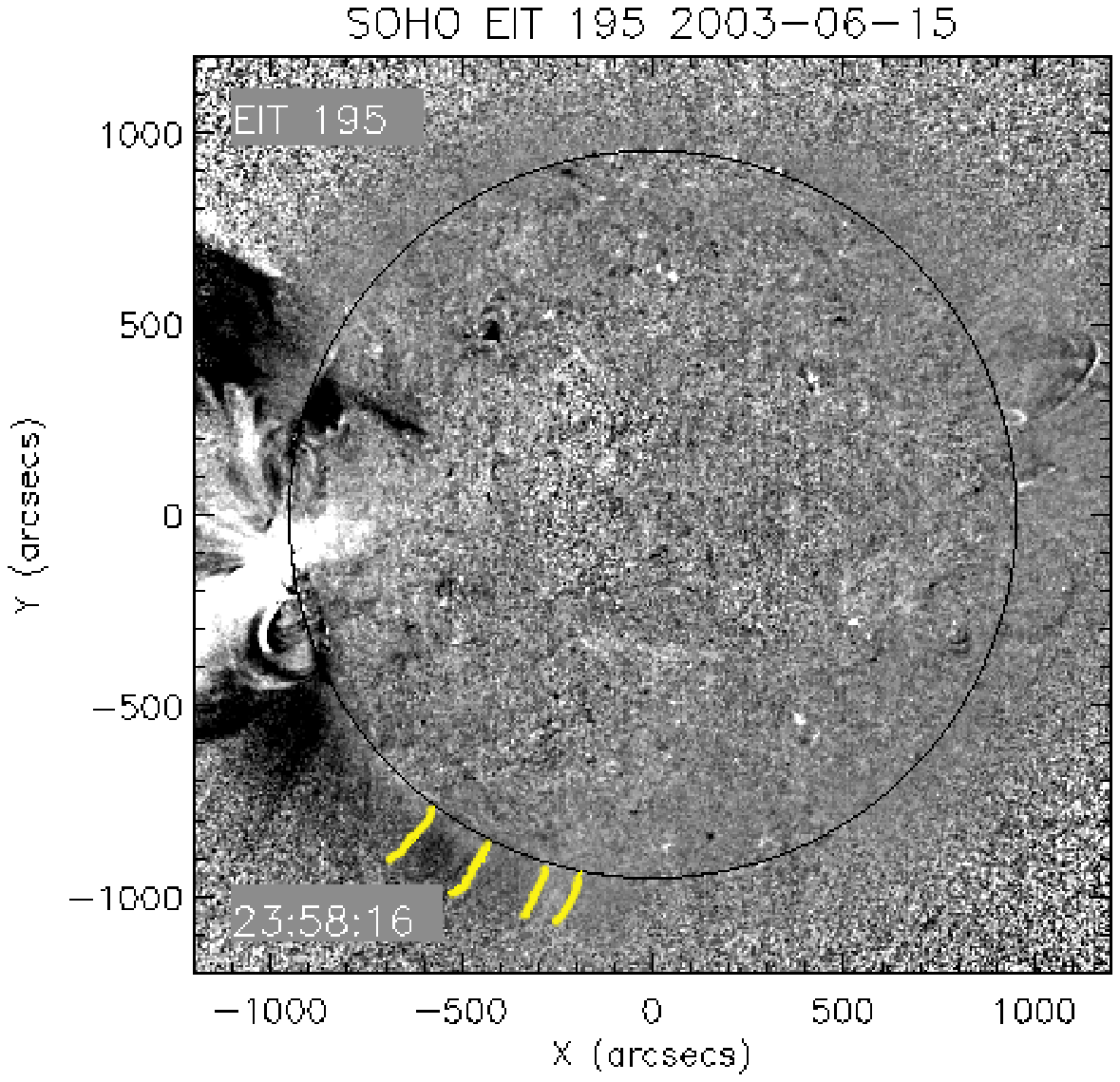}
            \hspace*{-0.08\textwidth}
            \includegraphics[width=0.53\textwidth,clip=]{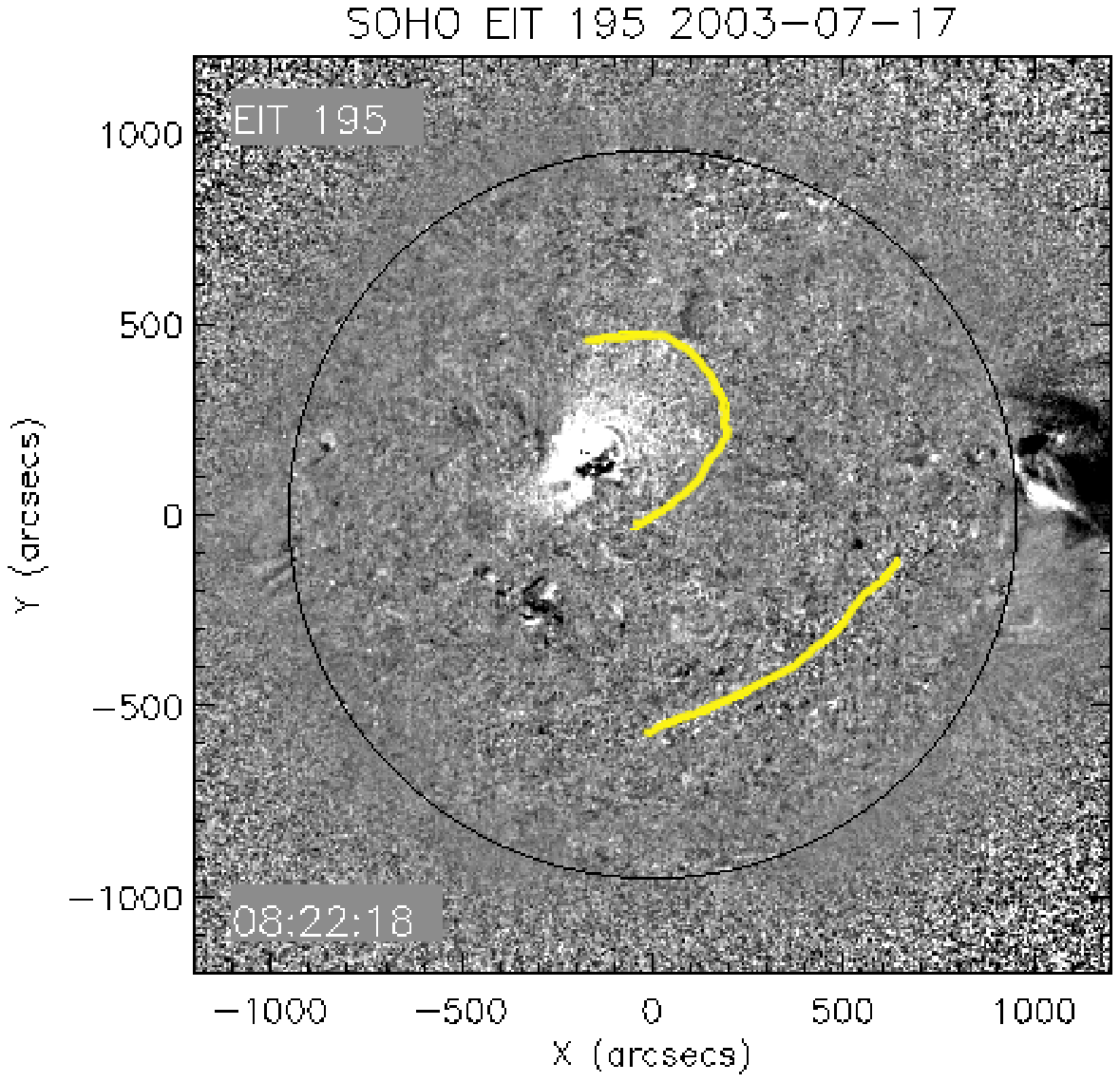}}
            \vspace*{-0.06\textwidth}
\centerline{\hspace*{0.03\textwidth}
            \includegraphics[width=0.53\textwidth,clip=]{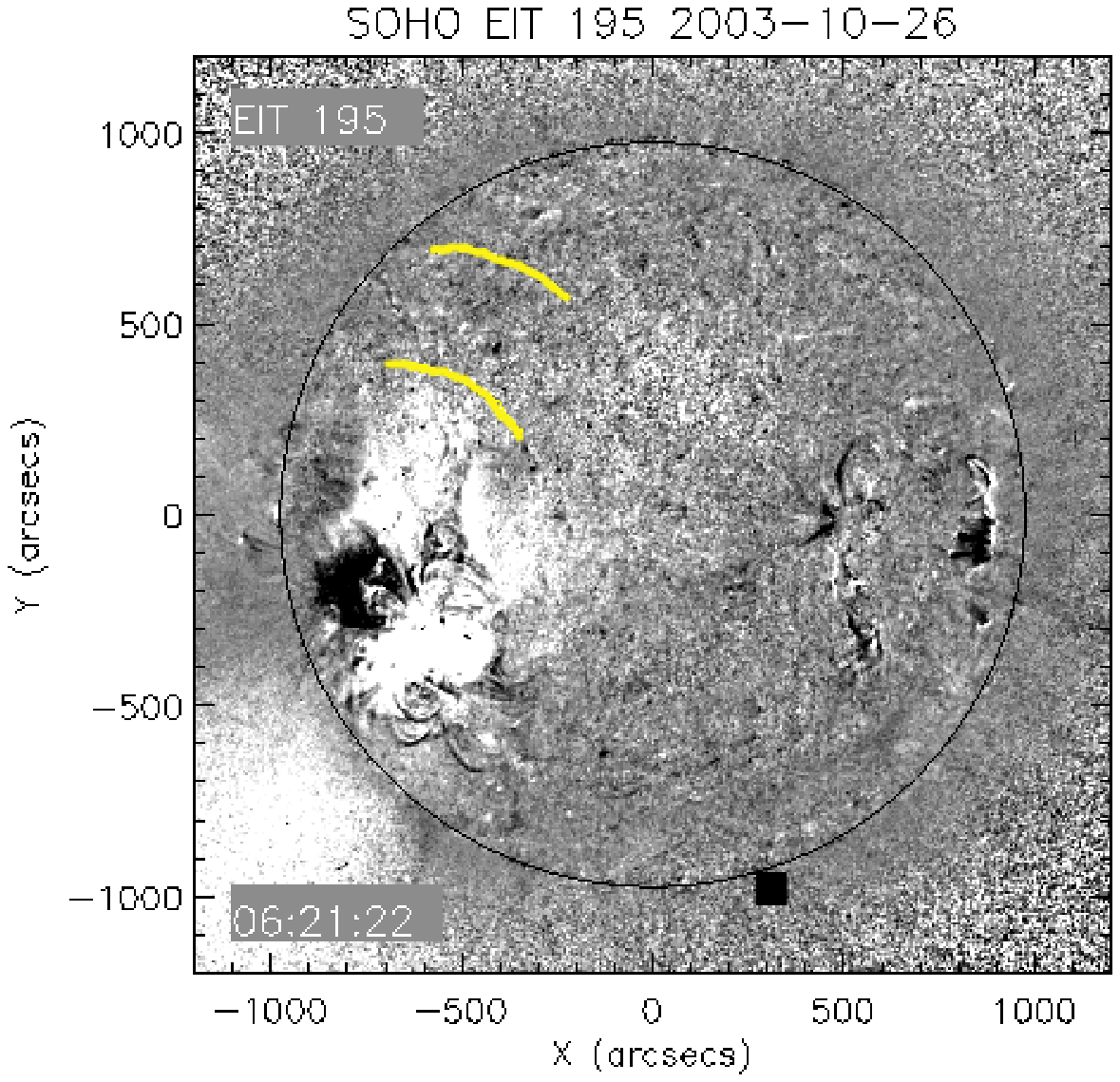}
            \hspace*{-0.08\textwidth}
            \includegraphics[width=0.53\textwidth,clip=]{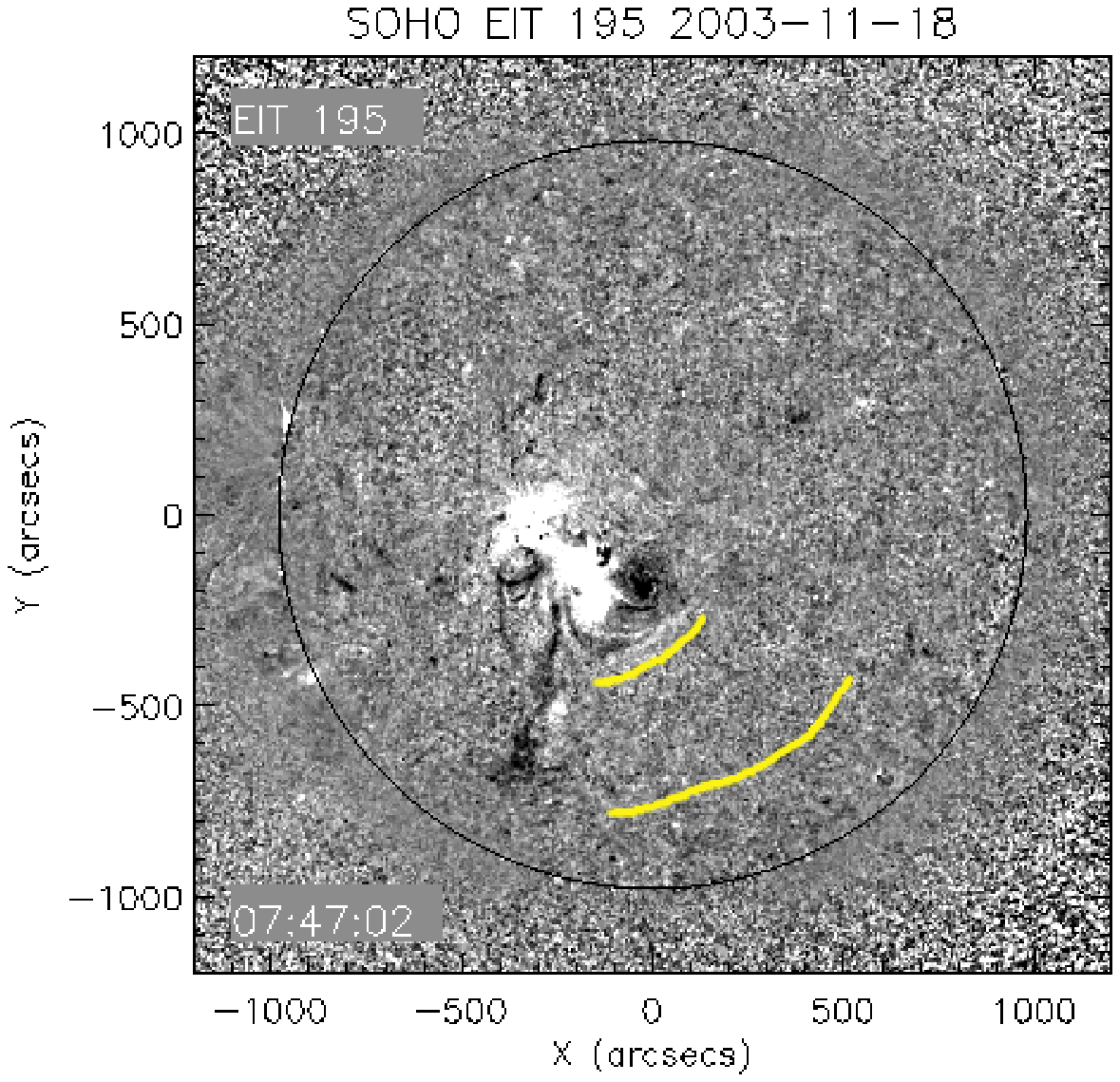}}
            \vspace*{-0.06\textwidth}
\centerline{\hspace*{0.03\textwidth}
            \includegraphics[width=0.53\textwidth,clip=]{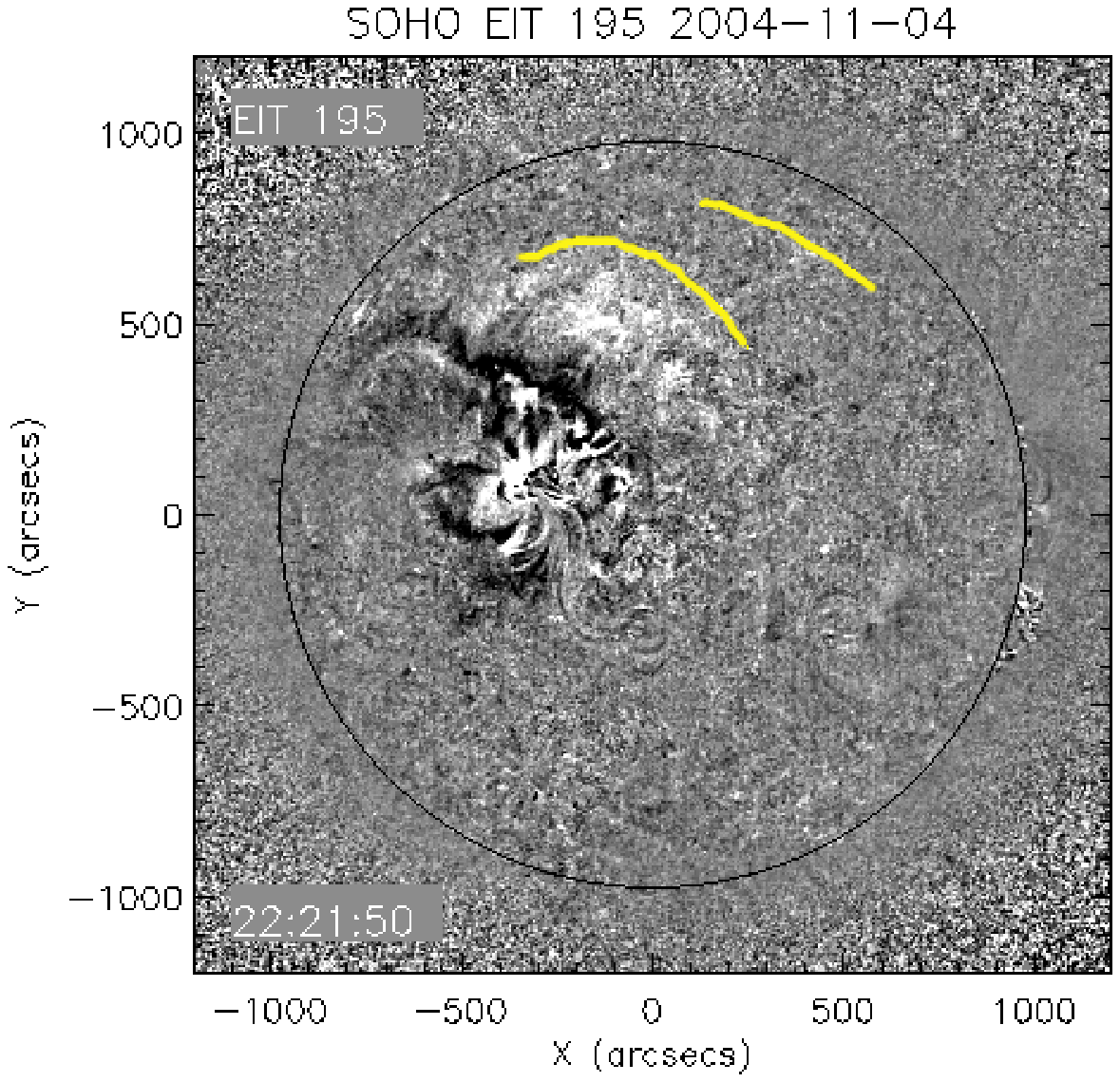}
            \hspace*{-0.08\textwidth}
            \includegraphics[width=0.53\textwidth,clip=]{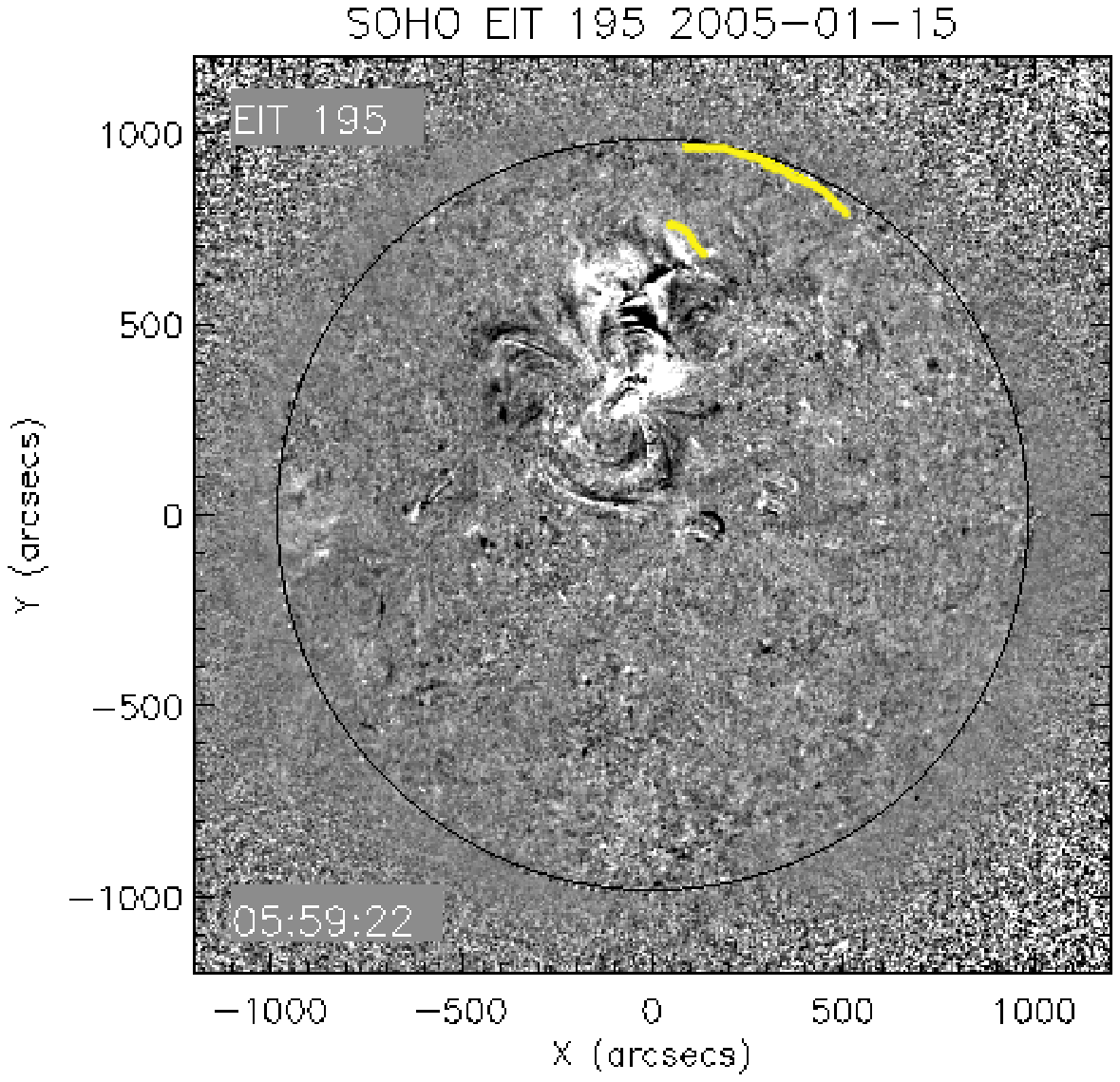}}
            \vspace*{-0.03\textwidth}
\caption{(continued)}
            \vspace*{0.05\textwidth}
\addtocounter{figure}{-1}
   \end{figure}

\begin{figure}[h!]
\centerline{\hspace*{0.03\textwidth}
            \includegraphics[width=0.53\textwidth,clip=]{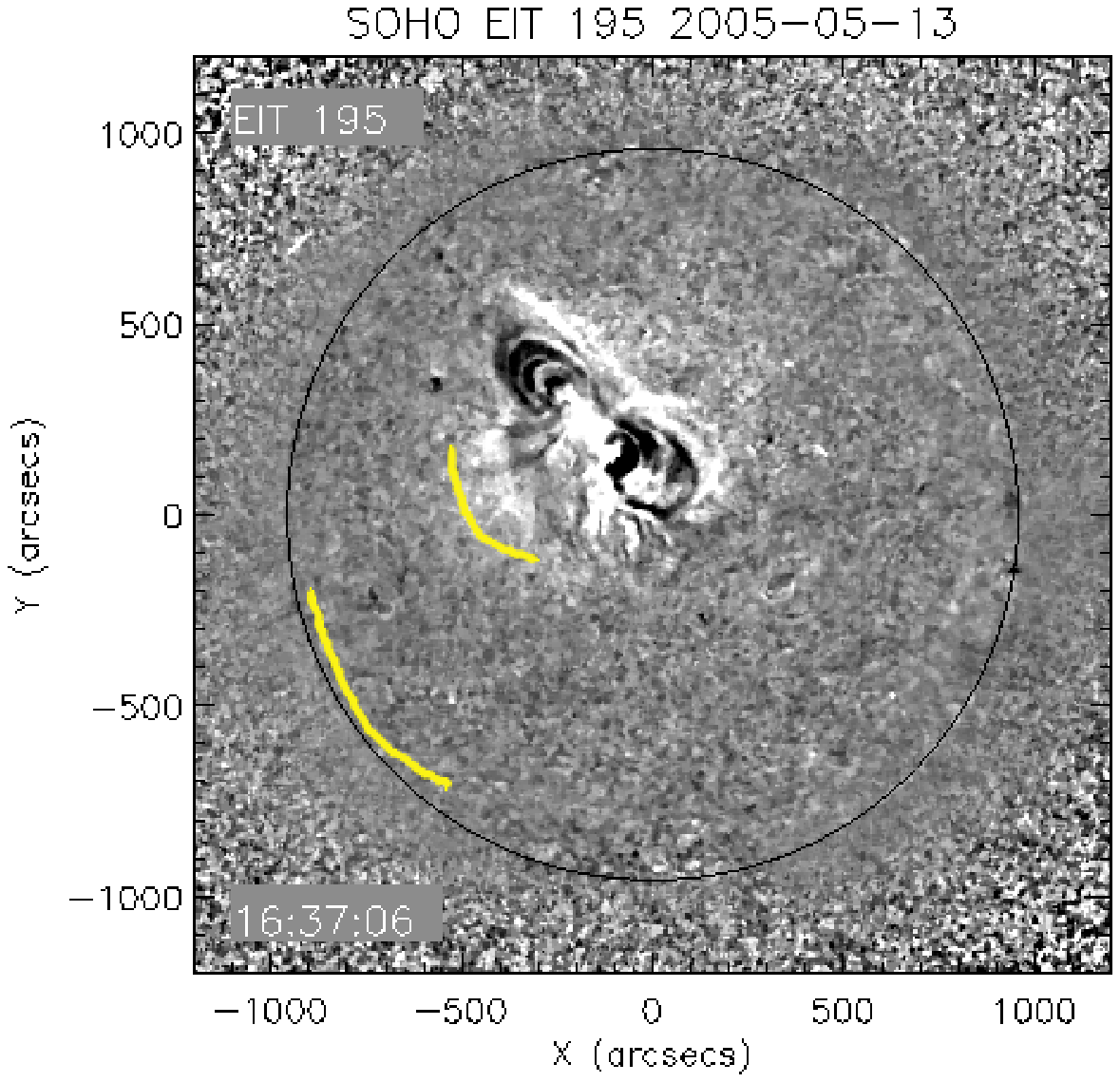}
            \hspace*{-0.08\textwidth}
            \includegraphics[width=0.53\textwidth,clip=]{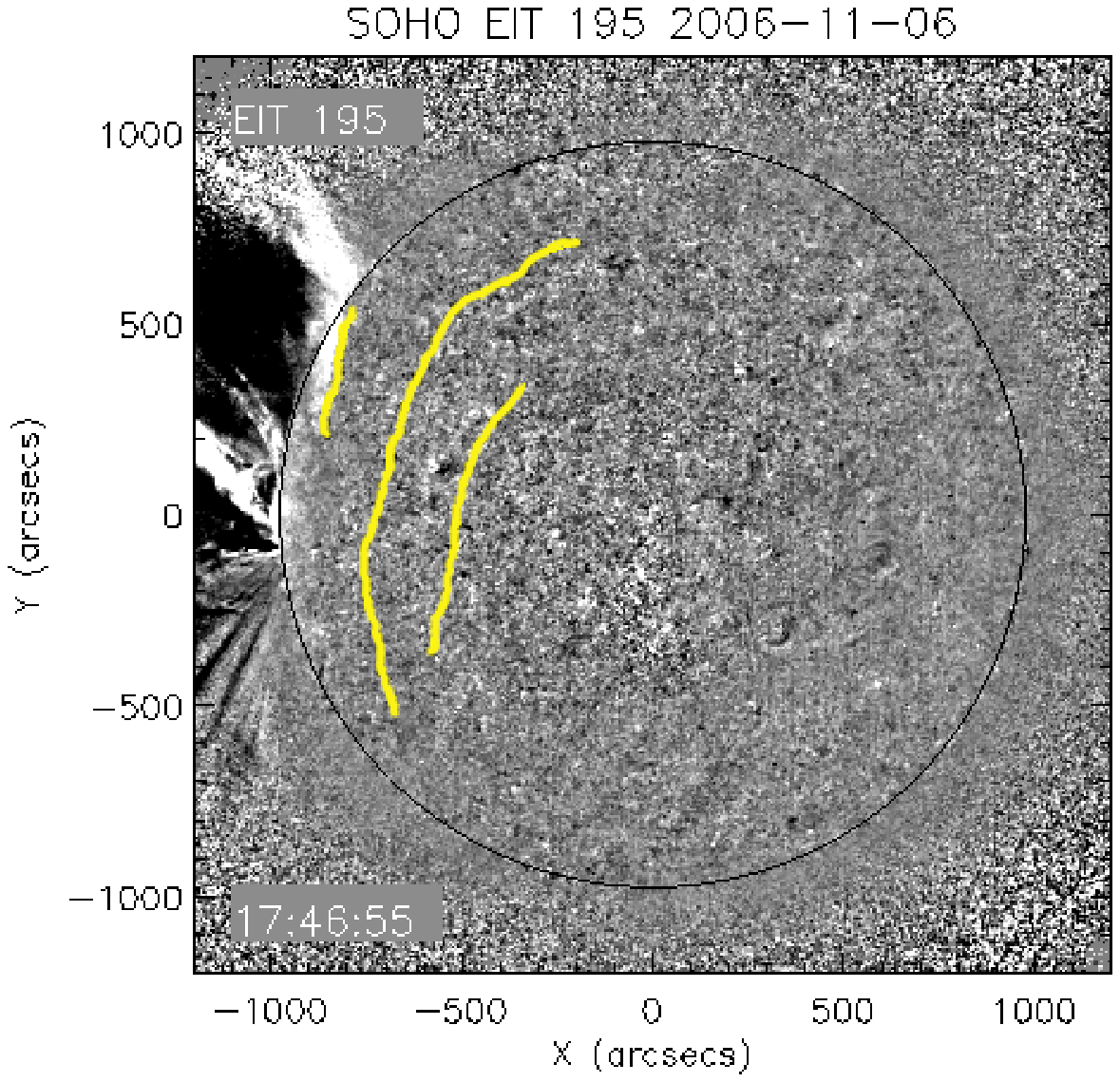}}
            \vspace*{-0.03\textwidth}
\caption{(continued)}
\addtocounter{figure}{-1}
   \end{figure}

\begin{table}[t!]
\caption[]{Properties of 26 eastern EIT waves for which at least two wave fronts could be identified. We list the event date, EIT onset time $[t_{\rm EIT,on}]$, averaged EIT speed $[v_{\rm av,EIT}]$, wave travel time from the flare location to the Parker-spiral longitude, arrival time of the EIT wave at the Parker spiral [$t_{\rm EIT}$], and a comment on the magnetic environment.\\
AR: active region; FE: filament eruption; {\it n}: next day, {\it l}: lower limit for the wave speed.}
\label{T-EIT_waves}
\tiny
\vspace{0.5cm}
\begin{tabular}{r@{ }l@{}rclll}
\hline
\multicolumn{2}{c}{Event}        & \multicolumn{4}{c}{EIT wave properties}     & Magnetic environment/activity at \\
\multicolumn{2}{c}{date}         & $t_{\rm EIT,on}$ & $v_{\rm av,EIT}$     & travel    & $t_{\rm EIT}$  & western helio-longitudes \\
\multicolumn{2}{c}{I/S/V}        &  {[}UT{]}        & {[}km$\,$s$^{-1}${]} & {[}min{]} & {[}UT{]}       &                  \\
%(1)      & (2)                 & (3)                 &  (4)    & (5)            & (6)                 \\
\hline
01 Apr &1997 S& 13:54   & 260 & 57 & 14:51$\pm$13 &          \\
24 Sep &1997 S& 02:41   & 340 & 50 & 03:31$\pm$10 & loop system at W-limb \\
29 Apr &1998 S& 16:11   & 235 & 79 & 17:30$\pm$14 &            \\
%990503 S&  -      & u   & NE              & -  &   -          & AR NW-limb\\
18 Jan &2000 S& 17:05   & 450 & 38 & 17:43$\pm$7  & loop system at center-W direction \\
17 Feb &2000 S& 20:17   & 480 & 28 & 20:45$\pm$7  & M2.5 (18:41) S26W14; EIT wave (18:47) \\
06 Jun &2000 I& 15:02   & 345 & 38 & 15:40$\pm$10 & same AR: X1.1 (13:30); FE $\gtrsim$14:00\\
10 Jul &2000 V& 22:05   & 345 & 57 & 23:02$\pm$10 & M1.9 (19:55) N16W43; FE $\gtrsim$20:00\\
29 Oct &2000 I& 01:47   & 595 & 25 & 02:16$\pm$6  & \\
25 Nov &2000 V& 01:00   & 565 & 39 & 01:39$\pm$6  & X1.8 (21:43) N21W14; EIT wave $\sim$22:00  \\
20 Jan &2001 S& 18:41   & 205 & 115 & 20:36$\pm$16 & second EIT wave $\sim$21:20 \\
15 Jun &2001 S& 10:05   & 550 & 39 & 10:44$\pm$6  & EIT dome? (10:12) \\
17 Sep &2001 S& 08:17   & 345$^l$& 33 & 08:50$\pm$10 & loop systems on W-limb\\
24 Sep &2001 I& 10:19   & 585$^{l}$& 26 & 10:45$\pm$6& slowly developing CME \\
09 Oct &2001 S& 10:49   & 250 & 51 & 11:40$\pm$13 & coronal restructuring in W-direction\\
28 Nov &2001 S& 16:29   & 255 & 57 & 17:26$\pm$13 & AR in the west\\
20 May &2002 I& 15:18   & 330$^l$& 72 & 16:30$\pm$10 & FE $\gtrsim$15:50 near E-limb\\
16 Aug &2002 S& 11:30   & 530 & 23 & 11:53$\pm$6  & large coronal restructuring \\
25 Apr &2003 S& 05:30   & 320 & 79 & 06:49$\pm$10 & ARs close to W-limb     \\
15 Jun &2003 I& 23:52   & 170$^{l}$ & 148 & 02:20$^{n}\pm$20 &  \\
17 Jul &2003 S& 08:16   & 540$^l$ & 27 & 08:43$\pm$6  & CME $\sim$08:00 at W-limb\\
26 Oct &2003 V& 06:15   & 330 & 61 & 07:16$\pm$10 & sigmoid/FE $\sim$06:10\\
%031028 V&  -      & s   &  -  &   -         & major event    \\
18 Nov &2003 S& 07:41   & 455 & 23 & 08:04$\pm$7  & flare/wave/FE? ($\sim$09:30) at E-limb  \\
%040107 S& 10:19   & s   & -  &   -         & ARs (disc center, W-limb)\\
04 Nov &2004 S& 22:16   & 390 & 42 & 22:58$\pm$9  & \\
%%041202 S& 23:54   & 420 & 27 & 00:21$^{n}\pm$8 & ARs in west\\
15 Jan &2005 S& 05:53   & 655$^{l}$ & 14 & 06:07$\pm$5  & \\
13 May &2005 S& 16:32   & 475 & 24 & 16:56$\pm$7  & AR at W-limb \\
06 Nov &2006 S& 17:41   & 325 & 99 & 19:20$\pm$10 & AR (center-W)\\
\hline
\end{tabular}
\footnotetext{}
\end{table}

\section{Results}         %%%%%%%%%%%%%%%%%%%%%%%%%%%%%%%%%%%%%%%% Section 3
      \label{S-Results}

\subsection{SEP Events and EIT Waves During Solar Cycle 23: Overall Association}
      \label{S-Assoc_rates}

\begin{table}[t!]
\caption[]{Association rates (Yes/No) given in percentages between particle events and EIT waves for the different categories of SEP events (ICME, vicinity and solar wind) calculated separately for western, eastern and the entire sample of events. Events with no EIT data and uncertain SEP--flare/CME association are dropped when calculating the association rates. The total number of events in each category is given in the last column of the table.}
\label{T-Assoc_rates}
\tiny
\vspace{0.1cm}
\begin{tabular}{lr@{}lr@{}lr@{}lc}
\hline
                    & \multicolumn{4}{c}{EIT waves}  & \multicolumn{2}{c}{Events with no}  & All \\
SEP events          & \multicolumn{2}{c}{Yes} & \multicolumn{2}{c}{No} & \multicolumn{2}{c}{EIT data}  & SEP \\
                    &              &          &        &               & \multicolumn{2}{c}{+ uncertain}                & events \\
                    &              &          &        &               & \multicolumn{2}{c}{SEPs--flare/CME} & \\ &              &          &        &               & \multicolumn{2}{c}{association} & \\
\hline
{\bf East}          & {\bf 91$\,$\%} &\,\,{\bf (29/32)}    &  {\bf 9$\,$\%} &\,\,{\bf (3/32)} & $\quad$$\quad${\bf 13}&+{\bf 3} &  {\bf 48} \\
$\quad$ICME         &  100$\,$\% &\,\,(5/5)          & 0$\,$\% &\,\,(0/5)            & $\quad$$\quad$2&+1 &  8 \\
$\quad$Vicinity     &  100$\,$\% &\,\,(4/4)          & 0$\,$\% &\,\,(0/4)            & $\quad$$\quad$2&+1 &  7 \\
$\quad$Solar wind   &  87$\,$\% &\,\,(20/23)         & 13$\,$\% &\,\,(3/23)          & $\quad$$\quad$9&+1 &  33 \\
&&&&\\
{\bf West}          &  {\bf 86$\,$\%} &\,\,{\bf (104/121)} &  {\bf 14$\,$\%} &\,\,{\bf (17/121)} & $\quad${\bf 10}& & {\bf 131} \\
$\quad$ICME         &  83$\,$\% &\,\,(19/23)         & 17$\,$\% &\,\,(4/23)          & $\quad$$\quad$3& &  26 \\
$\quad$Vicinity     &  83$\,$\% &\,\,(29/35)         & 17$\,$\% &\,\,(6/35)          & $\quad$$\quad$2& &  37\\
$\quad$Solar wind   &  89$\,$\% &\,\,(56/63)         & 11$\,$\% &\,\,(7/63)          & $\quad$$\quad$5& &  68 \\
&&&&\\
{\bf East + West}   &  {\bf 87$\,$\%} &\,\,{\bf (133/153)} &  {\bf 13$\,$\%} &\,\,{\bf (20/153)} & $\quad$$\quad${\bf 23}&+{\bf 3} &  {\bf 179} \\
$\quad$ICME         &  86$\,$\% &\,\,(24/28)         &  14$\,$\% &\,\,(4/28)  & $\quad$$\quad$5&+1  &  34 \\
$\quad$Vicinity     &  85$\,$\% &\,\,(33/39)         &  15$\,$\% &\,\,(6/39)    & $\quad$$\quad$4&+1  &  44 \\
$\quad$Solar wind   &  88$\,$\% &\,\,(76/86)         &  12$\,$\% &\,\,(10/86)   & $\quad$$\quad$14&+1 &  101 \\
\hline
\end{tabular}
\end{table}

We found a high association rate of about 87$\,$\% between all SEP events and the accompanying EIT waves (Table~\ref{T-Assoc_rates}). A very similar association rate is found for the western SEPs and a slightly higher one ($\approx$90$\,$\%) for the eastern ones. Each longitudinal group is further subdivided into the different categories according to the IP conditions, introduced in Section~\ref{S-IP}. One notices that all ICME/vicinity SEP events in the East are accompanied by EIT waves, whereas the association rate for the solar-wind events is the same for the eastern and western group (87\,--\,89\,\%). The association rate is slightly lower for the western ICME/vicinity SEP events (83$\,$\%) than for the eastern ones.

The reliability of the association rates calculated here depends on the correct SEP--parent activity identification adopted from \inlinecite{2010JGRA..11508101C}. During the detailed analysis of the eastern events however, we found that for three No-association cases the parent flare/CME might be a later event pair that is associated with an EIT disturbance. These three cases are dropped from the final evaluation, but when kept, one obtains 83$\,$\% (29/35) positive and 17$\,$\% (6/35) negative association, comparable to the western category. No such follow-up work is done for the western events and hence their correlation rates might be subject to change.

In order to evaluate the opposite association, starting with a set of EIT waves and finding the percentage of EIT disturbances associated with SEP events, we need a list of EIT waves during Solar Cycle 23. Such a complete list of EIT waves is still lacking. The only available list of EIT wave transients in Solar Cycle 23 is from \inlinecite{2009ApJS..183..225T} covering a period of about 15 months from March 1997 to June 1998, which reports 176 events. In 16 cases we found that the EIT wave is related to a SEP event. This leads to an association rate of 9$\,$\% (16/176) between EIT waves and SEP events. If one extrapolates this association rate to the entire Solar Cycle 23, it would mean that the majority of the EIT waves ($\approx$90$\,$\%) are not accompanied by a particle event.

In the following we focus on the SEP events associated with eruptive activity in the eastern solar hemisphere, where the standard solar-wind models predict no direct magnetic connection to a spacecraft near Earth. A comparison between the timing of the disturbance and of the SEP onset in these events provides a stringent test of the hypothesis that the SEPs are accelerated at the laterally expanding EIT disturbance. We excluded SEP events observed within SOHO data gaps, with no identified EIT wave or for which no EIT wave speed could be estimated, because the wave front was uncertain or only detected in a single image. SEP events observed either within or close to an ICME are addressed separately, because there may be a transient direct magnetic connection from the eastern hemisphere to the Earth through the magnetic field of the ICME.

\subsection{Eastern SEP Events Observed Within and in the Vicinity of an ICME} %%%%%%%%%%%%%%%%%%%%%%%%%%%%%%%%%%%%%%%%
      \label{S-ICMESEPs}

SEPs observed while the Earth is within or near an ICME propagate in strongly disturbed IP magnetic-field (IMF) conditions that are different from the PS model. \inlinecite{1991JGR....96.7853R} showed that SEP within ICMEs indeed have properties suggestive of a direct magnetic connection from the acceleration region in the eastern solar hemisphere to Earth. This is consistent with the finding \cite{2013SoPh..282..579M} that the peak intensity of ICME SEP events correlates more strongly with the parent solar flare properties than the peak intensity of SEP events in the standard solar wind.

Among the eastern SEP events of our sample 17$\,$\% (8/48) are ICME events (consistent with \opencite{1991JGR....96.7853R}) and about 14$\,$\% (7/48) are vicinity events. Among the ICME electron events associated with EIT waves, two were initially beamed, one showed moderate anisotropy, and for the other two the anisotropy could not be evaluated because of ion contamination or poorly defined electron onset.

\begin{figure}[t!]
\centerline{\hspace*{-0.05\textwidth}
            \includegraphics[width=0.55\textwidth,clip=]{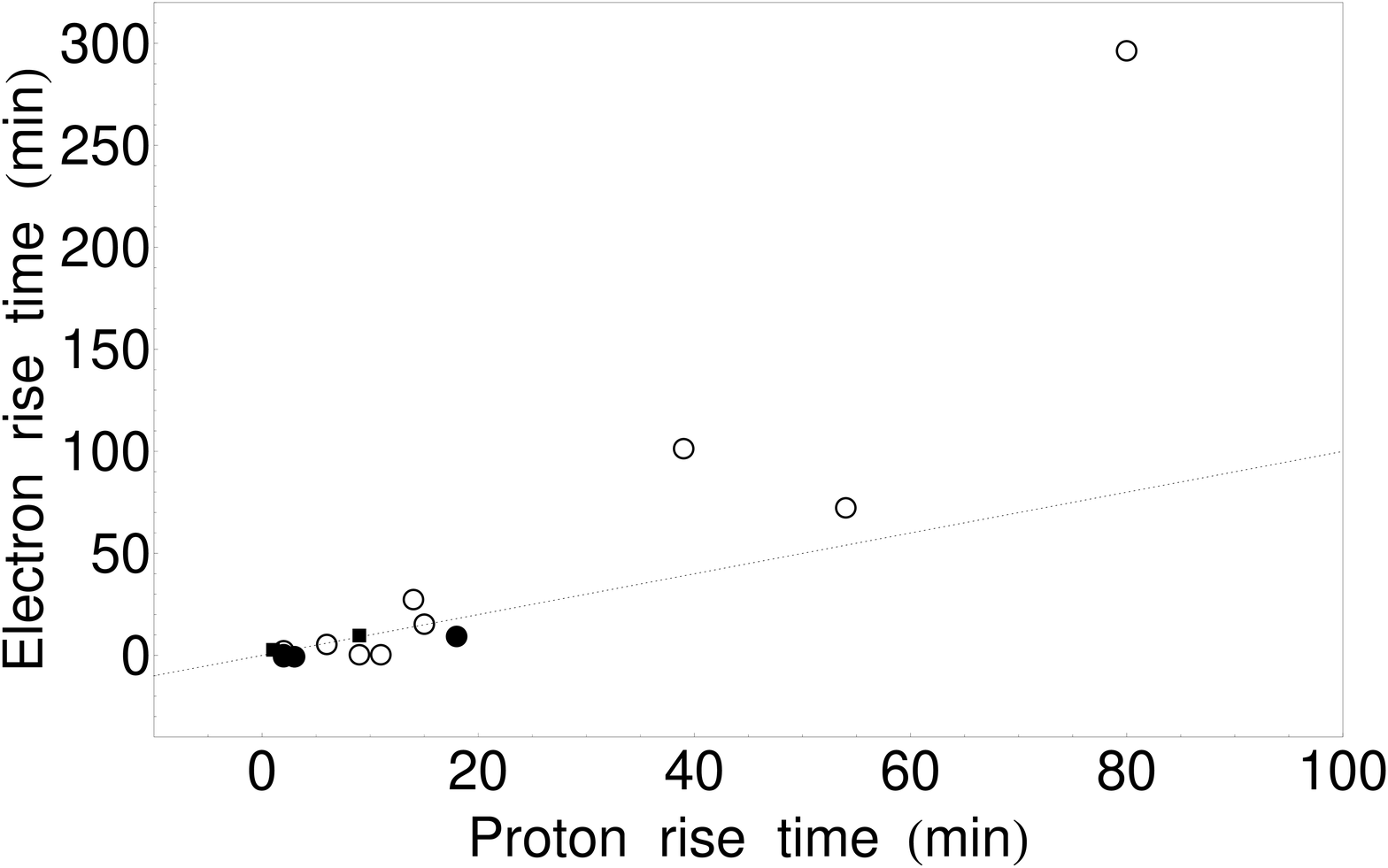}
            \hspace*{-0.05\textwidth}
            \includegraphics[width=0.55\textwidth,clip=]{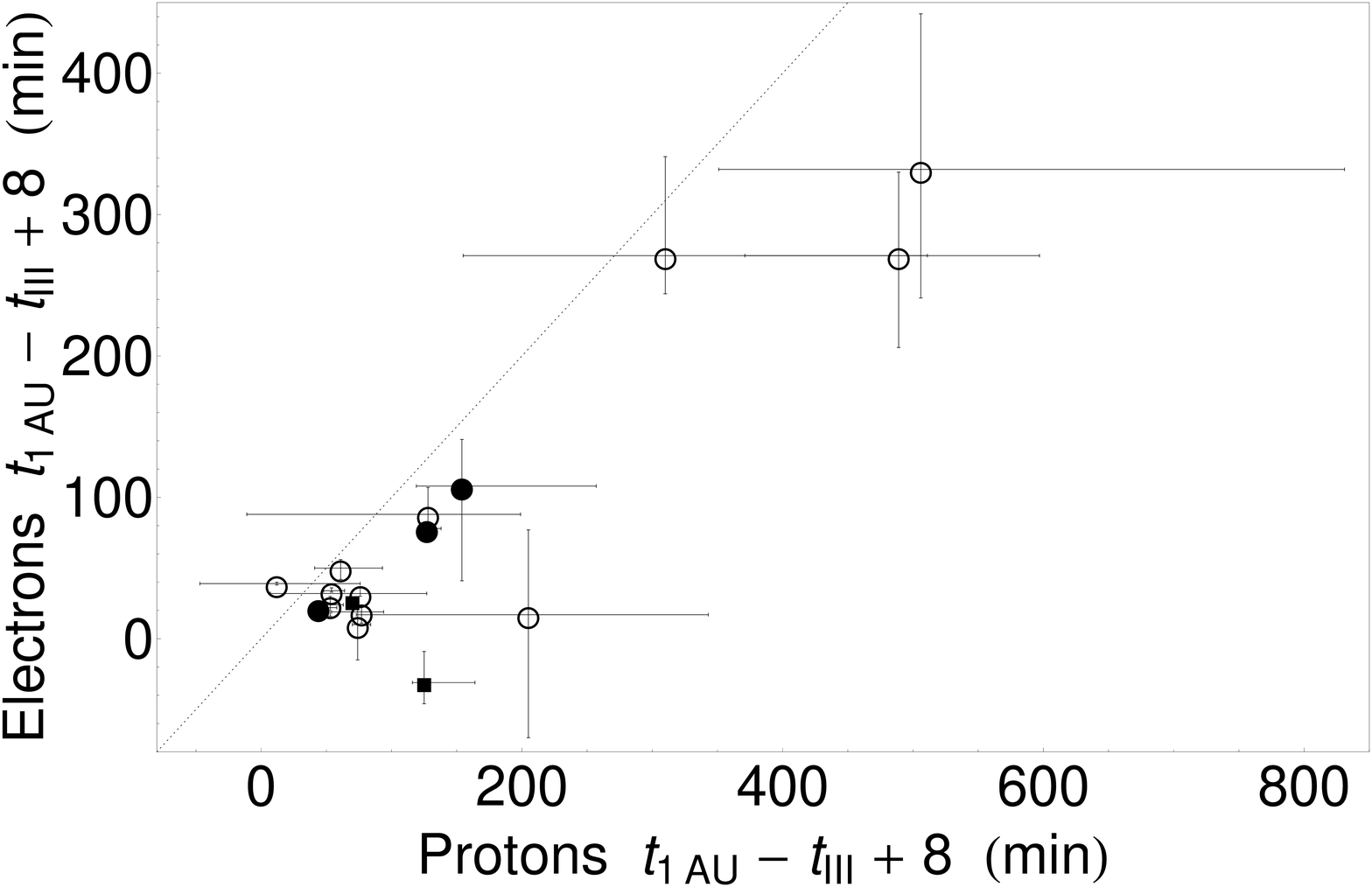}}
\caption{Scatter plot of the rise times (left) and the delay $t_{\rm 1AU}-t_{\rm III}+8\, {\rm minutes}$ (right) between protons and electrons. The lines have a slope of one. The open circles denote the solar-wind events, filled circles the ICME events and the filled squares the events in the vicinity of an ICME.}
\label{F-Rise_times}
\end{figure}

\begin{figure}[t!]
\centerline{\hspace*{-0.05\textwidth}
            \includegraphics[width=0.55\textwidth,clip=]{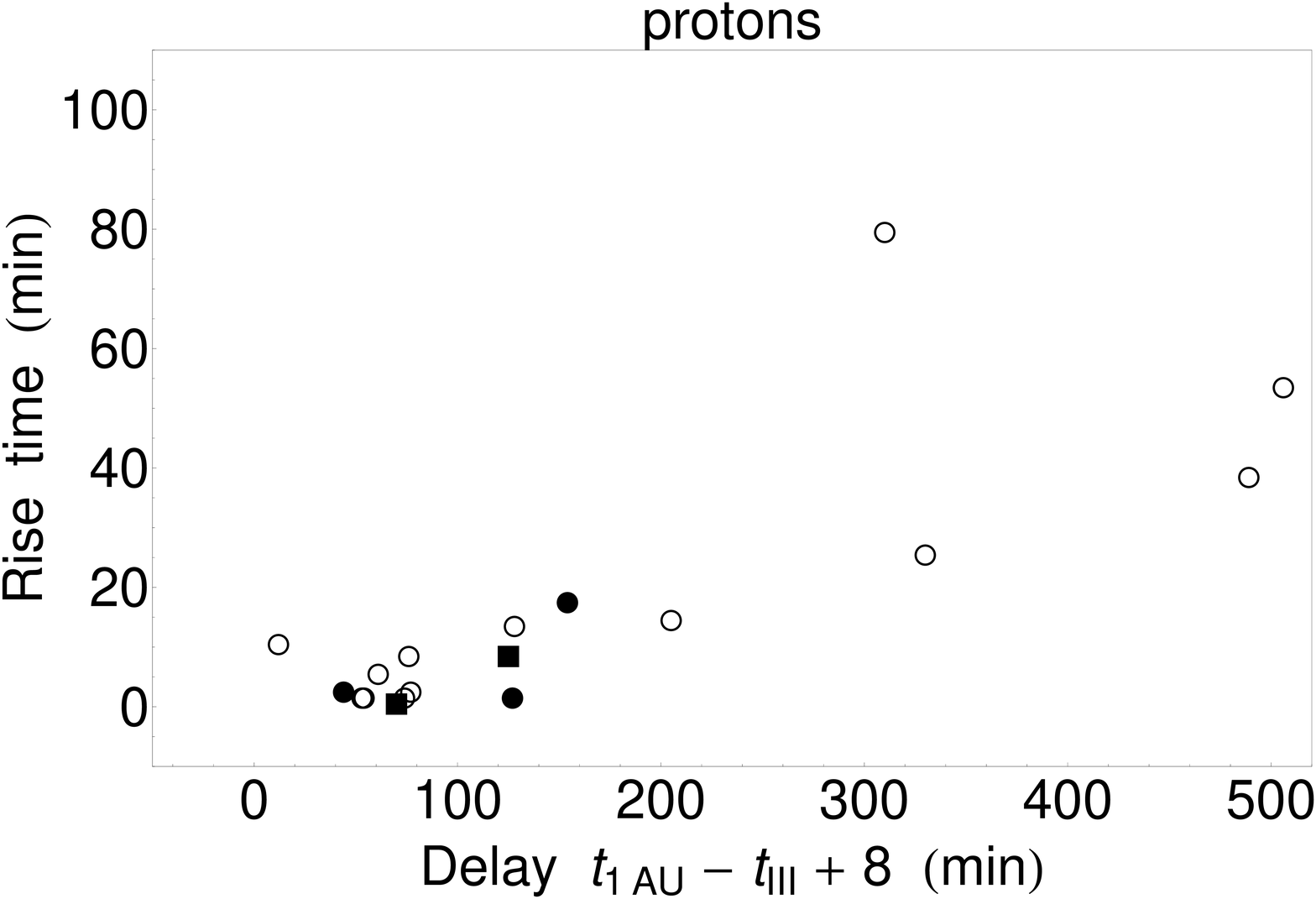}
            \hspace*{-0.05\textwidth}
            \includegraphics[width=0.55\textwidth,clip=]{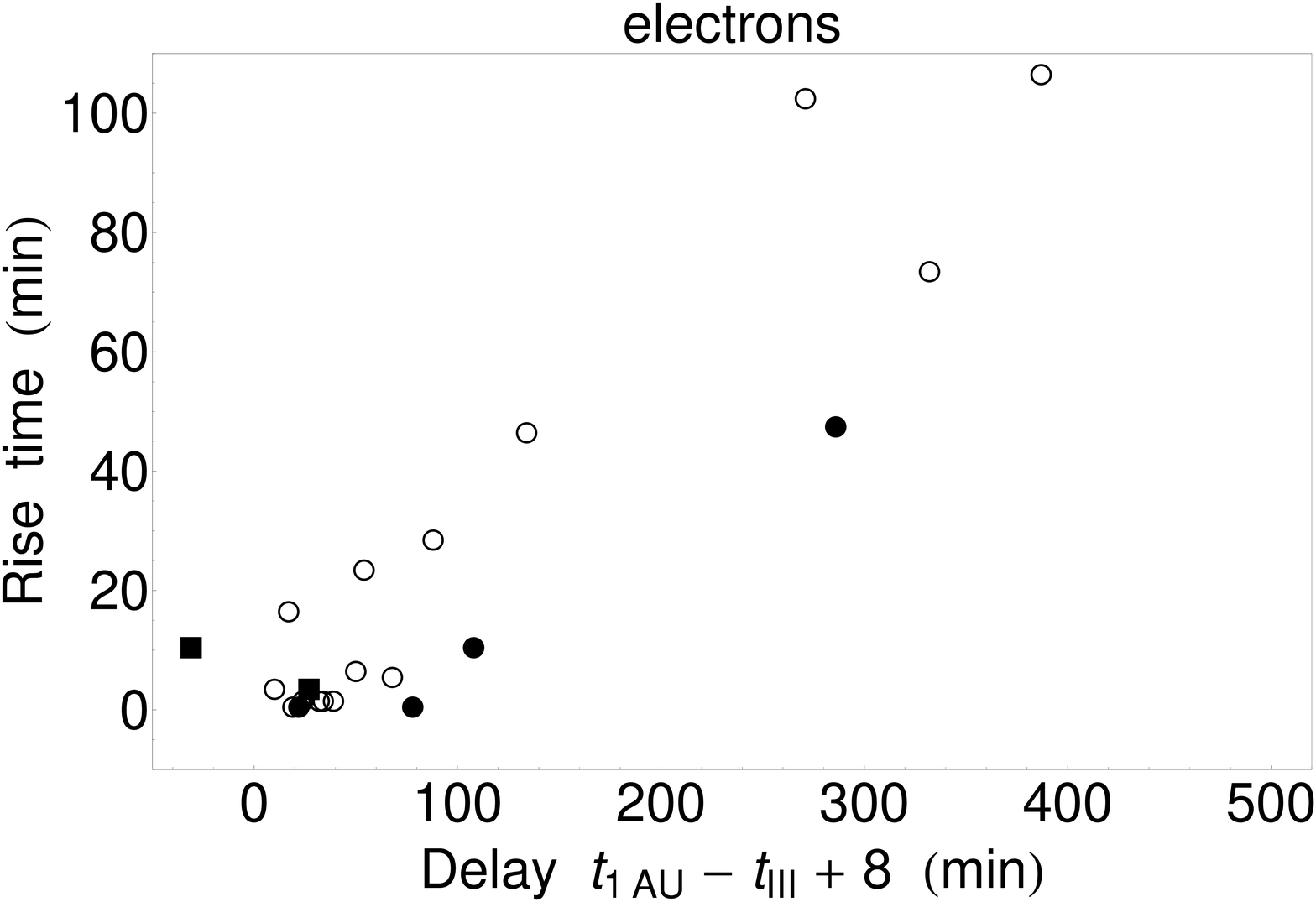}}
\caption{Scatter plot of the delay $t_{\rm 1AU}-t_{\rm III}+8\, {\rm minutes}$ and rise times for protons (left) and electrons (right). The open circles denote the solar-wind events, filled circles the ICME events and the filled squares the events in the vicinity of an ICME.}
\label{F-tIII_Rise}
\end{figure}

The early rise times and onset delays of the eastern SEP events are addressed in Figures~\ref{F-Rise_times} and \ref{F-tIII_Rise}, where the ICME events are represented by filled circles. The figures show all eastern events associated with EIT waves for which the onset and rise times could be measured. The onset time of the SEP event at the spacecraft, $[t_{\rm 1 \; AU}]$, is compared to the time of the first injection of electron beams into the IP medium, as revealed by the time of the first decametric-to-hectometric (DH) Type~III burst, [$t_{\rm III}$]. The time delay $t_{\rm 1 AU} - (t_{\rm III} -8 \; \rm minutes)$ is a combination of: i) the travel time of the particles from the Sun to the spacecraft, ii) possibly the delay between the first particle release onto the spacecraft-connected IMF line and the first electron release to space, in the case where the SEPs are released at the Sun onto different field lines in the course of time. If the spacecraft is not well connected to any of the IMF lines onto which the SEPs are injected, the delay can also reveal the time needed for the spacecraft to reach these field lines.

\subsection{Eastern SEP Events Observed in the Solar Wind} %%%%%%%%%%%%%%%%%%%%%%%%%%%%%%%%%%%%%%%%
      \label{S-SoWiSEPs}

Among the 33 eastern particle events propagating in the standard solar wind, both onset time at 1 AU $[t_{\rm 1\,AU}]$ and EIT wave speed $[v_{\rm av, EIT}]$ could be determined for 12 proton events and 15 electron events.

\subsubsection{SEP Onset Delays and Early Rise Times} %%%%%%%%%%%%%%%%%%%%%%%%%%%%%%%%%%%%%%%%
      \label{S-SoWiSEPsonset}

\begin{figure}[t!]
\centerline{\hspace*{-0.05\textwidth}
            \includegraphics[width=0.55\textwidth,clip=]{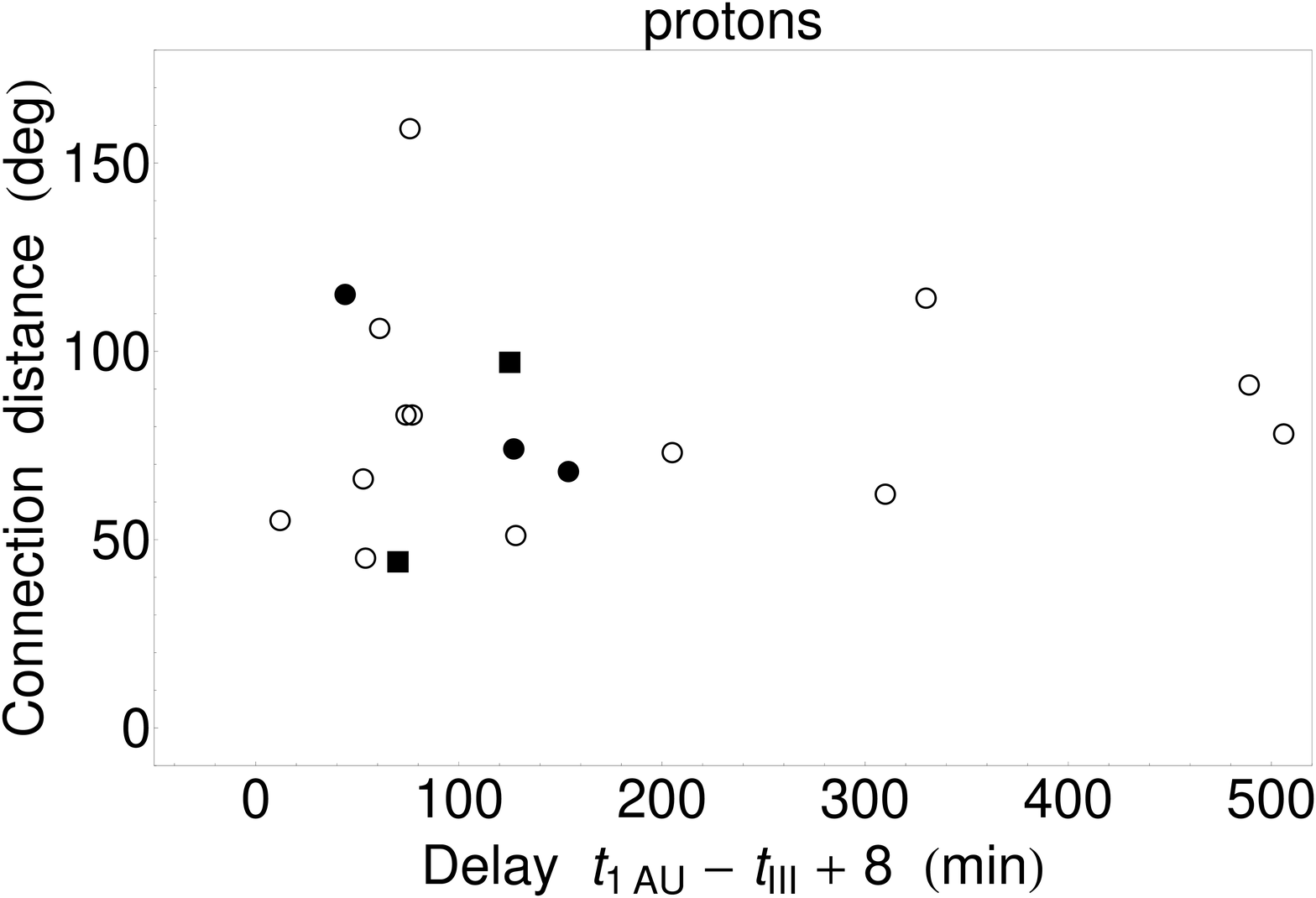}
            \hspace*{-0.05\textwidth}
            \includegraphics[width=0.55\textwidth,clip=]{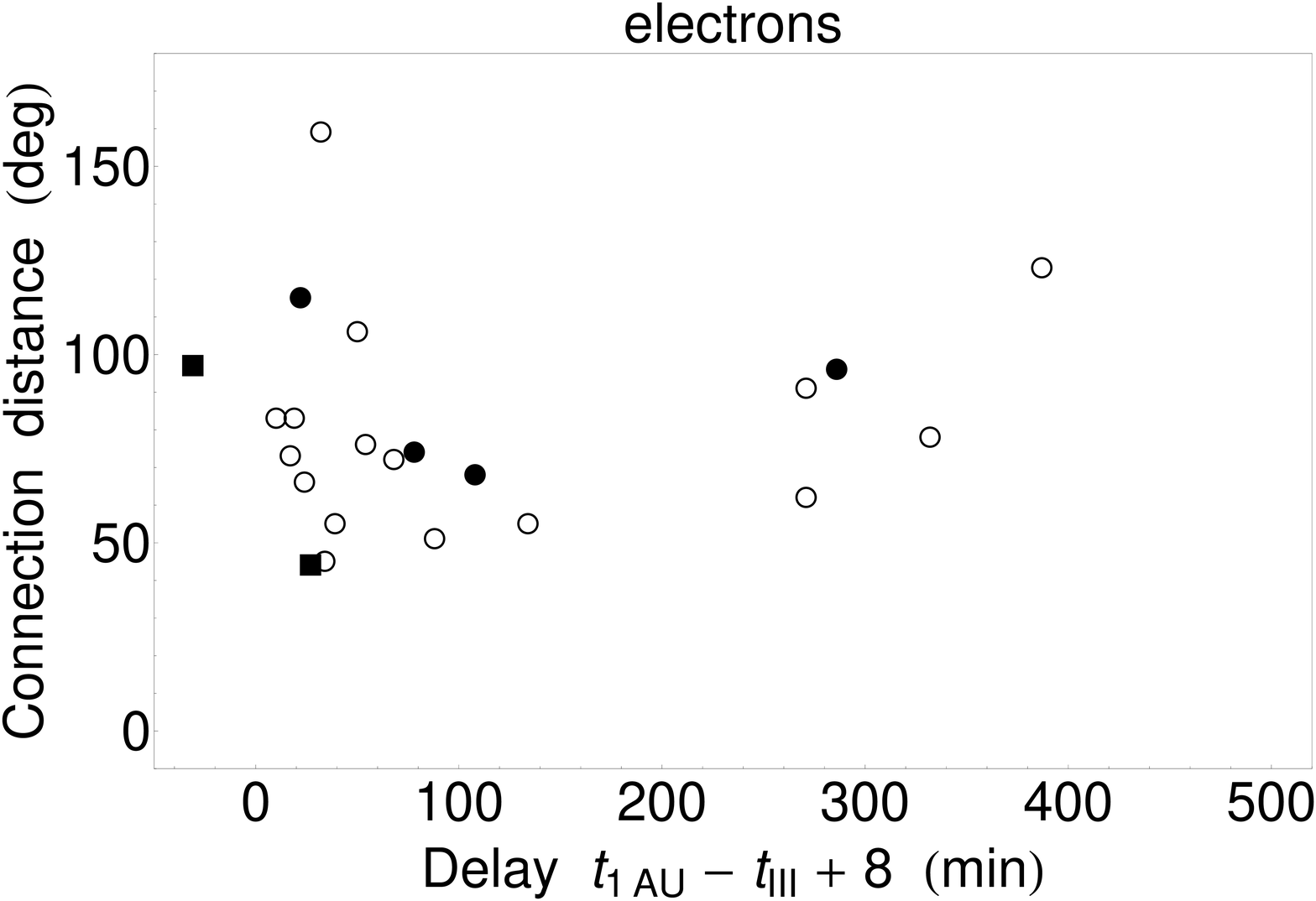}}
\caption{Scatter plot of the delay $t_{\rm 1AU}-t_{\rm III}+8\, {\rm minutes}$ and the modulus of the connection distance for protons (left) and electrons (right). The open circles denote the solar-wind events, filled circles the ICME events and the filled squares the events in the vicinity of an ICME.}
\label{F-tIII_Conn}
\end{figure}

On average, electrons and protons display similar trends in their rise times and onset delays, as shown by the scatter plots of Figure~\ref{F-Rise_times}. The onset delays in solar-wind events (open circles) are longer for protons than for electrons, but in general the delays are correlated. The longest delays are found in the solar-wind events. The rise times of electrons and protons are also correlated (left panel of Figure~\ref{F-Rise_times}), albeit with two outliers where the electron profiles rise exceptionally slowly. The longest rise times are again found in solar-wind events, although some of them have comparable rise times to ICME events. Figure~\ref{F-tIII_Rise} shows that the characteristic times of electrons and protons are also correlated. This suggests that similar processes contribute to generate the delays and rise times of the two SEP species. This likely points to an influence of interplanetary transport processes on the two characteristic times. The results are, however, inconsistent with a pure interpretation in terms of interplanetary transport, as advocated in early work on eastern SEP events (see the review by \opencite{1991PCS....21..243K}), since the onset delays are not ordered by the distance between the erupting active region and the spacecraft-connected PS, as shown in Figure~\ref{F-tIII_Conn}.

\subsubsection{SEP Onset Delays and EIT Wave Timing}

\begin{figure}[t!]
\centerline{\hspace*{-0.05\textwidth}
               \includegraphics[width=0.55\textwidth,clip=]{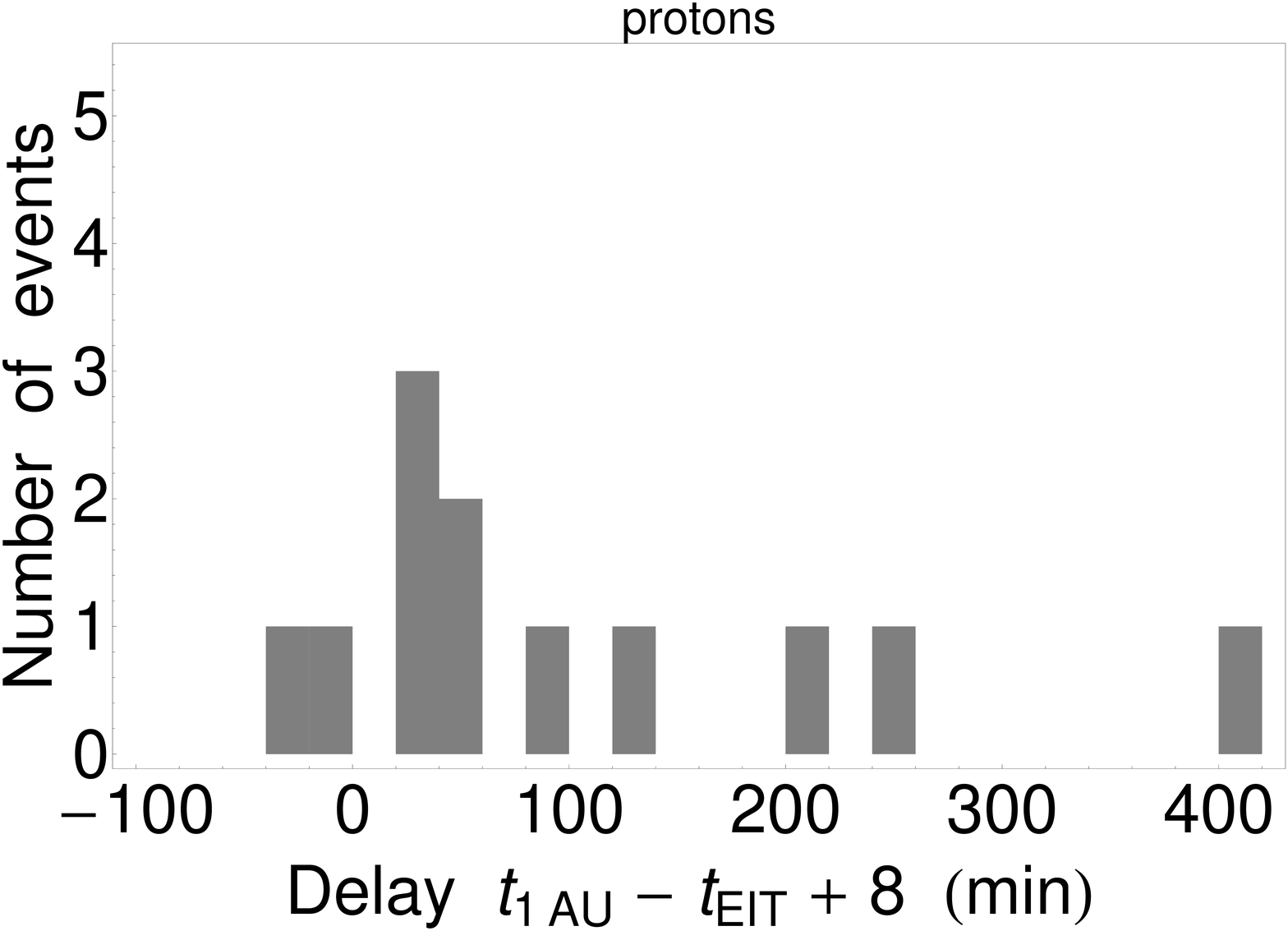}
               \hspace*{-0.05\textwidth}
               \includegraphics[width=0.55\textwidth,clip=]{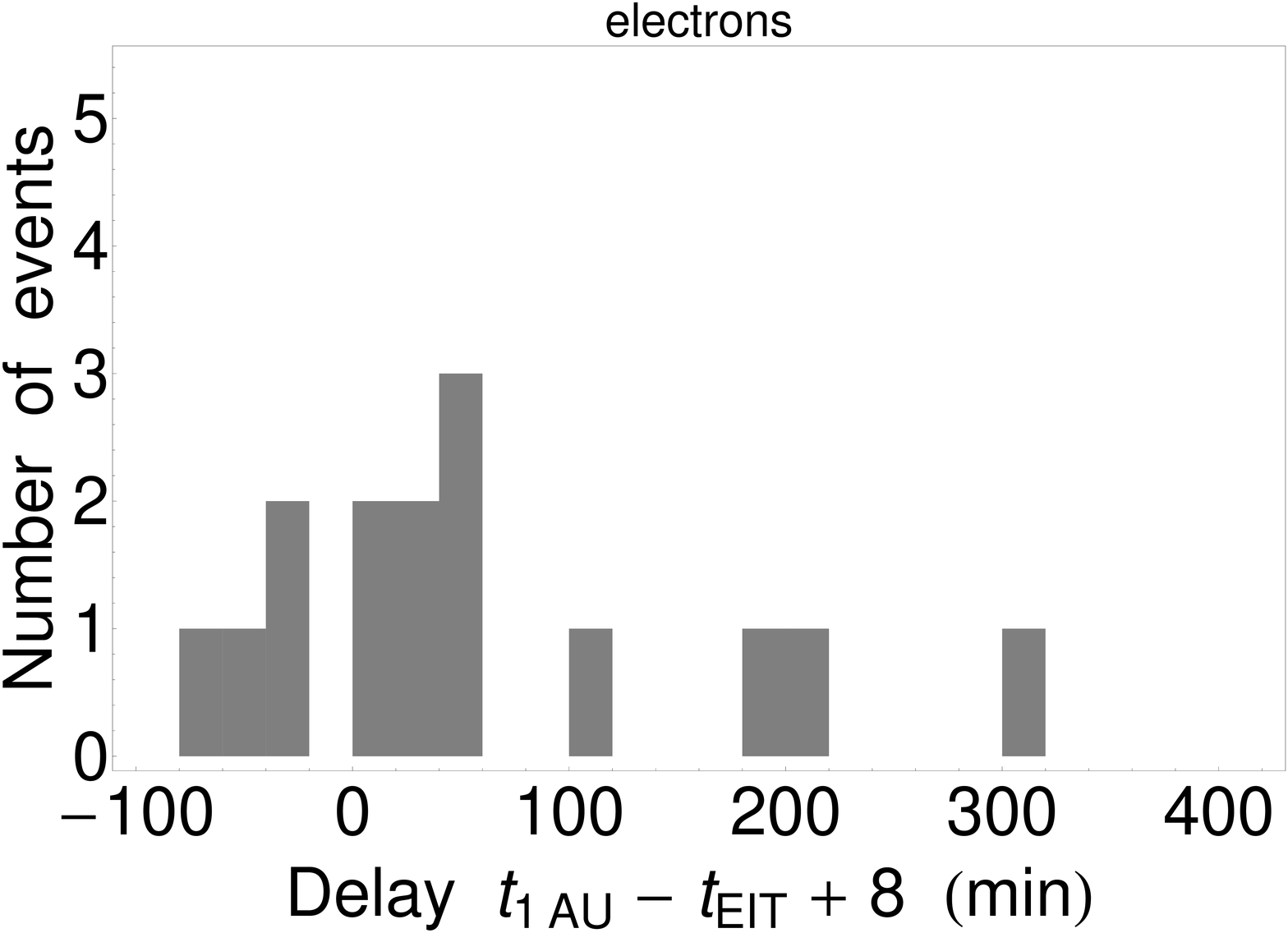}}
               \vspace*{-0.0\textwidth}
\caption{Histogram of the delay $t_{\rm 1 AU} - t_{\rm EIT} + 8$~minutes between for proton (left) and electron (right) events propagating in quiet solar-wind conditions.}
   \label{F-HistoDelay}
   \end{figure}

The overall behavior of onset time delays and rise times of electrons and protons, together with a longer delay of the protons, is consistent with a scenario of a delayed acceleration remote from the parent active region, followed by a species-dependent interplanetary travel time. We now turn to the question of whether the onset delays of the eastern solar-wind SEP events are consistent with acceleration occurring as the EIT wave reaches the foot of the PS field line connected to the spacecraft. In this scenario the time difference $t_{\rm 1 \; AU} - (t_{\rm EIT} -8 \; \rm minutes)$ would be the interplanetary travel time of the SEPs. The histograms of the delays $t_{\rm 1 \; AU} - (t_{\rm EIT} -8 \; \rm minutes)$ are displayed separately for protons and electrons in Figure~\ref{F-HistoDelay}. If the particles start to be released as the EIT wave intercepts the footpoint of the PS, the delays must be positive and cannot be shorter than the free-streaming time along the field line, which amounts to about 10~minutes for electrons at 1~MeV and 35~minutes for protons at 33~MeV. The actually measured distribution departs from this expectation: there is a wide spread from negative delays, where the SEP event is found to start before the EIT wave reaches the spacecraft-connected PS, to delays up to several hundreds of minutes. Electrons and protons show a similar behavior, but more electron than proton events have negative delays.

\begin{figure}[t!]
\centerline{\hspace*{-0.05\textwidth}
            \includegraphics[width=0.8\textwidth,clip=]{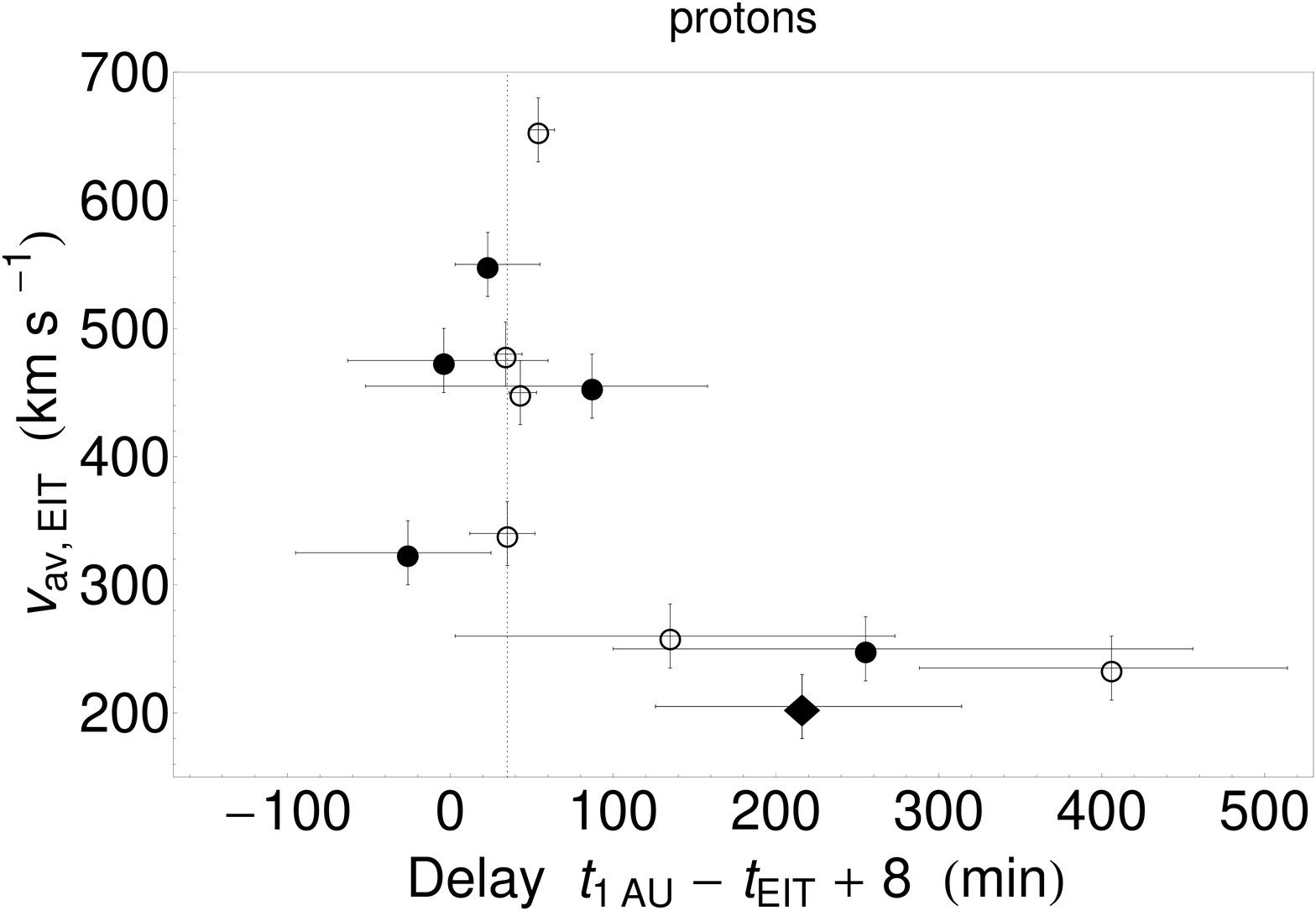}}
            \vspace*{-0.01\textwidth}
\centerline{\hspace*{-0.05\textwidth}
            \includegraphics[width=0.8\textwidth,clip=]{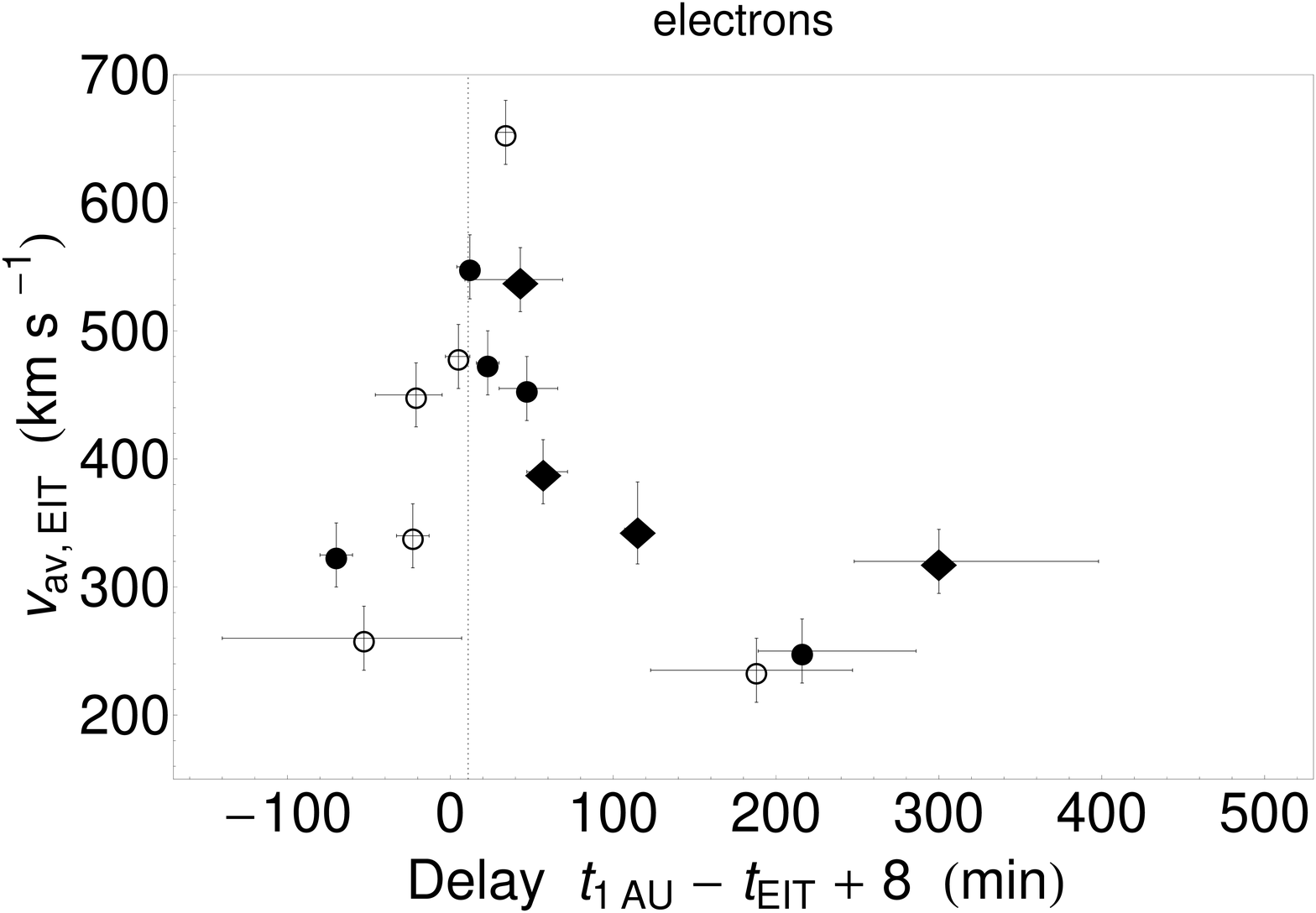}}
\caption{Scatter-plot for the delay $t_{\rm 1AU}-t_{\rm EIT} +8$~minutes {\it vs.} the average EIT wave speed for protons (top) and electrons (bottom) for solar-wind SEP events only. Events for which velocity dispersion is present are plotted with open circles, the non-dispersive events are given with filled circles and those events for which no conclusion on the dispersion could be made due to data gaps are plotted with filled diamonds.}
   \label{F-Errorbars}
   \end{figure}

The interpretation of time delays in terms of interplanetary travel times applies only to those events where the spacecraft has been staying on the field line onto which the SEPs are released ever since the start of the injection. This is not necessarily the case: the spacecraft may have entered a magnetic-flux tube that was already filled with SEPs. We identify the events where the first SEPs are observed by a significantly earlier start of the intensity rise of electrons than protons. Based on the particle onset times at 1 AU and their estimated uncertainties (Table~\ref{T-SEP_data}), only six solar-wind events show a significant velocity dispersion (01 April 1997\footnote{For this event, the electrons arrive four minutes after the protons but since the latter value is due to the large error bars obtained for the electron and proton onsets, we will keep this event in the category of dispersive events.}, 24 September 1997, 29 April 1998, 18 January 2000, 17 February 2000 and 15 January 2005). They are represented by empty circles in Figure~\ref{F-Errorbars}. Five events do not show evidence for a dispersive onset (2001 Jun 15, 2001 Oct 09, 2003 Nov 18, 2005 May 13 and 2006 Nov 06, plotted with filled circles). For five events the onset dispersion could not be assessed due to data gaps ({\it i.e.} we do not have information for both electrons and protons). These cases are plotted with filled diamonds.

Figure~\ref{F-Errorbars} presents the delay $t_{\rm 1 \; AU} - (t_{\rm EIT} -8 \; \rm minutes)$ {\it vs.} the averaged EIT speed. The top plot shows that for all six proton events with velocity dispersed onsets the delays are positive: namely equal to or larger than the scatter-free field-aligned travel time, which is denoted by the vertical line.  Within the statistical uncertainties they are consistent with the idea that the protons are released as the EIT wave reaches the footpoint of the PS through the spacecraft. For the electrons, however, 4/6 events with dispersive onsets have shorter delays than the free-streaming limit (bottom panel of Figure~\ref{F-Errorbars}). Only two have delays that are clearly positive. Both electron and proton SEP events associated with fast EIT waves ($\geq$400 km~s$^{-1}$) have short delays. Delays exceeding 100 minutes are only observed in cases with slow EIT waves.

\subsubsection{Electron Anisotropy}

The electron events in the standard solar wind were generally found to display no clear anisotropy. The anisotropy could not be identified in 5/20 events. Only one event was identified as beamed, three were irregular, four isotropic and seven had moderate anisotropy. Overall, the eastern SEP events propagating in the solar wind are seen to be less strongly focused than those propagating in ICMEs. The result can also be compared with the electron anisotropy in SEP events associated with activity in the western solar hemisphere. Among the western SEP events detected in the solar wind and studied here, only 18 are contained in the event list of the SEPServer project \cite{2013JSWSC...3A..12V}, identified from observations at energies above 50 MeV by the \textit{Energetic and Relativistic Nuclei and Electron experiment} (ERNE) onboard SOHO. The anisotropy characteristics derived from ACE/EPAM (Table 5 of \opencite{2013JSWSC...3A..12V}) in the same way as in the present article show that 12/18 events are beamed, the others have moderate anisotropy. None is labeled isotropic.

\section{Summary of Observational Results}
\label{S-Summary}

We conducted a survey of the association between SEP events and EIT waves in the Solar Cycle 23. For a more detailed investigation of the timing relationship between the EIT wave and the SEP event onset, we then focused on eastern solar events. Whenever possible we used SOHO/EIT imaging observations to extrapolate the arrival time of the EIT wave at the foot of the Earth-connected PS field line in the western hemisphere and tested the consistency with the onset time of the SEP events, both near-relativistic electrons and deka-MeV protons, at the SOHO spacecraft. The observational results are the following:
\begin{enumerate}[i)]
%\item SEP events associated with eruptive activity on the earthward solar hemisphere: \% from the west, \% from the east:
\item The large majority of SEP events are associated with EIT waves (87$\,$\%). The rate is comparable for events associated with eastern and western activity with an uncertainty of some percent due to the uncertain association of solar activity with some of the SEP events. The converse is not true: only a minority of EIT waves (about 10$\,$\%) are accompanied by SEP events.
\item Eastern SEP events are observed within ICMEs (17$\,$\%), in the standard solar wind (69$\,$\%) or in a situation where an ICME is within a day from the Earth and can have caused a major perturbation of the medium in which the SEPs travel. The latter category is ambiguously defined. But the fraction of events detected within ICMEs is consistent with the earlier findings of \inlinecite{1991JGR....96.7853R} of the order of 15$\,$\%.
\end{enumerate}

A more detailed study of the timing of SEPs (onset delay with respect to the parent solar activity and slope of the initial rise of the time profiles, characterized by the rise time $[t_{\rm r}]$) and EIT waves was carried out for the eastern SEP events detected in the standard solar wind:

\begin{enumerate}[i)]
\item SEP events from eastern eruptive activity that are observed in the standard solar wind tend to have a broader range of rise times and onset delays with respect to the parent eruptive activity than events observed in ICMEs.
\item The onset delays of electrons and protons are correlated. About half of the SEP events detected in the solar wind show no clear velocity-dispersed onset, as seen by a comparison of deka-MeV protons and near-relativistic electrons. In the others, the temporal profile of the proton intensities rises after the electrons.
\item The time constants of the early rise of the intensity--time profiles of electrons and protons are also correlated.
\item Both for electrons and protons, the onset delays are correlated with the rise times.
\item The onset delays -- and hence also the early rise times -- are not ordered by the connection distance, that is the longitudinal difference between the erupting active region and the spacecraft-connected PS.
\item The arrival of the first deka-MeV protons is consistent with the scenario of an acceleration in the low corona as the EIT wave reaches the footpoint of the nominal PS to the spacecraft. But in the majority of cases the first electrons are detected earlier than expected by the scenario.
\item With one exception, the pitch-angle distributions of near-relativistic electrons are weakly anisotropic or isotropic. This is different from events observed within ICMEs and electron events associated with activity in the western hemisphere.
\end{enumerate}

\section{Discussion}
\label{S-Disc}

The enigmatic solar origin of eastern SEP events ({\it e.g.} far away from the PS field line that connects the Sun with the Earth) has found several tentative interpretations. A recent one is the acceleration by a disturbance that originates in the parent solar eruption and intercepts in the course of its propagation through the corona the spacecraft-connected interplanetary magnetic-field line. The discovery of EIT waves and their identification with the large-scale propagation of MHD disturbances -- be it a real wave or the CME itself -- has prompted case studies and statistical analysis of the relationship with SEP event onsets.

Our analysis does not support this scenario. This conclusion is based mainly on the timing and anisotropy characteristics of the first near-relativistic electrons detected during an event. But the correlations between the parameters describing the timing of protons and electrons during this phase -- onset delay and early rise time -- suggest that both particle species are affected by closely related processes of acceleration and transport. We therefore believe that the failure to reproduce the observed electron onset casts doubt on the EIT-wave-acceleration scenario in the low corona for SEPs in general. A second major contradiction to the EIT-wave-acceleration scenario comes from the weak anisotropic or even isotropic pitch-angle distributions of the first near-relativistic electrons detected near Earth: if they were accelerated in the low corona, close to the spacecraft-connected PS, by the EIT wave, one would expect the initial pitch-angle distributions to become adiabatically focused during their interplanetary propagation, as is the case of most ICME events and western SEP events. This is clearly not observed.

About half of the SEP events detected in the solar wind show no clear velocity-dispersed onset, as seen by a comparison of deka-MeV protons and near-relativistic electrons. This number is large, partly because of the large error bars of the onset-time determinations. In others the lack of data prevented the assessment of a velocity dispersion. Some of these SEP events might be consistent with the EIT-wave-acceleration scenario, since their onsets occur after the earliest expected arrival of SEPs accelerated at the footpoint of the spacecraft-connected PS. However, they may also have been released much earlier, with the large onset delays being due to the time needed by the spacecraft to enter the flux tube filled with SEPs.

The results from the timing analysis performed here confirm those published in the literature: the timing of the SEP onset is consistent with acceleration at the EIT waves for protons ({\it e.g.} the case studies by \opencite{1999ApJ...510..460T}; \opencite{2012ApJ...752...44R}), but not for electrons \cite{1997ESASP.415..207B,1999ApJ...519..864K}, at least not in all cases. Our estimation of the travel time of the EIT wave from the erupting active region to the PS is actually an underestimation. This is because we used a constant speed derived from observations close to the active region, whereas the actual wave decelerates with increasing distance, {\it e.g.} \inlinecite{2008ApJ...681L.113V}, \inlinecite{2008ApJ...680L..81L}, and \inlinecite{2010AdSpR..45..527W}, and because we neglected latitudinal propagation. We therefore believe that the timing analysis favored the scenario. The incompatibility of several events with this scenario is therefore unlikely to be removed by a more sophisticated estimation or data with higher cadence.

Some of the observed inconsistencies could be removed if the acceleration did not occur in the low corona, but farther away from the Sun. This could on the one hand contribute to resolve the timing discrepancy of the electron onsets \cite{1999ApJ...519..864K}, if the lateral speed of the large-scale disturbance is faster in the high corona than in the low corona, where the EUV signature is emitted. In order to explain the weak initial anisotropy of the electrons, the transport processes must dominate the adiabatic focusing, {\it e.g.} if the acceleration occurred at greater distance from the Sun. We evaluated the distance of the CME front at the time when the electrons arrive at the spacecraft. Using the CME linear speed and time (considered here to be at distance of 1.5 R$_\odot$, {\it i.e.} at the inner diameter of the LASCO-C2 coronagraph), as reported in the LASCO CDAW database, we find that the CME fronts are at most at a heliocentric distance of 30~R$_\odot$. Adiabatic focusing should still operate on the way to the spacecraft. In addition, in the single case when the CME front is further than 30 solar radii from the Sun, the electron anisotropy is moderate, not isotropic. Hence, the distance of the CME front from the Sun does not appear to order the anisotropy characteristics of the near relativistic electrons. These anisotropy characteristics do not favor acceleration processes acting in small spatial regions, be it at the interface of the CME and the spacecraft-connected interplanetary magnetic-field line shown by the EUV emission or in front of a shock wave in the high corona. A spatially extended acceleration region, {\it i.e.} at a height-extended interface between the laterally expanding CME and the spacecraft-connected interplanetary magnetic-field line, might account for the absence of a clear anisotropy and also for the frequent absence of a clear velocity dispersion as shown by the comparison of the onsets of near-relativistic electrons and deka-MeV protons. The EIT wave would show only the lowest coronal parts of this height-extended region of interaction (see also \opencite{2012ApJ...752...44R}).

An alternative explanation of the weak or non-existent electron anisotropy and the velocity dispersion would be that the onset of an SEP event associated with an eastern active region is actually seen when the spacecraft enters an interplanetary flux tube that is already filled with SEPs. The scenario does not appear plausible as a general interpretation, because we observed onset-time delays up to a few hours. Much longer time delays between the parent solar activity and the onset time of the SEP events would be expected from at least some of the eastern events, given a rotation speed of the interplanetary flux tubes of about 14$^\circ$ per day. The idea of cross-field particle transport due to diffusion or to field-line wandering has gained new support from STEREO observations, {\it e.g.} \inlinecite{2010ApJ...709..912D}; \inlinecite{2012SoPh..281..281D}. The scenario accounts for moderate anisotropies as well as for a correlation between the onset delay of an SEP event and the slope of its rising intensity profile.

In summary, the present survey contradicts the role of an EIT wave, whatever its nature, as the principal SEP accelerator in cases where the particles come from poorly connected eruptive activity. SEPs from eastern activity that are observed in the standard solar wind show characteristic differences from those that are observed within an ICME or that are associated with western solar activity. Both western SEP events and events from the eastern hemisphere where SEPs travel within an ICME probably have a direct magnetic-field connection from the accelerator to the particle detector. This is probably not the case in the SEPs from eastern activity that are observed in the standard solar wind. The timing characteristics of deka-MeV protons and near-relativistic electrons and the anisotropy of the electrons suggest a combination of a spatially extended acceleration region and of interplanetary transport.

\appendix
\label{Appendix}

\section*{EIT Waves and Their Relationship with Other Coronal Phenomena}
      \label{S-EIT_flCME}

We compared the occurrence of the EIT disturbance with the reports of shock signatures in the corona, and we found in general a high correlation, see the contingency table: Table~\ref{T-waves_typeIIs}. The majority of EIT waves (88$\,$\%, 112/127) are associated with signatures of a shock wave in the corona. The same percentages are found for the western and eastern sample. However, we would like to note that the data set used is only a subset of all EIT waves in Solar Cycle 23, since we started with a SEP list. Moreover, the events for which no data were found (EIT or radio) are in general dropped from the analysis.

\begin{table}[t!]
\caption[]{Contingency table for EIT waves and Type-II radio bursts}
\label{T-waves_typeIIs}
\tiny
\vspace{0.1cm}
\begin{tabular}{lcccccc}
\hline
                                    & \multicolumn{6}{c}{EIT waves}         \\
                                    &\multicolumn{2}{c}{Eastern} & \multicolumn{2}{c}{Western} & \multicolumn{2}{c}{All events} \\
\multicolumn{1}{c}{Type-II bursts} & Yes  & No  & Yes  &  No  & Yes & No   \\
\hline
\multicolumn{1}{c}{Yes}            &  28  & 4  &  84  &  11  &  112  &  15  \\
\multicolumn{1}{c}{Not reported}   &  1   & 1  &  17  &   6  &  18   &  7   \\
\hline
\end{tabular}
\end{table}

The opposite association rate is not fully investigated. \inlinecite{2000A&AS..141..357K} presented a list of 21 shock signatures during 1997 where in 90$\,$\% of the cases the metric Type-II radio burst had an associated EIT wave. No correlation was found between the speed of the Type-II driver and the EIT wave, as the radio signature of the shock wave was found to be about three times faster than the EIT wave.

In addition, the onset time of the EIT disturbances $[t_{\rm EIT,on}]$ can be compared with the onset time of signatures of a shock wave propagating through the solar corona, {\it i.e.} metric Type-II radio bursts \cite{1985srph.book..333N}. For the onset time of the western EIT waves, we performed a time shift by six minutes from the time of the first observed front. We expect an underestimation of this time only for a minority of the events ({\it e.g.} when the EIT cadence was actually longer). For the Type-II signatures, we used the reported onset times by different radio observatories at metric wavelengths.

\begin{figure}[t!]
\centerline{\includegraphics[width=0.55\textwidth,clip=]{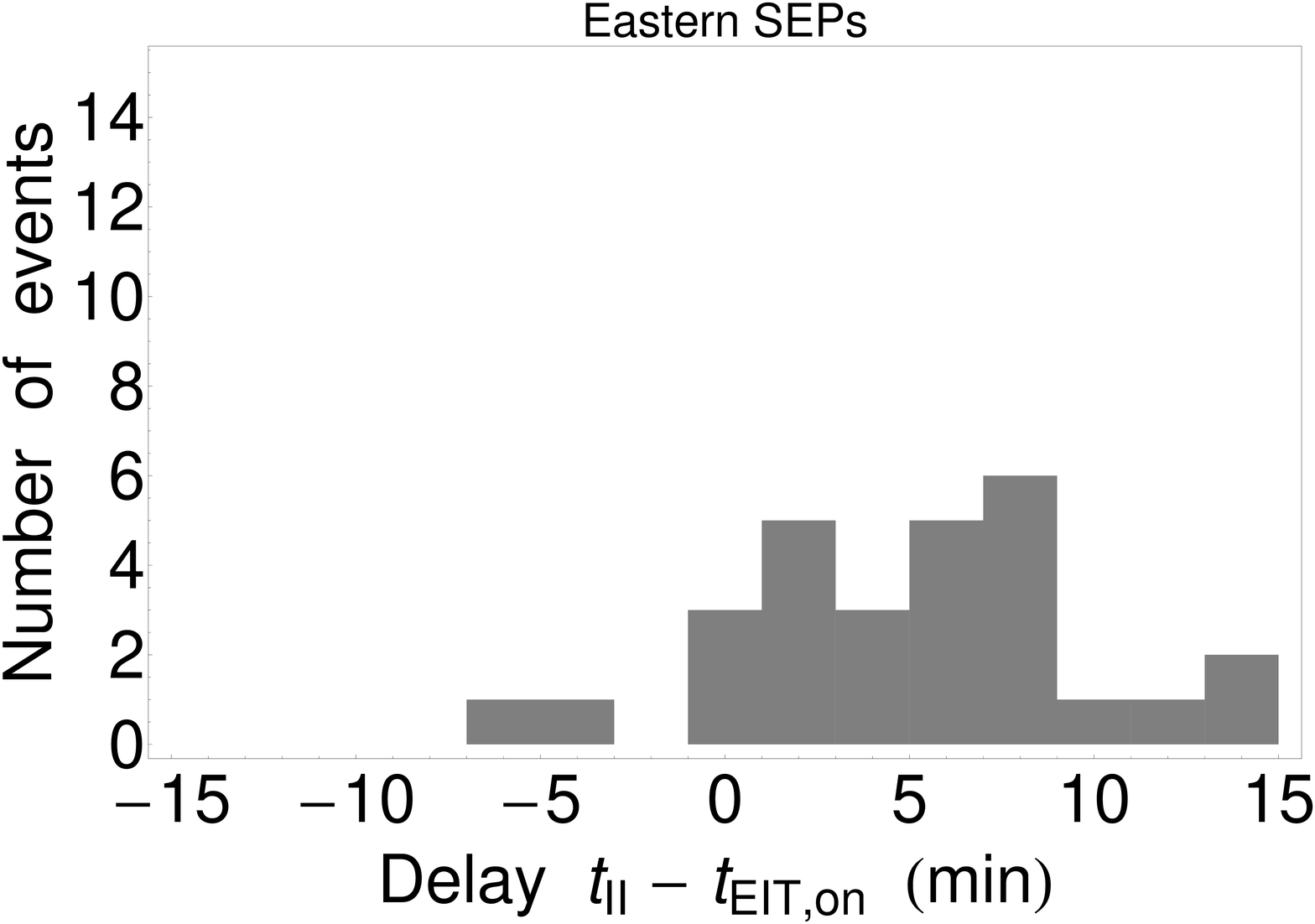}
               \hspace*{-0.05\textwidth}
               \includegraphics[width=0.55\textwidth,clip=]{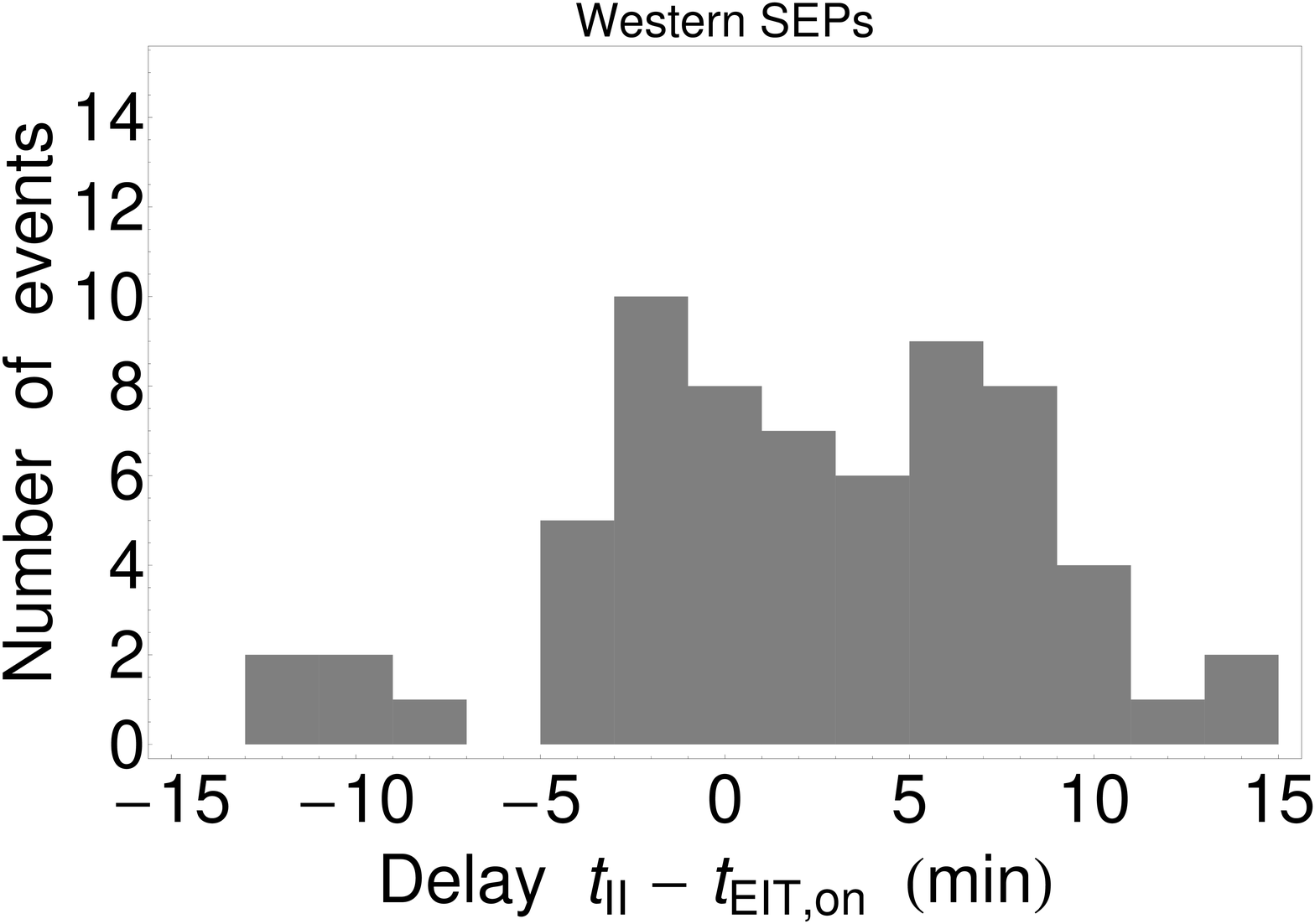}}
\caption{Histogram for the delay between the onset of the Type-II metric radio burst and the EIT wave onset, $t_{\rm II}-t_{\rm EIT,on}$ for eastern (left) and western events (right).}
   \label{F-DelayIIEIT}
   \end{figure}

The time difference $t_{\rm II}-t_{\rm EIT,on}$ is shown in Figure~\ref{F-DelayIIEIT}. The number of events is represented by the length of the bars in the histograms. A larger spread than the five minutes reported by \inlinecite{2010AdSpR..45..527W} is present. Due to the large uncertainty of $t_{\rm EIT,on}$ (of the order of the SOHO/EIT time cadence), the slight shift in the distributions towards positive delays (implying that the Type-II burst starts on average seven minutes after the onset of the EIT wave), is not significant. One can conclude that in a majority of the cases (88$\,$\%), at the time of the EIT disturbance a shock wave is also present in the corona.

\begin{figure}[t!]
\centerline{\includegraphics[width=0.5\textwidth,clip=]{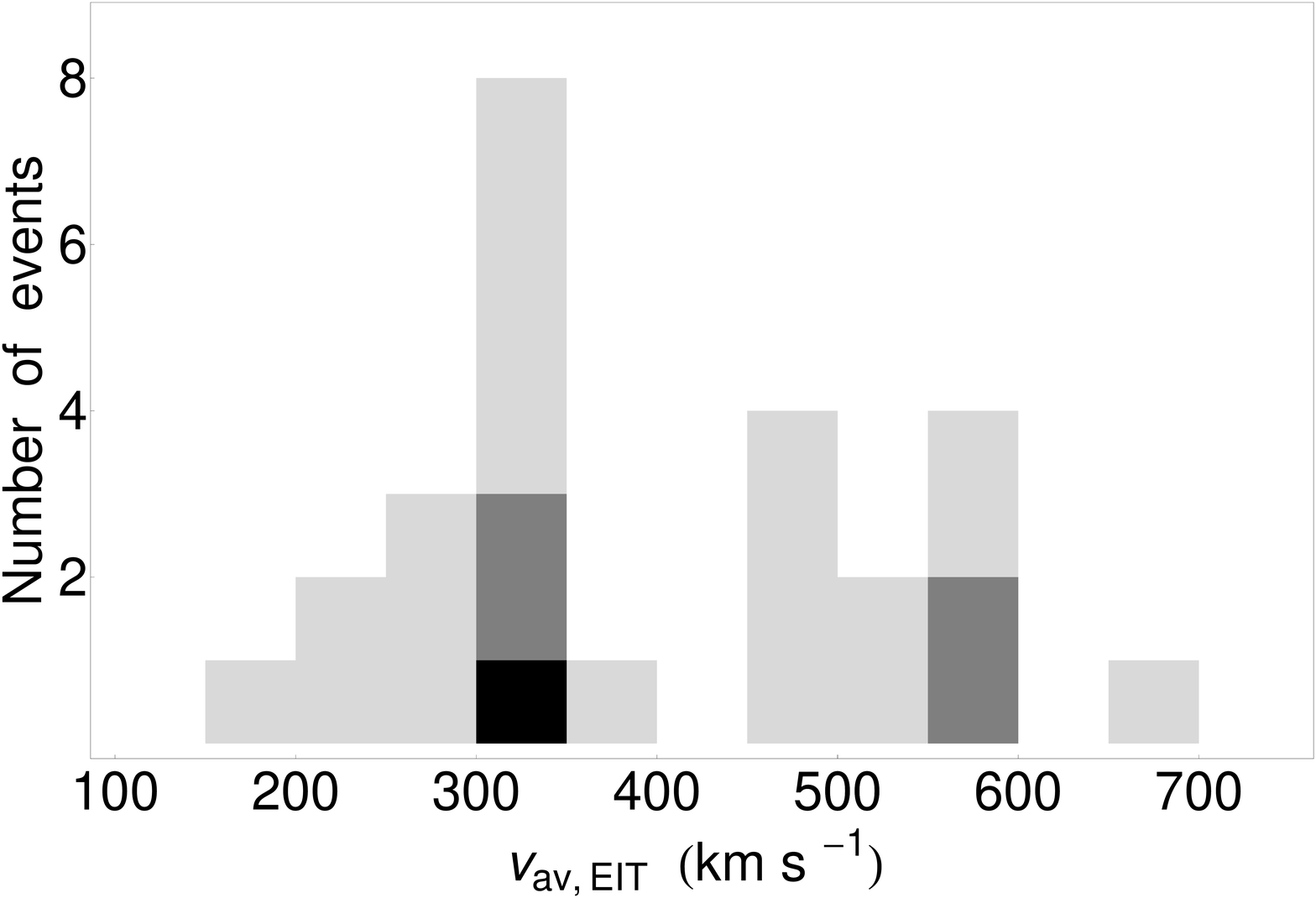}
            \hspace*{-0.05\textwidth}
            \includegraphics[width=0.5\textwidth,clip=]{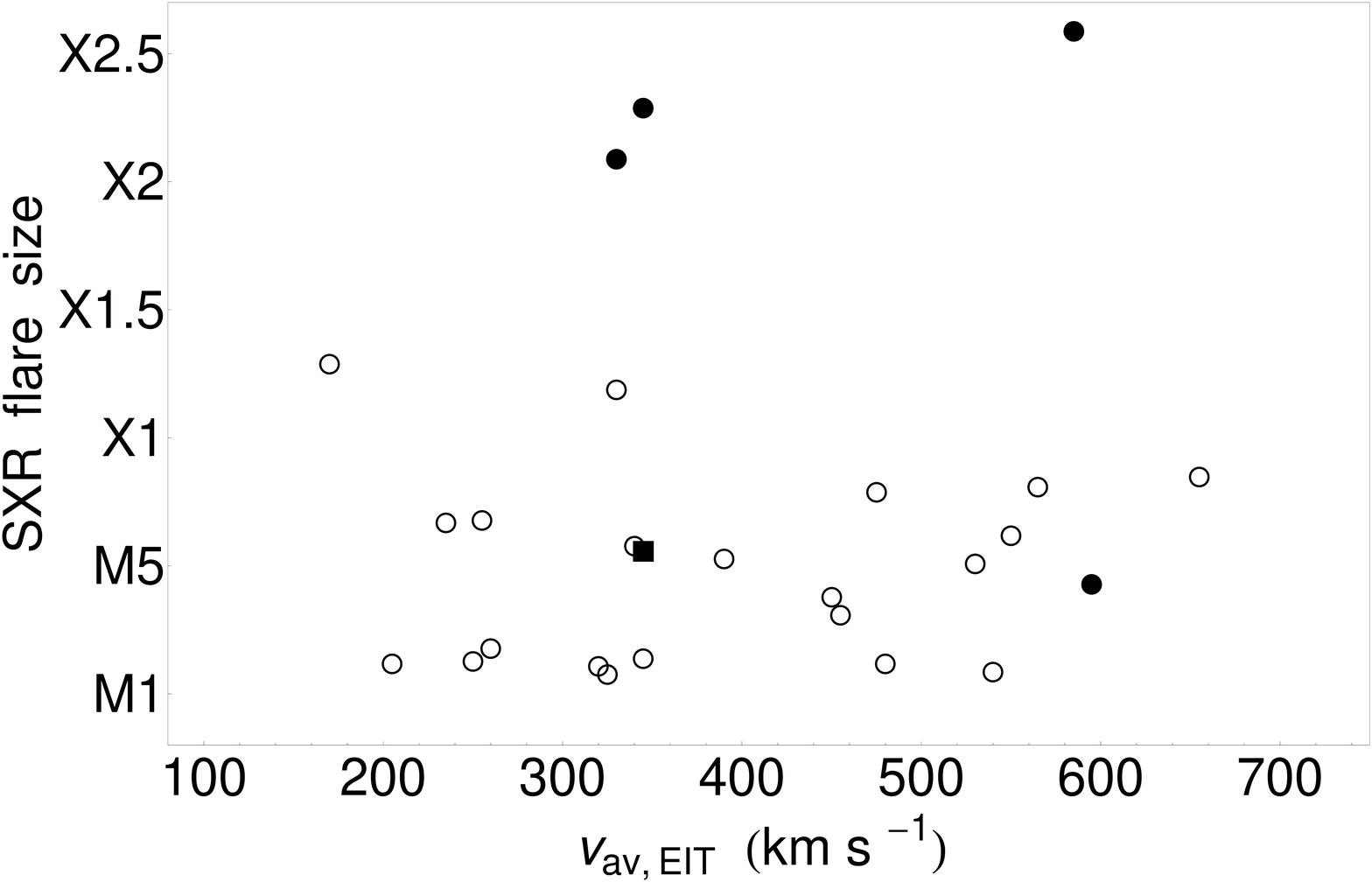}
}
             \vspace*{-0.01\textwidth}
\caption{Histogram of the EIT wave speed (left) and scatter plots between the EIT speed and the associated flare size (right). In the histogram, the light gray color denotes the solar-wind events, gray the ICME events and black the event in the vicinity of an ICME. The number of the events in each category is given by the bar length of specific color. In the scatter plot, the open circles denote the solar-wind events, filled circles the ICME events and the filled squares the events in the vicinity of an ICME.}
   \label{F-HistoEITspeed_flare}
   \end{figure}

For three events among the eastern events under study, only single (or uncertain) wave front(s) could be identified and hence no EIT wave speed could be obtained. Other three SEP events that are labeled with no EIT wave [N] may in fact be associated with a later EIT disturbance that accompanied another flare/CME. They are given with ``d'' in Table~\ref{T-East_events}. Since no conclusion can be made, these events are dropped from the present analysis. In summary, only 26 events were found with an associated EIT disturbance in at least two subsequent images. For those, the average EIT wave speed is in the range between 170 and 670 km$\,$s$^{-1}$. A histogram of the EIT wave speed is given in Figure~\ref{F-HistoEITspeed_flare}. The different colors indicate the events in different IP magnetic-field configuration, namely with light gray the solar-wind events, with dark gray the ICME events, and with black particle events propagating in the vicinity of an ICME. Recently, a kinematics classification was proposed by \inlinecite{2011A&A...532A.151W}, where all disturbances are divided into three groups: slow EIT signatures ($\lesssim$170 km$\,$s$^{-1}$, due to magnetic-field reconfiguration), signatures with constant speed (in the range of 170\,--\,320 km$\,$s$^{-1}$, interpreted as linear waves traveling at the local fast-mode speed), and fast disturbances ($\gtrsim$320 km$\,$s$^{-1}$, probably large amplitude waves or shocks). Under this classification the eastern EIT disturbances are all linear waves and/or shocks.

\begin{figure}[t!]
\centerline{\includegraphics[width=0.5\textwidth,clip=]{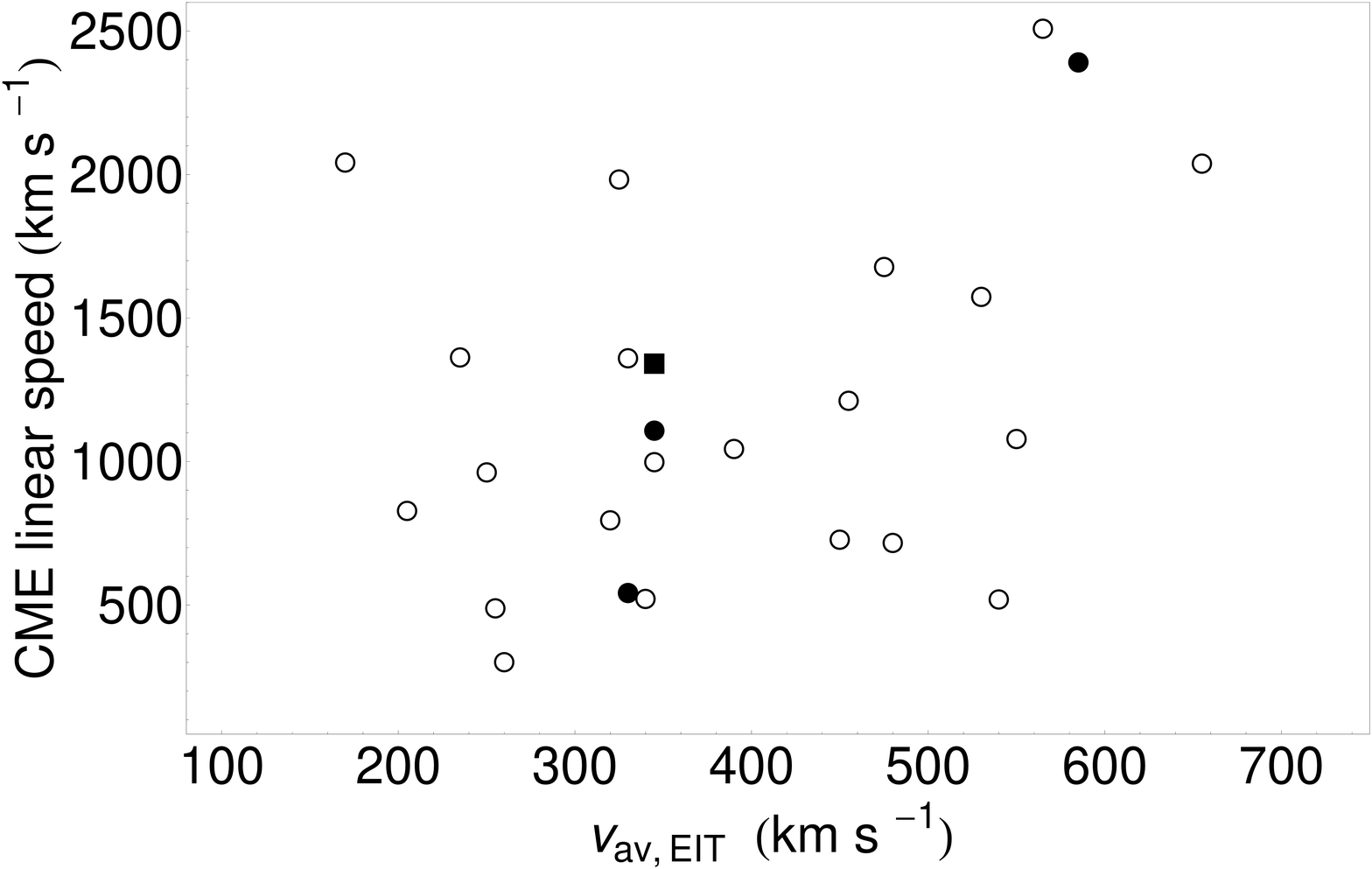}
            \hspace*{-0.05\textwidth}
            \includegraphics[width=0.5\textwidth,clip=]{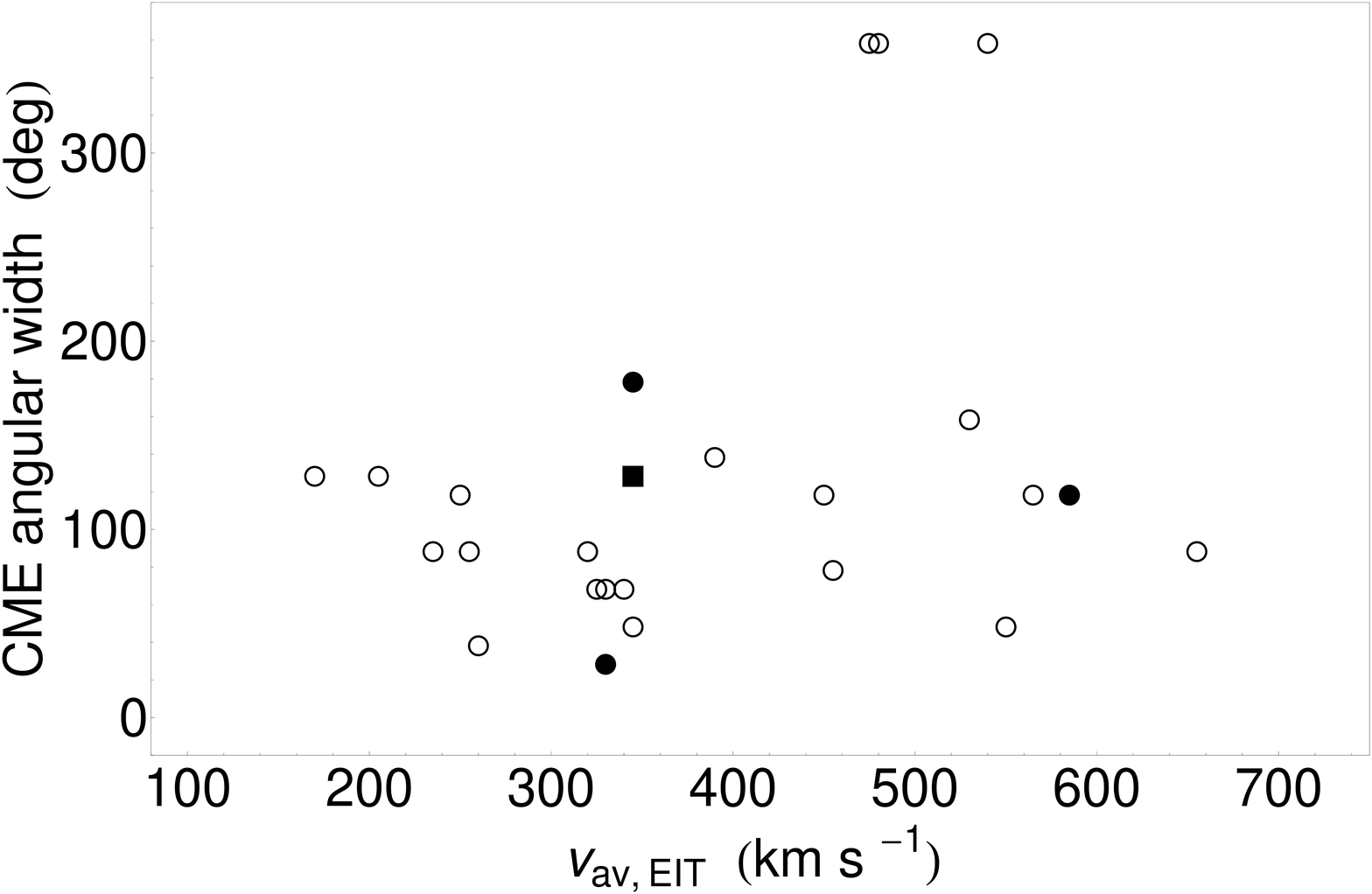}
}
             \vspace*{-0.01\textwidth}
\caption{Scatter plots between the averaged EIT speed and the associated CME linear speed (left) and angular width (right). The open circles denote the solar-wind events, filled circles the ICME events and the filled squares the events in the vicinity of an ICME.}
   \label{F-EITspeed_CME}
   \end{figure}

In addition, we present the scatter-plots between the propagation speed of the EIT disturbance and the properties of the associated flare SXR size (in Figure~\ref{F-HistoEITspeed_flare}, right) and CME linear speed and angular width (in Figure~\ref{F-EITspeed_CME}). We found that the averaged EIT wave speed is neither correlated with the flare, nor with the CME properties.

\acknowledgments
The authors thank the referee for her/his constructive comments that help us improve the paper. The exchange between the University of Graz and Paris Observatory was funded by the Austrian-French Programme {\rm Amadeus}/WTZ-\"{O}AD FR 17/2012. RM acknowledges a post-doctoral fellowship by the Paris Observatory. AV, IK and MT acknowledge the Austrian Science Fund (FWF): FWF P24092-N16 and FWF V195-N16. KLK acknowledges support from Centre National d'Etudes Spatiales (CNES) and the French Polar Institute (IPEV). Part of this work was done within the FP7-SEPServer and FP7-HESPE projects. Helpful discussions with A.~Klassen, B.~Klecker, G.~Mason, and M.~Wiedenbeck are acknowledged. The CME catalog is generated and maintained at the CDAW Data Center by NASA and the Catholic University of America in cooperation with the Naval Research Laboratory. SOHO is a project of international cooperation between ESA and NASA.

%%% BIBLIOGRAPHY %%%%%%%%%%%%%%%%%%%%%%%%%%%%%%%%%%%%%%%%%%%%%%%%%%%%%%%%%%%
\mbox{}~\\
     % format of references provided by the journal (.bst)
\bibliographystyle{spr-mp-sola}
%\bibliographystyle{spr-mp-sola-cnd} %% Alternative style: no title,
                                      % no concluding page.

     % name your Bibtex file containing your references (.bib)
\bibliography{sola_bibliography_example}

     % Checking: look if the file containing the ``\bibitem'' exits
     %           so check if the .bbl file exist (bibTeX compilation)
\IfFileExists{\jobname.bbl}{} {\typeout{}
\typeout{****************************************************}
\typeout{****************************************************}
\typeout{** Please run "bibtex \jobname" to obtain} \typeout{**
the bibliography and then re-run LaTeX} \typeout{** twice to fix
the references !}
\typeout{****************************************************}
\typeout{****************************************************}
\typeout{}}

\end{article}

\end{document}